\documentclass[twocolumn,aps,prl]{revtex4-1}

	\usepackage[usenames,dvipsnames]{xcolor}
	\usepackage{amsmath}
	\usepackage{amsfonts}
	\usepackage{amssymb}
	\usepackage[colorlinks=true,citecolor=blue,linkcolor=red]{hyperref}
	\usepackage{graphicx}
	\usepackage{bbold}					
	\usepackage[makeroom]{cancel}		
	\usepackage{multirow}				
	\usepackage[normalem]{ulem}        
	\usepackage{array}

	\newcolumntype{x}[1]{>{\centering\let\newline\\\arraybackslash\hspace{0pt}}p{#1}}

	\DeclareMathAlphabet{\mathbbold}{U}{bbold}{m}{n}

	\newcounter{subeqn} %
	\makeatletter
	\@addtoreset{subeqn}{equation}
	\makeatother

\definecolor{TB}{rgb}{1,0.5,0}
\definecolor{TB}{rgb}{0,0,0} 

\definecolor{ZX}{rgb}{0.4,0,1}
\definecolor{ZX}{rgb}{0,0,0}

\definecolor{HL}{rgb}{0.8,0,0.2}

\begin{document}
\title{Upper bound for the quantum coupling between free electrons and photons}

\date{\today}

\begin{abstract}
The quantum interaction between free electrons and photons is fundamental to free-electron based light sources and free-electron quantum optics applications. A large coupling between free electrons and photons is generally desired. In this manuscript, I obtain the upper bound for the quantum coupling between free electrons and photons. The upper bound has a straightforward expression and can be applied to a broad range of optical materials, especially widely used low-loss photonic materials.
The upper bound depends on the optical medium, the free-electron velocity, and the separation between the free electron and the optical medium. With simple structures, the numerically calculated coupling coefficient can reach $\sim$99\% of the upper bound.
This study provides simple and practical guidance to reach the strong coupling between free electrons and photons. 
\end{abstract}

\author{Zhexin Zhao$^1$}
\email[]{zhexin.zhao@fau.de}
\affiliation{$^1$Department of Physics, Friedrich-Alexander University (FAU) Erlangen-Nürnberg, Staudtstraße 1, 91058 Erlangen, Germany}

\maketitle

\emph{Introduction.--} The interaction between swift electrons and electromagnetic fields has attracted considerable research attention. Photon-induced near-field electron microscopy (PINEM) \cite{barwick2009photon} has been used to observe photonic \cite{wang2020coherent, kfir2020controlling}, plasmonic \cite{piazza2015simultaneous}, and polariton excitations \cite{kurman2021spatiotemporal}, utilizing the nanometer spatial resolution and femtosecond temporal resolution of the free-electron probe. 
Significant progress has been made in understanding and engineering the free-electron--light interaction.
In typical PINEM, where the light field is strong, the quantum effects can be manifested by describing the free electron quantum mechanically while treating the electromagnetic field classically
\cite{garcia2010multiphoton, park2010photon, feist2015quantum, pan2018spontaneousclassical, shiloh2022quantum}. Researchers have also studied the transition between the quantum effects in PINEM and the classical electron acceleration/deceleration \cite{zhou2019quantum, pan2019anomalous}.
Recently, the theory has been extended to treat both the electron and the photon quantum mechanically using quantum electrodynamics (QED) \cite{di2019probing, kfir2019entanglements, pan2019spontaneous, remez2019observing}. 
Moreover, macroscopic quantum electrodynamics (MQED) \cite{gruner1996green, dung1998three, rivera2020light}  has been applied to the study of free-electron--light interactions \cite{di2020electron, kfir2021optical, huang2023electron}, which can describe the interaction between free electrons and photonic excitations with a continuous spectrum.

Interesting physics happens when the free-electron--light interaction reaches the strong coupling regime \cite{kfir2019entanglements, ben2021shaping}, which can enable quantum applications including free-electron based photon sources \cite{feist2022cavity, huang2023electron} and quantum computing using free electrons \cite{karnieli2023jaynes, dahan2023creation, baranes2023free, karnieli2024PRXQuantum}. To reach the strong coupling regime, it is important to understand the ultimate limit of the free-electron--light coupling \cite{yang2018maximal}. In this manuscript, I present the upper bound of the quantum coupling coefficient that quantifies the coupling between free electrons and photons. 

In this manuscript, I first recap the Hamiltonian and scattering matrix description of the free-electron--light interaction, including two cases: (1) when the photonic excitations have a continuous spectrum as in the general case, and (2) when the photonic modes have a discrete spectrum as in the case of a lossless optical cavity. Then, I present the upper bound for the quantum coupling coefficient in both cases. I also discuss the connection between these two cases when the optical loss, including absorption and radiation, is low. I present numerical demonstration of the upper bounds and discuss its implication on the choice of the optical medium, the electron velocity, and the separation distance to achieve strong coupling.

\begin{figure}
    \centering
    \includegraphics[width=7cm]{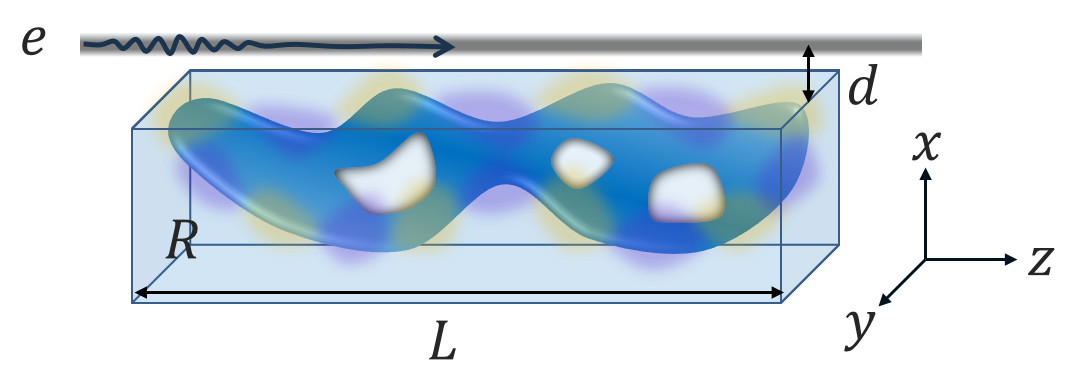}
    \caption{Illustration of the free-electron--light interaction. The optical medium (dark blue) is enclosed within a design region \textit{R} (light blue) with a minimal separation \textit{d} from the electron trajectory. The interaction length is \textit{L}. The yellow and purple coloring illustrates an optical mode.}
    \label{fig:schematic}
\end{figure}

\emph{The quantum coupling coefficient.--}  
Here, I \textit{recap} the Hamiltonian and scattering matrix that describe the quantum interaction between free electrons and photons, in the framework of MQED \cite{di2019probing, di2020electron, kfir2021optical, huang2023electron, gruner1996green, dung1998three, rivera2020light}, where the system is illustrated in Fig.\,\ref{fig:schematic}. 
The Hamiltonian describing the free-electron--light interaction is \cite{kfir2021optical, huang2023electron}:
\begin{equation}
    \label{eq:H}
    \hat{H} = \hat{H}_0 + \hat{V} = \hat{H}_p + \hat{H}_e + \hat{V},
\end{equation}
where $\hat{H}_p$ is the Hamiltonian of photonic excitations, $\hat{H}_e$ is the Hamiltonian of the free electron, $\hat{H}_0 = \hat{H}_p + \hat{H}_e$, and $\hat{V}$ describes the interaction. 

In MQED, quantized harmonic oscillators are assigned to each position, orientation and frequency, from which one can obtain the current operator and field operators \cite{gruner1996green, dung1998three, rivera2020light} (Supplemental Material (SM) Sec.\,I).
The Hamiltonian of the photonic excitation is
\begin{equation}
    \label{eq:H_p}
    \hat{H}_p = \int d^3 \boldsymbol{r} \int_0^{+\infty} d\omega \hbar \omega \hat{f}^\dagger_i(\boldsymbol{r},\omega) \hat{f}_i(\boldsymbol{r},\omega),
\end{equation}
where $\hat{f}_i^\dagger(\boldsymbol{r},\omega)$ and $\hat{f}_i(\boldsymbol{r},\omega)$ are creation and annihilation operators of the quantized harmonic oscillators. $\boldsymbol{r}$, $\omega$, and $i\in\{x,y,z\}$ represent position, frequency and orientation respectively. The repeated index $i$ will be summed up.

To describe the free electron, the non-recoil assumption is used \cite{kfir2019entanglements}. The electron is assumed to travel along z-direction (Fig.\,\ref{fig:schematic}), where the longitudinal dispersion relation is approximately linear near the reference energy $E_0$ and the reference momentum $\hbar k_0$, and the free-electron transverse wave function ($\phi_e(\boldsymbol{r}_\perp)$) is approximately unchanged. 
The free-electron Hamiltonian (SM Sec.\,I) is
\begin{equation}
    \label{eq:H_e}
    \hat{H}_e = \sum_k [E_0 + \hbar v (k-k_0)] \hat{c}_k^\dagger \hat{c}_k,
\end{equation}
where $v$ is the electron velocity, and $\hat{c}_k$ is the fermionic annihilation operator associated with the free-electron wave vector $k$ \cite{huang2023electron}.

The interaction Hamiltonian is \cite{di2019probing, huang2023electron} (SM Sec.\,I)
\begin{equation}
    \label{eq:V}
    \begin{split}
        \hat{V} = &\; - \int d^3\boldsymbol{r} \boldsymbol{\hat{J}}(\boldsymbol{r}) \cdot \boldsymbol{\hat{A}}(\boldsymbol{r}) \\ = & \; -\frac{q_e v}{L}\int d^3\boldsymbol{r} \sum_{k, q}e^{iqz} \hat{c}^\dagger_k \hat{c}_{k+q} \phi_e^*(\boldsymbol{r}_\perp) \phi_e(\boldsymbol{r}_\perp) \\ &\; \int_0^{+\infty} d\omega \big[\hat{A}_z(\boldsymbol{r},\omega) + H.C.\big],
    \end{split}
\end{equation}
where $q_e$ is the electron charge, $H.C.$ stands for Hermitian conjugate. $\boldsymbol{r}_\perp$ is the transverse position, and $\boldsymbol{r} = \boldsymbol{r}_\perp + z \boldsymbol{\hat{z}}$, where $\boldsymbol{\hat{z}}$ is a unit vector in the z-direction. As established in MQED, the quantized harmonic oscillators can drive the fields, such that the vector potential $\boldsymbol{\hat{A}}$ in the Coulomb gauge is connected with $\boldsymbol{\hat{f}}$ via \cite{gruner1996green, dung1998three, rivera2020light}
\begin{equation}
    \label{eq:A_operator_freq}
    \hat{A}_i(\boldsymbol{r},\omega) = \sqrt{\frac{\hbar}{\pi\epsilon_0}} \frac{\omega}{c^2} \int d^3 \boldsymbol{s} G_{ij}(\boldsymbol{r}, \boldsymbol{s},\omega)\sqrt{\epsilon_I(\boldsymbol{s},\omega)} \hat{f}_{j}(\boldsymbol{s},\omega),
\end{equation}
where $G_{ij}(\boldsymbol{r}, \boldsymbol{s}, \omega)$ is the Green's function, $\epsilon_I(\boldsymbol{s},\omega)$ is the imaginary part of the relative permittivity. I assume an isotropic non-magnetic and local medium, since most photonic systems belong to this category. 

The scattering matrix describing the interaction between free electrons and a general photonic system (SM Sec.\,I) is
\begin{equation}
    \label{eq:S_2}
    \hat{S} = e^{i\hat{\chi}} e^{\int_0^{+\infty}d\omega \big[ \hat{b}_{\frac{\omega}{v}}^\dagger \int dz \frac{i q_e}{\hbar} e^{-i\frac{\omega}{v}z}  \hat{A}_z(\boldsymbol{r}_{e\perp}+z\boldsymbol{\hat{z}}, \omega) -H.C.\big]},
\end{equation}
where the first term $\exp{(i\hat{\chi})}$ is a phase operator acting only on the free electron \cite{di2020electron}, and $\hat{b}$ is the electron ladder operator \cite{kfir2019entanglements, zhao2021quantum} 
\begin{equation}
    \label{eq:b_operator}
    \hat{b}_{q} = \sum_k \hat{c}^\dagger_k \hat{c}_{k+q}.
\end{equation}
For simplicity, I assume the transverse free-electron wave function is centered around $\boldsymbol{r}_{e\perp}$ and its spread is small, within which the vector potential is almost unchanged. (The influence of transverse distribution is discussed in the SM Sec.\,IV.)

When absorption and dispersion are negligible, the photonic structure can support discrete eigenmodes. In this case \cite{kfir2019entanglements}, the vector potential is 
\begin{equation}
    \label{eq:A_m_freq}
    \boldsymbol{\hat{A}}(\boldsymbol{r},\omega) = \sum_m \sqrt{\frac{\hbar}{2\omega_m\epsilon_0}} \boldsymbol{U}_m(\boldsymbol{r}) \hat{a}_m \delta(\omega - \omega_m),
\end{equation}
where $\omega_m$ is the eigenmode frequency, $\hat{a}_m$ is the annihilation operator for mode m, and $\boldsymbol{U}_m(r)$ is the normalized eigenmode distribution, such that
\begin{equation}
    \label{eq:Um_1}
    \nabla\times\nabla\times \boldsymbol{U}_m(\boldsymbol{r})  - \frac{\omega_m^2}{c^2} \epsilon(\boldsymbol{r}, \omega_m) \boldsymbol{U}_m(r) = 0,
\end{equation}
\begin{equation}
    \label{eq:Um_normalizatin}
    \int d^3 \boldsymbol{r} \epsilon(\boldsymbol{r}, \omega_m) U_{m,i}(\boldsymbol{r})U_{n,i}^*(\boldsymbol{r}) = \delta_{mn}.
\end{equation}
The scattering matrix becomes
\begin{equation}
    \label{eq:S_mode_m}
    \hat{S} = e^{i\hat{\chi}}e^{\sum_m g_{Qu,m}\hat{b}_\frac{\omega_m}{v}^\dagger  \hat{a}_m - H.C.},
\end{equation}
which is consistent with previous results in \cite{kfir2019entanglements}.
The quantum coupling coefficient for mode m is
\begin{equation}
    \label{eq:gQu_mode_m}
    g_{Qu,m} = \frac{i q_e}{\sqrt{2\hbar\omega_m\epsilon_0}} \int dz e^{-i\frac{\omega_m}{v}z} U_{m,z}(\boldsymbol{r}_{e\perp}+z\boldsymbol{\hat{z}}).
\end{equation}
Equation \ref{eq:gQu_mode_m} shows that the eigenmode plays an essential role in the quantum coupling coefficient. When the interaction length is long, the eigenmode should have a z-dependence of $\exp(i\frac{\omega_m}{v} z)$, which is referred to as the phase-matching condition \cite{kfir2019entanglements, kfir2021optical, henke2021integrated}.

The general scattering matrix (Eq.\,\ref{eq:S_2}) can be re-formulated in the following form \cite{huang2023electron}:
\begin{equation}
    \label{eq:S_omega}
    \hat{S} = e^{i\hat{\chi}}e^{\int_0^{+\infty} d\omega \big (g_{Qu}(\omega) \hat{b}^\dagger_{\frac{\omega}{v}} \hat{a}_\omega - H.C. \big)},
\end{equation}
where $g_{Qu}(\omega)$ is the quantum coupling coefficient when the optical excitation spectrum is continuous.
\begin{equation}
    \label{eq:gQu_omega_a}
    g_{Qu}(\omega)\hat{a}_\omega = \frac{iq_e}{\hbar}\int dz e^{-i\frac{\omega}{v}z}\hat{A}_z(\boldsymbol{r}_{e\perp}+z\hat{z},\omega).
\end{equation}
The explicit form of $g_{Qu}(\omega)$ is obtained by imposing the commutation relation $[\hat{a}_\omega, \hat{a}_{\omega'}^\dagger] = \delta(\omega - \omega')$ \cite{huang2023electron} (SM Sec.\,I). It has been shown that $|g_{Qu}(\omega)|^2$ is identical to the electron energy loss spectrum (EELS) $\Gamma(\omega)$ \cite{de2010optical, ritchie1957plasma, ritchie1988inelastic, di2021modulation}.
\begin{equation}
    \label{eq:gQu_omega_2}
    \begin{split}
    &|g_{Qu}(\omega)|^2 \\ = & \;\frac{q_e^2\omega^2}{\hbar \pi \epsilon_0 c^4} \int dz \int dz' e^{i\frac{\omega}{v}(z'- z)} \int d^3 \boldsymbol{s} \epsilon_I(\boldsymbol{s}, \omega)\\ & \;  G_{zi}(\boldsymbol{r}_{e\perp} + z\boldsymbol{\hat{z}}, \boldsymbol{s}, \omega)  G^*_{zi}(\boldsymbol{r}_{e\perp} + z'\boldsymbol{\hat{z}}, \boldsymbol{s}, \omega) \\
    = & \;\frac{q_e^2}{\hbar \pi \epsilon_0 c^2}  \int dz \int dz' \text{Re}\Big[ -i e^{i\frac{\omega}{v}(z'- z)}  \\ & \; G_{zz}(\boldsymbol{r}_{e\perp} + z\boldsymbol{\hat{z}}, \boldsymbol{r}_{e\perp} + z'\boldsymbol{\hat{z}},\omega) \Big] .
    \end{split}
\end{equation}

\emph{Upper bound of the quantum coupling coefficient.--}
I derive and discuss the upper bound for the quantum coupling coefficient with a discrete mode ($|g_{Qu,m}|^2$) and with modes in a continuum ($|g_{Qu}(\omega)|^2$). 

For the quantum coupling coefficient with a discrete mode, I rewrite (Eq.\,\ref{eq:gQu_mode_m}) using $\boldsymbol{E}_m$, an eigenmode distribution without normalization, since the magnitude of $\boldsymbol{E}_m$ does not affect $|g_{Qu, m}|^2$.
\begin{equation}
    \label{eq:gQu_mode_m_2}
    |g_{Qu, m}|^2 = \frac{q_e^2}{2\hbar\omega_m \epsilon_0} \frac{ |\int dz e^{-i\frac{\omega_m}{v}z} E_{m,z}(\boldsymbol{r}_{e\perp}+z\boldsymbol{\hat{z}})|^2 }{ \int d^3 \boldsymbol{r} \boldsymbol{E}_m^{\dagger}(\boldsymbol{r}) \epsilon(\boldsymbol{r}) \boldsymbol{E}_m(\boldsymbol{r}) }.
\end{equation}
The electric field is related to the polarization field using the free-space Green's function ($G_0$) \cite{miller2016fundamental}:
\begin{equation}
    \label{eq:E_G0_Einc}
    \boldsymbol{E}_m(\boldsymbol{r}) = \boldsymbol{E}_\text{inc}(\boldsymbol{r}) + \frac{\omega_m^2}{c^2 \epsilon_0} \int d^3\boldsymbol{s} G_0(\boldsymbol{r},\boldsymbol{s},\omega_m)\boldsymbol{P}_m(\boldsymbol{s}),
\end{equation}
where $\boldsymbol{E}_\text{inc}$ is the incident electric field, and $\boldsymbol{P}_m$ is the polarization field.
Since $\boldsymbol{E}_m$ is the eigenmode, $\boldsymbol{E}_\text{inc}(r)=\boldsymbol{0}$. Substituting Eq.\,\ref{eq:E_G0_Einc} into Eq.\,\ref{eq:gQu_mode_m_2} and integrating over $z$ in the numerator gives
\begin{equation}
    \label{eq:gQu_mode_m_3}
    |g_{Qu, m}|^2 = \frac{\omega_m}{2\hbar \epsilon_0} \frac{ \big|\int d^3 \boldsymbol{s} \boldsymbol{E}_{e0}^\dagger (\boldsymbol{s}, \omega_m) \boldsymbol{P}_m(\boldsymbol{s})\big|^2 }{ \int d^3 \boldsymbol{r}  \boldsymbol{E}_m^{\dagger}(\boldsymbol{r}) \epsilon(\boldsymbol{r}) \boldsymbol{E}_m(\boldsymbol{r}) },
\end{equation}
where $\boldsymbol{E}_{e0}$ is the electric field associated with the free electron \cite{de2010optical, tsang2000scattering} obtained as following
\begin{equation}
    \label{eq:Ee0}
    \begin{split}
    \boldsymbol{E}_{e0}(\boldsymbol{s}, \omega) & = i\omega \mu_0 q_e \int dz G_0(\boldsymbol{s}, \boldsymbol{r}_{e\perp}+z\boldsymbol{\hat{z}}, \omega)\boldsymbol{\hat{z}}  \exp\Big(i\frac{\omega}{v}z\Big) \\ & = -\frac{q_e e^{ik_e s_z}}{2\pi \epsilon_0 \omega} \big[ i\alpha_e^2 K_0(\alpha_e \rho)\boldsymbol{\hat{z}} - k_e \alpha_e K_1( \alpha_e \rho)\boldsymbol{\hat{\rho}} \big].
    \end{split}
\end{equation}
Here, $k_e = \omega/v$, $k = \omega/c$, $\alpha_e = \sqrt{k_e^2 - k^2}$, $\rho = |\boldsymbol{s}_\perp - \boldsymbol{r}_{e\perp}|$, $\boldsymbol{\hat{\rho}} = (\boldsymbol{s}_\perp - \boldsymbol{r}_{e\perp})/|\boldsymbol{s}_\perp - \boldsymbol{r}_{e\perp}|$, and $K_0$ and $K_1$ are the modified Bessel's functions of the second kind with order 0 and 1 respectively. 
Using $\boldsymbol{P}_m(\boldsymbol{r}) = \epsilon_0\chi(\boldsymbol{r})\boldsymbol{E}_m(\boldsymbol{r})$, where $\chi$ is the susceptibility,
\begin{equation}
    \label{eq:gQu_mode_m_f}
    |g_{Qu, m}|^2 = \frac{\omega_m \epsilon_0}{2\hbar} \frac{ \big|\int d^3 \boldsymbol{r} \boldsymbol{E}_{e0}^\dagger (\boldsymbol{r}, \omega_m) \chi(\boldsymbol{r}) \boldsymbol{E}_m(r)\big|^2 }{\int d^3 \boldsymbol{r} \boldsymbol{E}_m^{\dagger}(\boldsymbol{r}) \epsilon(\boldsymbol{r}) \boldsymbol{E}_m(\boldsymbol{r}) }.
\end{equation}
Using the Cauchy-Schwarz inequality ($\boldsymbol{a} = \epsilon^{-\frac{1}{2}}\chi \boldsymbol{E}_{e0}(\omega_m)$, $\boldsymbol{b}=\epsilon^{\frac{1}{2}}\boldsymbol{E}_m$, $|\boldsymbol{a}^\dagger \boldsymbol{b}|^2 / \boldsymbol{b}^\dagger \boldsymbol{b} \leq \boldsymbol{a}^\dagger \boldsymbol{a}$), I find an analytical upper bound for $|g_{Qu,m}|^2$ (SM Sec.\,II)
\begin{equation}
    \label{eq:gQu_mode_m_upper}
    |g_{Qu, m}|^2 \leq \frac{\omega_m\epsilon_0}{2\hbar}  \frac{|\chi|^2}{\epsilon} \int_R d^3 \boldsymbol{r} \boldsymbol{E}_{e0}^\dagger (\boldsymbol{r}, \omega_m) \boldsymbol{E}_{e0} (\boldsymbol{r}, \omega_m),
\end{equation}
where $\frac{|\chi|^2}{\epsilon} = \max_{\boldsymbol{r}} \frac{|\chi(\boldsymbol{r})|^2}{\epsilon(\boldsymbol{r})}$, \textit{R} emphasizes that the integration is over the minimal region containing the photonic structure (Fig.\,\ref{fig:schematic}). 
Using the explicit form for $\boldsymbol{E}_{e0}$ (Eq.\,\ref{eq:Ee0}) and assuming that the interaction length is $L$ and the design region \textit{R} is expanded by a transverse cross section $R_\perp$ and a longitudinal length $L$, the upper bound for $|g_{Qu,m}|^2$, denoted as $g_\text{ub}^2$, is:
\begin{equation}
    \label{eq:gQu_mode_m_upper_2}
    \begin{split}
    |g_{Qu, m}|^2 < g_\text{ub}^2 \equiv & \; \frac{q_e^2}{4\pi\hbar c \epsilon_0} \frac{|\chi|^2}{\epsilon} \frac{kL}{2\pi} \int_{R_\perp} d^2 \boldsymbol{r}_\perp \\ & \;\Big[\frac{\alpha_e^4}{k^2} K_0^2(\alpha_e \rho)  + \frac{k_e^2 \alpha_e^2}{k^2} K_1^2(\alpha_e \rho)\Big].
    \end{split}
\end{equation}

Equation \ref{eq:gQu_mode_m_upper_2} is the main result of this study. The first term is the fine structure constant. The second term describes the dependence on the optical medium. It shows that higher index photonic resonators can increase the upper bound. When the photonic medium is anisotropic or have Lorentz dispersion, only this term in the upper bound is different (SM Sec.\,II). 
The third term is the scaling with the interaction length \cite{kfir2019entanglements}. The integral, referred to as the geometric factor ($g^2_\text{geo} = \int_{R_\perp} d^2 \boldsymbol{r}_\perp \big[\frac{\alpha_e^4}{k^2} K_0^2(\alpha_e \rho) + \frac{k_e^2 \alpha_e^2}{k^2} K_1^2(\alpha_e \rho)\big]$), is a unit-less number that depends on the electron velocity and the separation between the electron beam and the photonic structure (Fig.\,\ref{fig:schematic}). 
Besides the dependence on material permittivity, the upper bound is scale invariant, i.e., it is invariant when the geometric parameters are scaled by a same factor as the free-space wavelength. Thus, it is straightforward to apply this upper bound to the free-electron--light interaction at arbitrary frequencies. 

\begin{figure}
    \centering
    \includegraphics[width=8.6cm]{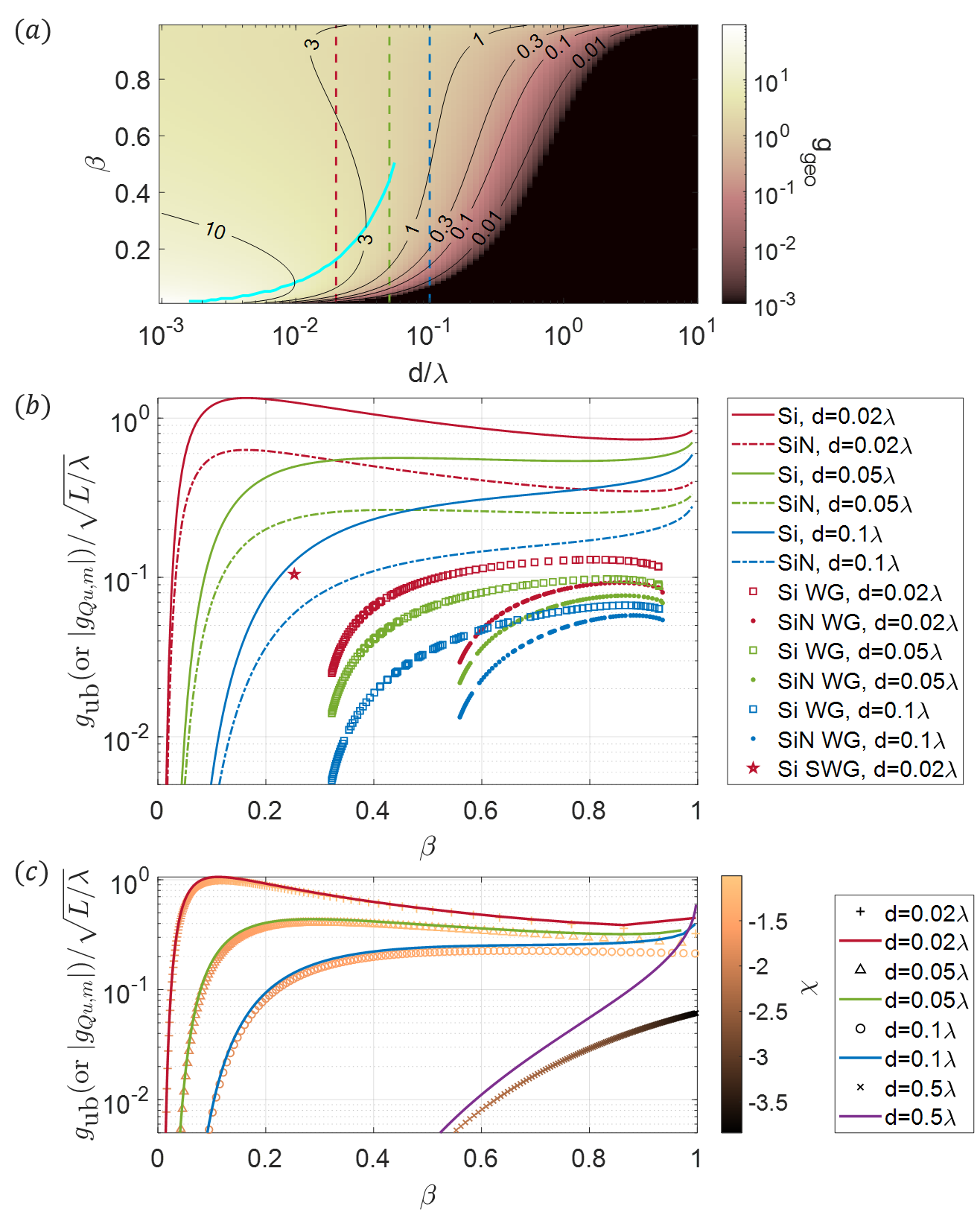}
    \caption{(a) $g_\text{geo}$ as a function of $d$ and $\beta$. Only $g_\text{geo} > 10^{-3}$ is shown. The cyan curve represents the sub-relativistic peak for each $d$. The red, green and blue dashed lines represent $d=0.02\lambda$, $0.05\lambda$, and $0.1\lambda$ respectively, where $\lambda=2\pi/k$. (b) The upper bound $g_\text{ub}$ normalized by the interaction length, as a function of $\beta$ for Si (solid curves) or SiN (dash-dot curves) with separation $d=0.02\lambda$ (red), $0.05\lambda$ (green), or $0.1\lambda$ (blue). The squares and dots represents interaction length normalized $|g_{Qu,m}|$ for Si and SiN waveguides respectively. The red star represents that for a Si SWG. (c) Interaction length normalized $|g_{Qu,m}|$ (markers) and the corresponding $g_\text{ub}$ (solid curves), for the SPP modes in a metallic hole with radius $d=0.02\lambda$ (red), $0.05\lambda$ (green), $0.1\lambda$ (blue), or $0.5\lambda$ (purple). The metal susceptibility is $\chi(\omega)=-\omega_p^2/\omega^2$ in the negligible absorption limit, shown by the marker color.}
    \label{fig:g_geo}
\end{figure}

In deriving the analytical upper bound for $|g_{Qu,m}|^2$, the Cauchy-Schwarz inequality is saturated when $\boldsymbol{E}_m(\boldsymbol{r})\propto \frac{\chi(\boldsymbol{r})}{\epsilon(\boldsymbol{r})}\boldsymbol{E}_{e0}(\boldsymbol{r}, \omega_m)$. This indicates the condition to approach the analytical upper bound. However, such an `optimal' field is typically not a solution to the Maxwell's equations.
Nevertheless, physical field distribution close to such an optimum can approach the upper bound.
With modes in dielectric hollow-core waveguides, $|g_{Qu,m}|$ can reach about 70\% of the upper bound (SM Sec.\,VII). Moreover, if the photonic medium has Lorentz dispersion, where $\epsilon(\omega) = \epsilon_B[1 + \omega_p^2/(\omega_0^2-\omega^2-i\omega\gamma_L)]$, and in the limit of negligible material absorption ($\gamma_L\rightarrow 0$), the upper bound for $|g_{Qu,m}|^2$ is (SM Sec.\,II)
\begin{equation}
    \label{eq:gQu_ub_Lorentz}
    |g_{Qu,m}|^2 < \frac{q_e^2}{2\pi\hbar c \epsilon_0} \frac{\big|\chi_B+\epsilon_B\frac{\omega_p^2}{\omega_0^2 - \omega_m^2}\big|^2}{\epsilon_B + \epsilon_B \frac{(\omega_0^2+\omega_m^2)\omega_p^2}{(\omega_0^2-\omega_m^2)^2}}\frac{kL}{2\pi} g_\textrm{geo}^2.
\end{equation}
With surface plasmon polariton (SPP) modes in a metallic hole, where the material dispersion is $\epsilon(\omega)=1 - \omega_p^2/(\omega^2+i\omega\gamma_L)$, $|g_{Qu,m}|$ can reach over 99\% of the upper bound in the limit of $\gamma_L\rightarrow 0$ (Fig.\,\ref{fig:g_geo}(c) and SM Sec.\,VII), which implies the tightness of the analytical upper bound.
 
The upper bound of $|g_{Qu}(\omega)|^2$, which is identical to the maximal free-electron
energy loss spectral probability, has been derived in \cite{yang2018maximal}. For completeness, its explicit form is shown here (derivation in SM Sec.\,II):
\begin{equation}
    \label{eq:gQu_omega_bound}
    \begin{split}
    |g_{Qu}(\omega)|^2 \leq & \; \frac{q_e^2}{4\pi \hbar \epsilon_0 c} \frac{2}{\pi\omega} \Bigg[ \frac{|\chi|^2}{\chi_I}\Bigg]_\omega \frac{kL}{2\pi} g_\textrm{geo}^2.
    \end{split}
\end{equation}

Next, I discuss the connection between $g_{Qu}(\omega)$ and $g_{Qu,m}$, and the connection between their upper bounds, when the material loss and dispersion are small. In this case, the Green's function can be decomposed with the eigenmodes, and I assume the decay rate ($\gamma_d$) of the eigenmodes is small (SM Sec. III).
\begin{equation}
    \label{eq:G_freq_mode_decomp}
    G(\boldsymbol{r}, \boldsymbol{r}',\omega) = \sum_m \frac{c^2}{-\omega^2 - i\omega\gamma_d + \omega_m^2}\boldsymbol{U}_m(\boldsymbol{r}) \boldsymbol{U}_m^\dagger(\boldsymbol{r}').
\end{equation}
Substitute this model decomposed Green's function (Eq.\,\ref{eq:G_freq_mode_decomp}) into Eq.\,\ref{eq:gQu_omega_2}, I get
\begin{equation}
    \label{eq:gQu_omega_mode}
    |g_{Qu}(\omega)|^2 \approx \frac{q_e^2}{4 \pi\hbar \epsilon_0 } \sum_m \frac{\gamma_d}{\omega_m} \frac{ \Big| \int dz e^{-i\frac{\omega}{v}z}U_{m,z}(\boldsymbol{r}_\perp+z\boldsymbol{\hat{z}})\Big|^2}{(\omega-\omega_m)^2 + \frac{\gamma_d^2}{4}}.
\end{equation}
Equation \ref{eq:gQu_omega_mode} indicates that $|g_{Qu}(\omega)|$ contains peaks centered around eigen frequencies ($\omega_m$). 
When these peaks are separated in frequency, such that the line width of each peak is much smaller than the peak separation and the eigenmodes are non-degenerate, the integration of the peak around $\omega_m$ is (SM Sec.\,III)
\begin{equation}
    \label{eq:gQu_omega_gQu_relation}
    \int_{\omega_m-\frac{\delta\omega}{2}}^{\omega_m+\frac{\delta\omega}{2}} d\omega |g_{Qu}(\omega)|^2 \approx |g_{Qu,m}|^2,
\end{equation}
where the bandwidth $\delta\omega$ covers \textit{only} the peak around $\omega_m$. This connection (Eq.\,\ref{eq:gQu_omega_gQu_relation}) shows that although a low absorption and high quality factor increase the on-resonant coupling coefficient ($|g_{Qu}(\omega_m)|^2$), the integration over the resonant bandwidth is determined by the model coupling coefficient $|g_{Qu,m}|^2$. 
Furthermore, the on-resonance $|g_{Qu}(\omega_m)|^2$ has a resonant enhancement $~1/\gamma_d$. Since $\gamma_d$ is the ratio between the absorption power and the stored energy, which scales with $\chi_I$ and $\epsilon$ respectively, the scaling of $\gamma_d$ with material response is $\gamma_d \sim \frac{\omega \chi_I}{\epsilon}$ when the absorption dominates the loss. This explains the different material dependences in the upper bounds for $|g_{Qu}(\omega)|^2$ and $|g_{Qu,m}|^2$. 

When multiple modes are involved in the free-electron--light interaction and the total effect is of interest, one needs to sum up the coupling ($|g_{Qu,m}|^2$) from multiple modes. Furthermore, when free electrons interact with extended optical systems, such as waveguides, the coupling with a spatial-temporal mode also depends on the effective number of modes (SM Sec.\,V), such that dispersion engineering becomes important \cite{karnieli2024strong}.

\emph{Numerical demonstration.--}
I show numerical examples for the upper bound of $|g_{Qu, m}|$, i.e., $g_\text{ub}$ in Eq.\,\ref{eq:gQu_mode_m_upper_2}. Since the scaling with material properties and interaction length are clear in the analytical expression, the numerical demonstration focuses on the geometric parameter, $g_\text{geo}$, which depends on the electron velocity and the separation between the electron beam and the photonic structure. 
I study the case where the photonic medium can be anywhere within a half space separated from the electron beam by a distance $d$ (Fig.\,\ref{fig:schematic})
\cite{henke2021integrated, yang2018maximal}. $g_\text{geo}$ depends only on the normalized electron velocity ($\beta=v/c$) and the separation distance normalized by the free-space optical wavelength ($d/\lambda$), as shown in Fig.\,\ref{fig:g_geo}(a). 


Figure \ref{fig:g_geo}(a) shows that the geometric factor decreases with the separation between the free-electron trajectory and the photonic medium. When the separation is small ($d/\lambda<0.06$), the geometric factor is peaked at a sub-relativistic velocity (the cyan curve in Fig.\,\ref{fig:g_geo}(a)), besides the divergence as $\beta \rightarrow 1$. 
To illustrate the influence of the optical medium and the free-electron velocity on the upper bound of the coupling coefficient, I plot, in Fig.\,\ref{fig:g_geo}(b), the upper bound ($g_\text{ub}$), normalized by the interaction length, as a function of $\beta$ for silicon (Si) and silicon nitride (SiN) photonic structures at 3 separation distances: $d=0.02\lambda$, $d=0.05\lambda$, and $d=0.1\lambda$. $g_\text{ub}$ for Si is twice that for SiN, due to its higher permittivity. For deep sub-wavelength separation between the free electron and the optical medium, $g_\text{ub}$ has a prominent peak at low electron velocity. For instance, with $d=0.02\lambda$, $L=\lambda$, and a silicon optical medium, $g_\text{ub}>1$ at a sub-relativistic velocity $0.1<\beta<0.4$, which is promising to reach strong coupling. Such a sub-relativistic peak in free-electron--light coupling is consistent with previous studies when $d$ is deep sub-wavelength \cite{liebtrau2021spontaneous}. Although the large $g_\text{ub}$ at small $d$ can be diminished with a large free-electron beam size, the analytical upper bound is approximately valid when $d$ is larger than the transverse electron beam size (SM Sec.\,IV). Furthermore, with a maximal interaction length estimated from the diffraction of a transversely confined free electron \cite{karnieli2024strong}, one can find the ultimate upper bound of the quantum coupling coefficient (SM Sec.\,IV). 

To study how far the typical free-electron--light interaction systems are from the upper bound, I numerically simulate $|g_{Qu,m}|$ when free electrons interact with guided modes in a Si or SiN waveguide (WG) (Fig.\,\ref{fig:g_geo}(b)) \cite{henke2021integrated, zhao2018design}, where the electron velocity matches the phase velocity of the guided mode (SM Sec.\,VI). I find that the simulated $|g_{Qu,m}|$ is within one order of magnitude difference from the upper bound for a large range of $\beta$ and $d$. Moreover, Si waveguides generally have larger $|g_{Qu,m}|$ than SiN waveguides, which is consistent with the upper bound prediction. When the electron velocity is lower than $c/\sqrt{\epsilon}$, one can utilize the guided modes in a grating to satisfy the phase matching condition. As an example, I numerically study the guided mode in a Si sub-wavelength grating (SWG) \cite{halir2015waveguide} and find $|g_{Qu,m}|=0.1$ with $\beta=0.25$, $d=0.02\lambda$, and $L=\lambda$ (red star in Fig.\,\ref{fig:g_geo}(b)), which can reach strong coupling with $L>95\lambda$ (SM Sec.\,VI). Nevertheless, there is a large room for future structure optimization to improve the free-electron--light coupling, especially with high index dielectric and small separation distances.

Furthermore, under the guidance of the mathematical `optimal' field to saturate the upper bound, one can find structures where $|g_{Qu,m}|$ is close to the upper bound. As a simple example, I calculate $|g_{Qu,m}|$ when free electrons interact with SPP modes in a metallic hole with hole radius $d$ (Fig.\,\ref{fig:g_geo}(c)) (details in SM Sec.\,VII). The metal susceptibility is described by a Drude model $\chi(\omega)=-\omega_p^2/\omega^2$ in the limit of negligible absorption. I find that $|g_{Qu,m}|$ is close to the upper bound (Eq.\,\ref{eq:gQu_ub_Lorentz} with $\omega_0=0$), especially when the electron-metal separation ($d$) is small ($d<0.1\lambda$), since the physical fields are close to the mathematical `optimal' field. In certain cases, more than 99\% of the upper bound is reached. Moreover, the large $|g_{Qu,m}|$ suggests that, when the absorption is low, the metallic hole is a promising system for strong free-electron--light interaction \cite{adamo2009light}.

\emph{Conclusion.--} In conclusion, I study the upper bound for the quantum coupling coefficient describing the interaction between free electrons and photons. I derive the analytical upper bound when the photonic excitation has a continuous spectrum and when the photonic excitation has a discrete spectrum, and I discuss the connections between these two cases. 
I provide numerical results to show the dependence of the upper bound on the optical medium, the free-electron velocity, and the separation between the free electron and the optical medium. This study establishes the fundamental upper bound of free-electron--photon coupling and provides guidance to the choice of system parameters to reach the strong coupling  regime.

\begin{acknowledgments}
\emph{Acknowledgments.--} I thank Prof. Peter Hommelhoff, Prof. Owen Miller, Dr. Zeyu Kuang, Mr. Zhaowei Dai, Dr. Tom{\'a}{\v s} Chlouba, Dr. Aviv Karnieli, Prof. Ido Kaminer, Dr. Xiao-Qi Sun, and Hommelhoff group members for helpful discussions and suggestions. This work is supported by ERC Adv. Grant AccelOnChip (884217) and the Gordon
and Betty Moore Foundation (GBMF11473).
\end{acknowledgments}


\emph{Note added.--} During the completion of this manuscript, I became aware of related work \cite{xie2024maximal, di2024toward}. 

\bibliography{DLA_quantum.bib}{}
\bibliographystyle{apsrev4-1}  
\end{document}


\title{Upper bound for the quantum coupling between free electrons and photons -- Supplemental Material}

\date{\today}

\begin{abstract}
In the supplemental material, I recapitulate the detailed derivation of the quantum coupling coefficient for the interaction between free electrons and photons, in the framework of macroscopic quantum electrodynamics (MQED). I derive the upper bound of the coupling coefficient when the free electron interacts with general photonic excitations, and supplement the main text with the derivation of the upper bound when the free electron interacts with discrete photonic modes. I also derive the upper bound of the coupling coefficient when the optical medium is dispersive. Then, I show the connections between the formalism for general photonic medium and that in the lossless limit. Moreover, I present additional numerical data about the upper bound and discuss the influence of the transverse distribution of the free electron. With the maximal interaction length estimated from the angular spread of a transversely confined free electron, I show the ultimate upper bound and point out the large range of parameters that could potentially achieve the strong coupling regime of the free-electron--light interaction. Afterwards, I present numerical examples of an electron interacting with silicon (Si) or silicon nitride (SiN) waveguides and the guided mode in a Si sub-wavelength grating (SWG). I discuss the quantum coupling coefficient for the SWG. Finally, I study two examples of free-electron--light coupling: (1) with guided modes in a dielectric hollow-core waveguide, and (2) with surface plasmon polariton modes in a metallic hole, to investigate how tight the analytical upper bounds are. I find that the coupling coefficient can reach about 70\% of the upper bound with the dielectric hollow-core waveguide and above 99\% of the upper bound with the metallic hole.
\end{abstract}

\author{Zhexin Zhao$^1$}
\email[]{zhexin.zhao@fau.de}
\affiliation{$^1$Department of Physics, Friedrich-Alexander University (FAU) Erlangen-Nürnberg, Staudtstraße 1, 91058 Erlangen, Germany}

\maketitle

\tableofcontents

\section{The quantum coupling coefficient for a general optical medium}
\label{sec:gQu_general}

In this section, I recap the derivation of the Hamiltonian and scattering matrix that describe the interaction between free electrons and photons for a general medium, which may be lossy and dispersive. The description of the interaction between free electrons and the photonic modes has been successfully developed in recent studies \cite{di2019probing, di2020electron, kfir2021optical, huang2023electron}, in the framework of macroscopic quantum electrodynamics (MQED) \cite{gruner1996green, dung1998three, rivera2020light}. 

\subsection{Macroscopic quantum electrodynamics (MQED)}
\label{sec:MQED}
In the framework of MQED \cite{gruner1996green, dung1998three, rivera2020light}, a quantized harmonic oscillator is assigned to each location, orientation and frequency. From these quantized harmonic oscillator, one get the current operator that can be regarded as the quantum Langevin source to drive the electromagnetic field \cite{rivera2020light}. The field (and vector potential) operators can be obtained from the current operator using the classical Green's function. In this way, one can quantize the electromagnetic field in a general optical medium, including a lossy and dispersive medium. For simplicity, the material is assumed to be isotropic, non-magnetic, and local, since the majority of photonic resonant systems belong to this category. The MQED framework can be generalized to include anisotropic, magnetic or non-local media \cite{rivera2020light}. 

The creation ($\hat{f}_i^\dagger(\boldsymbol{r},\omega)$) and annihilation ($\hat{f}_i(\boldsymbol{r},\omega)$) operators associated with each quantum harmonic oscillator satisfy the following commutation relations \cite{dung1998three}
\begin{equation}
    \label{eq:f_commu_1}
    [\hat{f}_i(\boldsymbol{r},\omega), \hat{f}^\dagger_j(\boldsymbol{r}', \omega')] = \delta_{ij} \delta(\boldsymbol{r}-\boldsymbol{r}') \delta(\omega - \omega'),
\end{equation}
\begin{equation}
    \label{eq:f_commu_2}
    [\hat{f}_i(\boldsymbol{r},\omega), \hat{f}_j(\boldsymbol{r}', \omega')] = 0 = [\hat{f}^\dagger_i(\boldsymbol{r},\omega), \hat{f}^\dagger_j(\boldsymbol{r}', \omega')],
\end{equation}
where $i$ and $j$ denote the orientation ($i, j\in \{ x, y, z \} $), $\boldsymbol{r}$ and $\boldsymbol{r}'$ denote the 3-dimensional position, and $\omega$ and $\omega'$ denote the frequency.
The Hamiltonian of the electromagnetic field is \cite{dung1998three}
\begin{equation}
    \label{eq:H_em_MQED}
    H_p = \int d^3 \boldsymbol{r} \int_0^{+\infty} d\omega \hbar \omega \hat{f}_i^\dagger(\boldsymbol{r},\omega) \hat{f}_i(\boldsymbol{r},\omega).
\end{equation}

The current operator ($\boldsymbol{\hat{j}}$) is \cite{dung1998three}
\begin{equation}
    \label{eq:j_MQED}
    \boldsymbol{\hat{j}}(\boldsymbol{r},\omega) = \sqrt{\frac{\hbar\omega^2\epsilon_0}{\pi} \epsilon_I(\boldsymbol{r},\omega)} \boldsymbol{\hat{f}}(\boldsymbol{r},\omega),
\end{equation}
where $\epsilon_I$ is the imaginary part of the permittivity. This definition of the current operator is consistent with the fluctuation-dissipation theorem \cite{joulain2005surface}. 

The electric field operator ($\boldsymbol{\hat{E}}$) is related to the current operator in the same manner that the classical electric field is related to the current.
\begin{equation}
    \label{eq:E_operator_freq}
    \nabla \times \nabla \times \boldsymbol{\hat{E}}(\boldsymbol{r},\omega) - \frac{\omega^2}{c^2} \epsilon(\boldsymbol{r}) \boldsymbol{\hat{E}}(\boldsymbol{r}, \omega) = i\omega\mu_0 \boldsymbol{\hat{j}}(\boldsymbol{r},\omega),
\end{equation}
\begin{equation}
    \label{eq:E_operator_G_freq}
    \boldsymbol{\hat{E}}(\boldsymbol{r},\omega) = i\omega \mu_0 \int d^3 \boldsymbol{r}' G(\boldsymbol{r},\boldsymbol{r}',\omega) \boldsymbol{\hat{j}}(\boldsymbol{r}', \omega),
\end{equation}
where the Green's function $G$ is the same as in the  classical theory:
\begin{equation}
    \label{eq:G_freq_def}
     \Big[\nabla \times \nabla \times - \frac{\omega^2}{c^2} \epsilon(\boldsymbol{r}
     ) \Big ] G(\boldsymbol{r}, \boldsymbol{r}', \omega) = \hat{I}\delta(\boldsymbol{r}-\boldsymbol{r}'),
\end{equation}
where $\hat{I}_{ij} = \delta_{ij}$.

In the temporal gauge, the scalar potential vanishes, and \cite{di2019probing, di2020electron}
\begin{equation}
    \label{eq:A_E_freq}
    \boldsymbol{\hat{A}}(\boldsymbol{r},\omega) = \frac{1}{i\omega}\boldsymbol{\hat{E}}(\boldsymbol{r},\omega).
\end{equation}
From Eqs.\,\ref{eq:j_MQED}, \ref{eq:E_operator_G_freq} and \ref{eq:A_E_freq}, the vector potential is
\begin{equation}
    \label{eq:A_operator_freq}
    \hat{A}_i(\boldsymbol{r},\omega) = \sqrt{\frac{\hbar}{\pi\epsilon_0}} \frac{\omega}{c^2} \int d^3 \boldsymbol{s} G_{ij}(\boldsymbol{r},\boldsymbol{s},\omega)\sqrt{\epsilon_I(\boldsymbol{s},\omega)} \hat{f}_{j}(\boldsymbol{s},\omega),
\end{equation}
where the repeated sub index will be summed. 
The Fourier transformation convention for the quantum operators is adopted from Ref.\cite{dung1998three}. For instance,
\begin{equation}
    \label{eq:A_Fourier_trans}
    \boldsymbol{\hat{A}}(\boldsymbol{r}) = \int_0^{+\infty} d\omega \Big[\boldsymbol{\hat{A}}(\boldsymbol{r},\omega) + H.C.\Big],
\end{equation}
where $H.C.$ stands for Hermitian conjugate.
The Fourier transformation from time to frequency domain is therefore:
\begin{equation}
    \label{eq:A_Fourier_inverse}
    \boldsymbol{\hat{A}}(\boldsymbol{r}, \omega) = \frac{1}{2\pi}\int dt \boldsymbol{\hat{A}}(\boldsymbol{r}, t) e^{i\omega t}
\end{equation}

\subsection{Hamiltonian describing the free-electron--light interaction}
\label{sec:Hamiltonian_electron_light}
The Hamiltonian describing the free-electron--light interaction is \cite{huang2023electron}:
\begin{equation}
    \label{eq:H}
    \hat{H} = \hat{H}_0 + \hat{V} = \hat{H}_e + \hat{H}_p + \hat{V},
\end{equation}
where $\hat{H}_e$ is the Hamiltonian of the free electron, $\hat{H}_p$ is the Hamiltonian of the electromagnetic field (photon), $\hat{H}_0 = \hat{H}_e + \hat{H}_p$, and $\hat{V}$ describes the interaction. 
The Hamiltonian of the electromagnetic field is given by Eq.\,\ref{eq:H_em_MQED}.

The non-recoil assumption for the free electron is adopted, where the free-electron transverse wave function is approximately unchanged \cite{kfir2019entanglements} and the dispersion relation with respect to the longitudinal momentum is approximately linear near the reference energy ($E_0$) and the reference momentum ($\hbar k_0$) \cite{di2019probing}. The non-recoil assumption can be justified in many free-electron--light interaction systems such as scanning electron microscopes (SEMs) and transmission electron microscopes (TEMs), where free electrons have predominantly a momentum in the longitudinal direction, and the photon energy is much less than the free-electron kinetic energy. Thus, the free-electron annihilation operator is 
\begin{equation}
    \label{eq:psi_e_1}
    \hat{\Psi}_e(\boldsymbol{r}) = \sum_k \hat{c}_k \frac{1}{\sqrt{L}} e^{ikz}\phi_e(\boldsymbol{r}_\perp),
\end{equation}
where $\boldsymbol{r} = \boldsymbol{r}_\perp + z \boldsymbol{\hat{z}}$ ($\hat{z}$ is a unit vector along z.), and $L$ is introduced due to the box quantization of the free-electron wave function. The choice of $L$ is arbitrary, since in the continuum limit ($L\rightarrow\infty$), $\sum_k \rightarrow \frac{L}{2\pi}\int dk$ and $L$ will be cancelled out.  $\phi_e(r_\perp)$ describes the transverse wave function of the free electron. It is normalized, i.e., 
\begin{equation}
    \label{eq:phi_e_normalization}
    \int d^2 \boldsymbol{r}_\perp \phi_e^*(\boldsymbol{r}_\perp) \phi_e (\boldsymbol{r}_\perp) = 1 .
\end{equation}
$\hat{c}_k$ is the annihilation operator associated with wave vector $k$. It satisfies the anti-commutation relation:
\begin{equation}
    \label{eq:ck_anticommutation_1}
    \{ \hat{c}_k, \hat{c}^\dagger_{k'} \} = \delta_{k,k'},
\end{equation}
\begin{equation}
    \label{eq:ck_anticommutation_2}
    \{ \hat{c}_k, \hat{c}_{k'} \} = 0 = \{ \hat{c}_k^\dagger, \hat{c}_{k'}^\dagger \},
\end{equation}

Since the spin effect is negligible in typical free-electron--light interaction, the Klein-Gordon Hamiltonian is used to describe the free electrons:
\begin{equation}
    \label{eq:H_e_Klein_Gordon}
    \hat{H}_e = \int d^3 \boldsymbol{r} \hat{\Psi}^\dagger_e(\boldsymbol{r}) \sqrt{m^2c^4 + c^2 \hat{p}^2} \hat{\Psi}_e(\boldsymbol{r}).
\end{equation}
With the assumption of linear dispersion near the reference energy ($E_0$) and momentum ($\hbar k_0$),
\begin{equation}
    \label{eq:E_e_linear}
    \sqrt{m^2c^4 + c^2 p^2} \approx E_0 + \hbar v (k-k_0).
\end{equation}
Thus, 
\begin{equation}
    \label{eq:H_e_linear}
    \hat{H}_e = \sum_k [E_0 + \hbar v (k-k_0)] \hat{c}_k^\dagger \hat{c}_k
\end{equation}

The interaction $\hat{V}$ is \cite{huang2023electron}
\begin{equation}
    \label{eq:V_1}
    \hat{V} = - \int d^3\boldsymbol{r} \boldsymbol{\hat{J}}(\boldsymbol{r}) \cdot \boldsymbol{\hat{A}}(\boldsymbol{r}),
\end{equation}
where $\hat{J}$ is the current operator associated with the free electron.
\begin{equation}
    \label{eq:J_e_def}
    \boldsymbol{\hat{J}}(\boldsymbol{r}) = \frac{q_e}{\gamma_e m} \hat{\Psi}_e^\dagger(\boldsymbol{r}) \boldsymbol{\hat{p}} \hat{\Psi}_e(\boldsymbol{r}),
\end{equation}
where $q_e$ is the electron charge, and $\gamma_e$ is the relativistic factor $\gamma_e = [1 - v^2/c^2]^{-1/2}$. 
Using the non-recoil assumption, 
\begin{equation}
    \label{eq:J_e_2}
    \boldsymbol{\hat{J}}(\boldsymbol{r}) = \frac{q_e v}{L} \sum_{k, q}e^{iqz} \hat{c}^\dagger_k \hat{c}_{k+q} \phi_e^*(\boldsymbol{r}_\perp) \phi_e(\boldsymbol{r}_\perp) \boldsymbol{\hat{z}} .
\end{equation}
Thus,
\begin{equation}
    \label{eq:V_2}
    \hat{V} = -\frac{q_e v}{L}\int d^3\boldsymbol{r} \sum_{k, q}e^{iqz} \hat{c}^\dagger_k \hat{c}_{k+q} \phi_e^*(\boldsymbol{r}_\perp) \phi_e(\boldsymbol{r}_\perp) \hat{A}_z(\boldsymbol{r})
\end{equation}
Using Eqs.\,\ref{eq:A_operator_freq} and \ref{eq:A_Fourier_trans} for $\hat{A}$, the interaction Hamiltonian is
\begin{equation}
    \label{eq:V_3}
    \begin{split}
    \hat{V} = & -\frac{q_e v}{L}\int d^3\boldsymbol{r} \sum_{k, q}e^{iqz} \hat{c}^\dagger_k \hat{c}_{k+q} \phi_e^*(\boldsymbol{r}_\perp) \phi_e(\boldsymbol{r}_\perp) \int_0^{+\infty} d\omega \big[\hat{A}_z(\boldsymbol{r},\omega) + H.C.\big] \\ = & -\frac{q_e v}{L}\int d^3\boldsymbol{r} \sum_{k, q}e^{iqz} \hat{c}^\dagger_k \hat{c}_{k+q} \phi_e^*(\boldsymbol{r}_\perp) \phi_e(\boldsymbol{r}_\perp) \int_0^{+\infty} d\omega \sqrt{\frac{\hbar}{\pi \epsilon_0}} \frac{\omega}{c^2} \\
    &\int d^3 \boldsymbol{s} \Big[ G_{zm}(\boldsymbol{r},\boldsymbol{s},\omega) \sqrt{\epsilon_I(\boldsymbol{s},\omega)} \hat{f}_m(\boldsymbol{s}, \omega) + H.C.\Big].
    \end{split}
\end{equation}
I also repeat $\hat{H}_0$ (Eqs.\,\ref{eq:H_em_MQED} and \ref{eq:H_e_linear}) here:
\begin{equation}
    \label{eq:H0}
    \hat{H}_0 = \sum_k [E_0 + \hbar v (k-k_0)] \hat{c}_k^\dagger \hat{c}_k + \int d^3 \boldsymbol{r} \int_0^{+\infty} d\omega \hbar \omega \hat{f}_m^\dagger(\boldsymbol{r},\omega) \hat{f}_m(\boldsymbol{r},\omega).
\end{equation}

\subsection{Interaction Hamiltonian in the interaction picture}
\label{sec:interaction_picture}
From the Hamiltonian (Eqs.\,\ref{eq:H}, \ref{eq:V_3}, and \ref{eq:H0}), one can get the scattering matrix $\hat{S}$ describing the free-electron--light interaction, formally expressed as \cite{zhao2021quantum, weinberg2005quantum}
\begin{equation}
    \label{eq:S_formal}
    \hat{S} = \mathcal{T} \exp \Big[ \frac{1}{i\hbar} \int_{-\infty}^{+\infty} \hat{V}_I(t) dt\Big],
\end{equation}
where $\mathcal{T}$ means time ordering, and $\hat{V}_I$ is the interaction Hamiltonian in the interaction picture. 
\begin{equation}
    \label{eq:VI}
    \hat{V}_I(t) = \exp\Big({\frac{i}{\hbar}\hat{H}_0 t}\Big) \hat{V} \exp\Big(-\frac{i}{\hbar}\hat{H}_0 t \Big)
\end{equation}
Equation \ref{eq:VI} can be simplified using the Baker-Campbell-Hausdorff formula
\begin{equation}
    \label{eq:BCH}
    e^{X} Y e^{-X} = \sum_{m=0}^{m=\infty} \frac{1}{m!}[X, Y]_m,
\end{equation}
where $[X, Y]_m = [X, [X, Y]_{m-1}]$ and $[X, Y]_0 = Y$. 

\begin{equation}
    \label{eq:VI_calculation_H_decom}
    [\frac{it}{\hbar}\hat{H}_0, \hat{V}] = \frac{it}{\hbar}[\hat{H}_e, \hat{V}] + \frac{it}{\hbar}[\hat{H}_p, \hat{V}]
\end{equation}
Since $\hat{H}_\text{e}$ contains only free-electron creation and annihilation operators, 
\begin{equation}
    \label{eq:VI_calculation_He}
    \frac{it}{\hbar}[\hat{H}_e, \hat{V}] = -\frac{i q_e vt}{\hbar L}\int d^3\boldsymbol{r} \Big[ \hat{H}_\text{e}, \sum_{k, q}e^{iqz} \hat{c}^\dagger_k \hat{c}_{k+q} \Big] \phi_e^*(\boldsymbol{r}_\perp) \phi_e(\boldsymbol{r}_\perp) \hat{A}_z(\boldsymbol{r}).
\end{equation}
The calculation of the commutation relations can be simplified using Leibniz rules. 
Since
\begin{equation}
    \label{eq:VI_calculation_He_2}
    \begin{split}
     \Big[ \hat{H}_e, \sum_{k, q}e^{iqz} \hat{c}^\dagger_k \hat{c}_{k+q} \Big] & = \sum_{k',k,q}[E_0 + \hbar v (k'-k_0)]e^{iqz} [\hat{c}^\dagger_{k'}\hat{c}_{k'}, \hat{c}^\dagger_k \hat{c}_{k+q}] \\
     & = \sum_{k',k,q}[E_0 + \hbar v (k'-k_0)]e^{iqz} [\delta_{k',k} \hat{c}^\dagger_{k'}\hat{c}_{k+q} - \delta_{k', k+q} \hat{c}^\dagger_k \hat{c}_{k'} ] \\
     & = \sum_{k,q}(-\hbar v q) e^{iqz} \hat{c}^\dagger_{k}\hat{c}_{k+q},
     \end{split}
\end{equation}
\begin{equation}
    \label{eq:VI_calculation_He_3}
    \frac{it}{\hbar}[\hat{H}_\text{e}, \hat{V}] = -\frac{q_e v}{L}\int d^3\boldsymbol{r} \sum_{k,q}(-i v qt) e^{iqz} \hat{c}^\dagger_{k}\hat{c}_{k+q} \phi_e^*(\boldsymbol{r}_\perp) \phi_e(\boldsymbol{r}_\perp) \hat{A}_z(\boldsymbol{r}).
\end{equation}
Similarly, $\hat{H}_p$ contains only photon creation and annihilation operators. Thus,
\begin{equation}
    \label{eq:VI_calculation_Hem_1}
    \frac{it}{\hbar}[\hat{H}_p, \hat{V}] = -\frac{i q_e vt}{\hbar L}\int d^3 \boldsymbol{r} \sum_{k, q}e^{iqz} \hat{c}^\dagger_k \hat{c}_{k+q} \phi_e^*(\boldsymbol{r}_\perp) \phi_e(\boldsymbol{r}_\perp) \Big[ \hat{H}_p, \hat{A}_z(\boldsymbol{r}) \Big].
\end{equation}
\begin{equation}
    \label{eq:VI_calculation_Hem_2}
    \begin{split}
        \Big[ \hat{H}_p, \hat{A}_z(\boldsymbol{r}) \Big]
        = & \int_0^{+\infty}d\omega' \int_0^{+\infty} d\omega \int d^3 \boldsymbol{r}' \int d^3 \boldsymbol{s} \hbar \omega' \sqrt{\frac{\hbar}{\pi \epsilon_0}} \frac{\omega}{c^2} \sqrt{\epsilon_I(\boldsymbol{s}, \omega)} \\ & \Big\{G_{zm}(\boldsymbol{r},\boldsymbol{s},\omega)\big[\hat{f}_i^\dagger(\boldsymbol{r}',\omega')\hat{f}_i(\boldsymbol{r}',\omega'), \hat{f}_m(\boldsymbol{s},\omega)\big] + G^*_{zm}(\boldsymbol{r},\boldsymbol{s},\omega)\big[\hat{f}_i^\dagger(\boldsymbol{r}',\omega')\hat{f}_i(\boldsymbol{r}',\omega'), \hat{f}_m^\dagger(\boldsymbol{s},\omega)\big] \Big\} \\
        = & \int_0^{+\infty}d\omega' \int_0^{+\infty} d\omega \int d^3 \boldsymbol{r}' \int d^3 \boldsymbol{s} \hbar \omega' \sqrt{\frac{\hbar}{\pi \epsilon_0}} \frac{\omega}{c^2} \sqrt{\epsilon_I(\boldsymbol{s}, \omega)} \\
        & \Big[G_{zm}(\boldsymbol{r},\boldsymbol{s},\omega) (-1)\delta_{im} \delta(\boldsymbol{r}'-\boldsymbol{s})\delta(\omega'-\omega) \hat{f}_i(\boldsymbol{r}',\omega') + G^*_{zm}(\boldsymbol{r},\boldsymbol{s},\omega) \delta_{im}\delta(\boldsymbol{r}'-\boldsymbol{s})\delta(\omega'-\omega) \hat{f}_i^\dagger(\boldsymbol{r}',\omega')\Big] \\
        = & \int_0^{+\infty} d\omega \int d^3 \boldsymbol{s} \hbar \omega \sqrt{\frac{\hbar}{\pi \epsilon_0}} \frac{\omega}{c^2} \sqrt{\epsilon_I(\boldsymbol{s}, \omega)} \Big[G_{zm}(\boldsymbol{r},\boldsymbol{s},\omega) (-1) \hat{f}_m(\boldsymbol{r},\omega) + G^*_{zm}(\boldsymbol{r},\boldsymbol{s},\omega)\hat{f}_m^\dagger(\boldsymbol{r},\omega)\Big] \\
        = & \int_0^{+\infty} d\omega  \hbar\omega [-\hat{A}_z(\boldsymbol{r}, \omega) + \hat{A}^\dagger_z(\boldsymbol{r}, \omega)]
    \end{split}
\end{equation}
\begin{equation}
    \label{eq:VI_calculation_Hem_3}
    \begin{split}
        \frac{it}{\hbar}[\hat{H}_p, \hat{V}] = & -\frac{q_e v}{L}\int d^3\boldsymbol{r} \sum_{k, q}e^{iqz} \hat{c}^\dagger_k \hat{c}_{k+q} \phi_e^*(\boldsymbol{r}_\perp) \phi_e(\boldsymbol{r}_\perp) \int_0^{+\infty} d\omega \big[(-i\omega t)\hat{A}_z(\boldsymbol{r},\omega) + (i\omega t) \hat{A}^\dagger_z(\boldsymbol{r}, \omega)\big].
    \end{split}
\end{equation}
In summary, from Eqs.\,\ref{eq:VI_calculation_He_3} and \ref{eq:VI_calculation_Hem_3}, 
\begin{equation}
    \label{eq:VI_calculation_H0}
    \begin{split}
    \Big[\frac{it}{\hbar}\hat{H}_0, \hat{V}\Big] = & -\frac{q_e v}{L}\int d^3\boldsymbol{r} \sum_{k, q}e^{iqz} \hat{c}^\dagger_k \hat{c}_{k+q} \phi_e^*(\boldsymbol{r}_\perp) \phi_e(\boldsymbol{r}_\perp) \\ & \int_0^{+\infty} d\omega \big[(-ivqt-i\omega t)\hat{A}_z(\boldsymbol{r},\omega) + (-ivqt + i\omega t) \hat{A}^\dagger_z(\boldsymbol{r}, \omega)\big].
    \end{split}
\end{equation}
Therefore, the interaction Hamiltonian in the interaction picture ($\hat{V}_I(r)$) is:
\begin{equation}
    \label{eq:VI_final}
    \begin{split}
    \hat{V}_I(t) = & -\frac{q_e v}{L}\int d^3\boldsymbol{r} \sum_{k, q}e^{iqz} \hat{c}^\dagger_k \hat{c}_{k+q} \phi_e^*(\boldsymbol{r}_\perp) \phi_e(\boldsymbol{r}_\perp) \int_0^{+\infty} d\omega \big[e^{-ivqt-i\omega t}\hat{A}_z(\boldsymbol{r},\omega) + e^{-ivqt + i\omega t} \hat{A}^\dagger_z(\boldsymbol{r}, \omega)\big].
    \end{split}
\end{equation}

\subsection{Scattering matrix}
\label{sec:scattering_matrix}
One can get the explicit form of the scattering matrix, shown formally in Eq.\,\ref{eq:S_formal}, using the Magnus expansion:
\begin{equation}
    \label{eq:S_magnus}
    \hat{S} = \exp\Big( \sum_{m=1}^{\infty} \hat{\Omega}_m \Big),
\end{equation}
where 
\begin{equation}
    \label{eq:S_Omega1}
    \hat{\Omega}_1 = \frac{1}{i\hbar} \int_{-\infty}^{+\infty} \hat{V}_I(t)dt,
\end{equation}
\begin{equation}
    \label{eq:S_Omega2}
    \hat{\Omega}_2 = \frac{1}{2(i\hbar)^2} \int_{-\infty}^{+\infty} dt_1 \int_{-\infty}^{t_1} dt_2 \big[\hat{V}_I(t_1), \hat{V}_I(t_2)\big],
\end{equation}
etc. Since $[\hat{V}_I(t_1), \hat{V}_I(t_2)]$ contains no creation or annihilation operators for the electromagnetic field (photon), so do all higher order terms, the scattering matrix takes the form \cite{di2020electron}
\begin{equation}
    \label{eq:S_2}
    \hat{S} = \exp(i\hat{\chi}) \hat{U} = \exp(i\hat{\chi})\exp \Big[ \frac{1}{i\hbar} \int_{-\infty}^{+\infty} \hat{V}_I(t)dt\Big],
\end{equation}
where the first term is a phase operator acting only on the free-electron wave function \cite{di2020electron}. 

Using Eq.\,\ref{eq:VI_final}, one can get
\begin{equation}
    \label{eq:Omega1_2}
    \hat{\Omega}_1 = -\frac{q_e v}{i\hbar L}\int d^3\boldsymbol{r} \sum_{k, q}e^{iqz} \hat{c}^\dagger_k \hat{c}_{k+q} \phi_e^*(\boldsymbol{r}_\perp) \phi_e(\boldsymbol{r}_\perp) \int_0^{+\infty} d\omega \big[ 2\pi\delta(\omega + vq) \hat{A}_z(\boldsymbol{r},\omega) + 2\pi \delta(\omega - vq) \hat{A}^\dagger_z(\boldsymbol{r}, \omega)\big].
\end{equation}
Take the continuous limit of $q$, i.e.,
\begin{equation}
    \frac{2\pi}{L}\sum_q \rightarrow \int dq.
\end{equation}
Then,
\begin{equation}
    \label{eq:Omega1_3}
    \hat{\Omega}_1 = -\frac{q_e}{i\hbar} \int d^3 \boldsymbol{r} \int_0^{+\infty} d\omega \phi_e^*(\boldsymbol{r}_\perp) \phi_e(\boldsymbol{r}_\perp) \Big[\sum_k e^{-i\frac{\omega}{v}z}\hat{c}^\dagger_k \hat{c}_{k-\frac{\omega}{v}} \hat{A}_z(\boldsymbol{r},\omega) + \sum_k e^{i\frac{\omega}{v}z}\hat{c}^\dagger_k \hat{c}_{k+\frac{\omega}{v}} \hat{A}_z^\dagger(\boldsymbol{r},\omega) \Big].
\end{equation}
The electron ladder operator \cite{kfir2019entanglements, zhao2021quantum, kfir2021optical, huang2023electron} is defined as
\begin{equation}
    \label{eq:b_operator}
    \hat{b}_{q} = \sum_k \hat{c}^\dagger_k \hat{c}_{k+q}.
\end{equation}
Its Hermitian conjugate is 
\begin{equation}
    \label{eq:b_dagger}
    \hat{b}_q^\dagger = \sum_k \hat{c}_{k+q}^\dagger \hat{c}_k = \sum_k \hat{c}_{k}^\dagger \hat{c}_{k-q}.
\end{equation}
Thus,
\begin{equation}
    \label{eq:Omega1_4}
    \hat{\Omega}_1 = -\frac{q_e}{i\hbar} \int d^3 \boldsymbol{r} \int_0^{+\infty} d\omega \phi_e^*(\boldsymbol{r}_\perp) \phi_e(\boldsymbol{r}_\perp) \Big[ e^{-i\frac{\omega}{v}z} \hat{b}_{\frac{\omega}{v}}^\dagger \hat{A}_z(\boldsymbol{r},\omega) + e^{i\frac{\omega}{v}z}\hat{b}_{\frac{\omega}{v}}\hat{A}_z^\dagger(\boldsymbol{r}, \omega)\Big].
\end{equation}
Therefore, one obtains the scattering matrix for the interaction between the free electron and photonic excitations in a general medium:
\begin{equation}
    \label{eq:S_3}
    \hat{S} = \exp({i\hat{\chi}})\exp\Big\{{-\frac{q_e}{i\hbar} \int d^3 \boldsymbol{r} \phi_e^*(\boldsymbol{r}_\perp) \phi_e(\boldsymbol{r}_\perp) \int_0^{+\infty} d\omega  \big[ e^{-i\frac{\omega}{v}z} \hat{b}_{\frac{\omega}{v}}^\dagger \hat{A}_z(\boldsymbol{r},\omega) + e^{i\frac{\omega}{v}z}\hat{b}_{\frac{\omega}{v}}\hat{A}_z^\dagger(\boldsymbol{r}, \omega)\big]}\Big\}.
\end{equation}

Furthermore, the transverse spread of the free-electron wave function is typically small such that the change of $g_{Qu}(\boldsymbol{r}_\perp, \omega)$ is negligible, i.e., the transverse wave function is tightly confined around $\boldsymbol{r}_{e\perp}$. Then, one can reformulate Eq.\,\ref{eq:S_3} into the following form \cite{di2021modulation, kfir2021optical}
\begin{equation}
    \label{eq:S_5}
    \hat{S} = \exp({i\hat{\chi}})\exp\Big\{{\int_0^{+\infty} d\omega \big [g_{Qu}(\boldsymbol{r}_{e\perp}, \omega) \hat{b}^\dagger_{\frac{\omega}{v}} \hat{a}_\omega - g_{Qu}^*(\boldsymbol{r}_{e\perp}, \omega) \hat{b}_{\frac{\omega}{v}} \hat{a}^\dagger_\omega \big]}\Big\}.
\end{equation}
Comparing Eq.\,\ref{eq:S_5} with Eq.\,\ref{eq:S_3}, one can find the definition of the quantum coupling coefficient, i.e., 
\begin{equation}
    \label{eq:gQu_a}
    g_{Qu}(\boldsymbol{r}_{e\perp}, \omega)\hat{a}_\omega = \frac{i q_e}{\hbar}\int dz e^{-i\frac{\omega}{v}z}\hat{A}_z(\boldsymbol{r}_{e\perp}+z\boldsymbol{\hat{z}},\omega).
\end{equation}
In the main text, I omit $\boldsymbol{r}_{e\perp}$ in $g_{Qu}(\boldsymbol{r}_{e\perp}, \omega)$ for the simplicity of the notation. Here, I keep the explicit notation of $\boldsymbol{r}_{e\perp}$ and will discuss the influence of a finite transverse spread of the free-electron wave function in Sec.\,\ref{sec:trans}.

\subsection{Expression of the quantum coupling coefficient}
\label{sec:gQu_omega}
The explicit expression of $g_{Qu}(\boldsymbol{r}_{e\perp}, \omega)$ can be obtained by enforcing the commutation relation of $\hat{a}_\omega$ as:
\begin{equation}
    \label{eq:a_omega_commutation}
    [\hat{a}_\omega, \hat{a}_{\omega'}^\dagger] = \delta(\omega - \omega').
\end{equation}
The derivation is as following, where Eqs.\,\ref{eq:f_commu_1}, \ref{eq:A_operator_freq}, \ref{eq:gQu_a} and \ref{eq:a_omega_commutation} are utilized.
\begin{equation}
    \label{eq:a_omega_commu_2}
    \begin{split}
    |g_{Qu}(\boldsymbol{r}_{e\perp}, \omega)|^2 \delta(\omega - \omega') = & \; |g_{Qu}(\boldsymbol{r}_{e\perp}, \omega)|^2[\hat{a}_\omega, \hat{a}_{\omega'}^\dagger] \\ = & \; \frac{q_e^2}{\hbar^2}\int dz \int dz' e^{i\frac{\omega'}{v}z' - i\frac{\omega}{v}z} \Big[\hat{A}_z(\boldsymbol{r}_{e\perp}+z\boldsymbol{\hat{z}},\omega), \hat{A}_z^\dagger(\boldsymbol{r}_{e\perp}+z'\boldsymbol{\hat{z}},\omega')\Big] \\
    = & \; \frac{q_e^2}{\hbar^2} \int dz \int dz' e^{i\frac{\omega'}{v}z' - i\frac{\omega}{v}z} \frac{\hbar}{\pi \epsilon_0} \frac{\omega \omega'}{c^4} \int d^3 \boldsymbol{s} \int d^3 \boldsymbol{s}' G_{zi}(\boldsymbol{r}_{e\perp} + z\boldsymbol{\hat{z}}, \boldsymbol{s}, \omega) \\ & \; G^*_{zj}(\boldsymbol{r}_{e\perp}+z'\boldsymbol{\hat{z}}, \boldsymbol{s}', \omega')\sqrt{\epsilon_I(\boldsymbol{s},\omega)} \sqrt{\epsilon_I(\boldsymbol{s}',\omega')} \Big[\hat{f}_i(\boldsymbol{s},\omega), \hat{f}^\dagger_j(\boldsymbol{s}', \omega') \Big] \\
    = &\; \frac{q_e^2}{\hbar \pi \epsilon_0} \int dz \int dz' e^{i\frac{\omega'}{v}z' - i\frac{\omega}{v}z} \frac{\omega\omega'}{c^4} \int d^3 \boldsymbol{s} \int d^3 \boldsymbol{s}' G_{zi}(\boldsymbol{s}_{e\perp} + z\boldsymbol{\hat{z}}, \boldsymbol{s}, \omega) \\ & \; G^*_{zj}(\boldsymbol{r}_{e\perp}+z'\boldsymbol{\hat{z}}, \boldsymbol{s}', \omega')\sqrt{\epsilon_I(\boldsymbol{s},\omega)} \sqrt{\epsilon_I(\boldsymbol{s}',\omega')} \delta_{ij} \delta(\boldsymbol{s}-\boldsymbol{s}')\delta(\omega-\omega') \\
    = & \; \delta(\omega - \omega') \frac{q_e^2\omega^2}{\hbar \pi \epsilon_0 c^4} \int dz \int dz' e^{i\frac{\omega}{v}(z'- z)} \\ & \; \int d^3 \boldsymbol{s}  G_{zi}(\boldsymbol{r}_{e\perp} + z\boldsymbol{\hat{z}}, \boldsymbol{s}, \omega)  G^*_{zi}(\boldsymbol{r}_{e\perp} + z'\boldsymbol{\hat{z}}, \boldsymbol{s}, \omega) \epsilon_I(\boldsymbol{s}, \omega) 
    \end{split}
\end{equation}
Thus, 
\begin{equation}
    \label{eq:gQu_omega}
    |g_{Qu}(\boldsymbol{r}_{e\perp}, \omega)|^2 = \frac{q_e^2\omega^2}{\hbar \pi \epsilon_0 c^4} \int dz \int dz' e^{i\frac{\omega}{v}(z'- z)} \int d^3 \boldsymbol{s}  G_{zi}(\boldsymbol{r}_{e\perp} + z\boldsymbol{\hat{z}}, \boldsymbol{s}, \omega)  G^*_{zi}(\boldsymbol{r}_{e\perp} + z'\boldsymbol{\hat{z}}, \boldsymbol{s}, \omega) \epsilon_I(\boldsymbol{s}, \omega) .
\end{equation}
Equation \ref{eq:gQu_omega} will be used to calculate its upper bound. 

Using the identity \cite{dung1998three}
\begin{equation}
    \label{eq:ImG_identity}
    \int d^3 \boldsymbol{s} G_{im}(\boldsymbol{r}, \boldsymbol{s}, \omega) G^*_{jm}(\boldsymbol{r}', \boldsymbol{s}, \omega) \epsilon_I(\boldsymbol{s}, \omega) = \frac{c^2}{\omega^2}\frac{1}{2i}\Big[ G_{ij}(\boldsymbol{r}, \boldsymbol{r}',\omega) - G^*_{ji}(\boldsymbol{r}', \boldsymbol{r}, \omega)\Big],
\end{equation}
one can get another form for $|g_{Qu}(\boldsymbol{r}_{e\perp}, \omega)|^2$ \cite{huang2023electron}:
\begin{equation}
    \label{eq:gQu_omega_2}
    \begin{split}
    |g_{Qu}(\boldsymbol{r}_{e\perp}, \omega)|^2 = & \;\frac{q_e^2}{\hbar \pi \epsilon_0 c^2} \int dz \int dz' e^{i\frac{\omega}{v}(z'- z)}  \frac{1}{2i}\Big[ G_{zz}(\boldsymbol{r}_{e\perp} + z\boldsymbol{\hat{z}}, \boldsymbol{r}_{e\perp} + z'\boldsymbol{\hat{z}},\omega) - G^*_{zz}(\boldsymbol{r}_{e\perp} + z'\boldsymbol{\hat{z}}, \boldsymbol{r}_{e\perp} + z\boldsymbol{\hat{z}}, \omega)\Big] \\
    = &\; \frac{q_e^2}{\hbar \pi \epsilon_0 c^2}  \frac{1}{2i}\Big[  \int dz \int dz' e^{i\frac{\omega}{v}(z'- z)} G_{zz}(\boldsymbol{r}_{e\perp} + z\boldsymbol{\hat{z}}, \boldsymbol{r}_{e\perp} + z'\boldsymbol{\hat{z}},\omega) \\
    &\; - \int dz \int dz' e^{- i\frac{\omega}{v}(z'- z)} G_{zz}^*(\boldsymbol{r}_{e\perp} + z\boldsymbol{\hat{z}}, \boldsymbol{r}_{e\perp} + z'\boldsymbol{\hat{z}},\omega) \Big] \\
    = & \; \frac{q_e^2}{\hbar \pi \epsilon_0 c^2}  \int dz \int dz' \text{Re}\Big[ -i e^{i\frac{\omega}{v}(z'- z)} G_{zz}(\boldsymbol{r}_{e\perp} + z\boldsymbol{\hat{z}}, \boldsymbol{r}_{e\perp} + z'\boldsymbol{\hat{z}},\omega) \Big] .
    \end{split}
\end{equation}
This result is the same as the energy-loss spectrum ($\Gamma$) in the electron energy loss spectroscopy (EELS) experiment \cite{huang2023electron}, i.e., 
\begin{equation}
    \label{eq:qGu_omega_Gamma}
    |g_{Qu}(\boldsymbol{r}_{e\perp}, \omega)|^2 = \Gamma(\boldsymbol{r}_{e\perp}, \omega).
\end{equation}


\section{The upper bound of the quantum coupling coefficient}
\label{sec:gQu_omega_bound}

In this section, I provide the derivation for the upper bound of $|g_{Qu}(\boldsymbol{r}_{e\perp}, \omega)|^2$, in the most general case. The derivation is similar to \cite{yang2018maximal}. I also present additional details in the derivation for the upper bound of $|g_{Qu, m}|^2$, when the photonic medium  is non-dispersive and supports discrete modes. Moreover, I provide a heuristic derivation for the upper bound of $|g_{Qu, m}|^2$, when the dispersion of the photonic medium can be described by a Lorentz model.

\subsection{The upper bound of $|g_{Qu}(\boldsymbol{r}_{e\perp}, \omega)|^2$}
\label{subsec:gQu_omega_bound}

The upper bound of $|g_{Qu}(\boldsymbol{r}_{e\perp}, \omega)|^2$ can be derived based on the explicit form shown in Eq.\,\ref{eq:gQu_omega}. Using the symmetry of the Green's function for non-magnetic medium
\begin{equation}
    \label{eq:G_symmetry}
    G_{ij}(\boldsymbol{r}, \boldsymbol{r}',\omega) = G_{ji}(\boldsymbol{r}', \boldsymbol{r}, \omega),
\end{equation}
Eq.\,\ref{eq:gQu_omega} can be re-written as
\begin{equation}
    \label{eq:gQu_omega_3}
    \begin{split}
    |g_{Qu}(\boldsymbol{r}_{e\perp}, \omega)|^2 = & \; \frac{1}{\pi\hbar} \int d^3 \boldsymbol{s} \Big[ -\frac{i\omega q_e}{\epsilon_0 c^2} \int dz' G_{iz}(\boldsymbol{s}, \boldsymbol{r}_{e\perp}+z'\boldsymbol{\hat{z}}, \omega) e^{-i\frac{\omega}{v}z'} \Big]^* \\ & \, \epsilon_0 \epsilon_I(\boldsymbol{s}, \omega) \Big[ -\frac{i\omega q_e}{\epsilon_0 c^2} \int dz G_{iz}(\boldsymbol{s}, \boldsymbol{r}_{e\perp} + z\boldsymbol{\hat{z}}, \omega) e^{-i\frac{\omega}{v} z}\Big] .
    \end{split}
\end{equation}
The term inside the bracket can be regarded as the full field ($\boldsymbol{E}_e$) generated by the current associated with the free electron with velocity $-v\boldsymbol{\hat{z}}$. 
\begin{equation}
    \label{eq:Ee_full}
    \bar{E}_{e,i}(\boldsymbol{r}', \omega) = -\frac{i\omega q_e}{\epsilon_0 c^2} \int dz G_{iz}(\boldsymbol{r}', \boldsymbol{r}_{e\perp} + z\boldsymbol{\hat{z}}, \omega) e^{-i\frac{\omega}{v} z}
\end{equation}
Equivalently, this field can be viewed as the full field when the incident field is the near field ($\boldsymbol{\bar{E}}_{e0}$) of the free electron with velocity $-v\boldsymbol{\hat{z}}$ in the free space.
\begin{equation}
    \label{eq:Ee0_2}
    \begin{split}
        \bar{E}_{e0,i}(\boldsymbol{r}', \omega) & = -\frac{i\omega q_e}{\epsilon_0 c^2} \int dz G_{0, iz}(\boldsymbol{r}', \boldsymbol{r}_{e\perp} + z\boldsymbol{\hat{z}}, \omega) e^{-i\frac{\omega}{v} z} \\ & = \frac{q_e}{2\pi \omega \epsilon_0} \exp (-ik_e z') \big[ i\alpha_e^2 K_0(\alpha_e \rho)\boldsymbol{\hat{z}} +k_e \alpha_e K_1( \alpha_e \rho)\boldsymbol{\hat{\rho}} \big]
    \end{split}
\end{equation}
where $k_e = \omega/v$, $k = \omega/c$, $\alpha_e = \sqrt{k_e^2 - k^2}$, $\rho = |\boldsymbol{r}'_\perp - \boldsymbol{r}_{e\perp}|$, $\boldsymbol{\hat{\rho}} = (\boldsymbol{r}'_\perp - \boldsymbol{r}_{e\perp})/|\boldsymbol{r}'_\perp - \boldsymbol{r}_{e\perp}|$, and $K_0$ and $K_1$ are the modified Bessel's functions of the second kind with order 0 and 1 respectively.
Therefore, Eq.\,\ref{eq:gQu_omega_3} is 
\begin{equation}
    \label{eq:gQu_omega_4}
    |g_{Qu}(\boldsymbol{r}_{e\perp}, \omega)|^2 = \frac{1}{\pi\hbar} \int d^3 \boldsymbol{r}' \boldsymbol{\bar{E}}_e^\dagger (\boldsymbol{r}', \omega) \epsilon_0 \epsilon_I(\boldsymbol{r}', \omega) \boldsymbol{\bar{E}}_e(\boldsymbol{r}', \omega).
\end{equation}
Recall the optical absorption \cite{miller2016fundamental}
\begin{equation}
    \label{eq:absorption}
    P_\textbf{abs} = \frac{\epsilon_0 \omega}{2} \int d^3 \boldsymbol{r} \boldsymbol{E}^\dagger(\boldsymbol{r}, \omega) \epsilon_I(\boldsymbol{r}, \omega) \boldsymbol{E}(\boldsymbol{r}, \omega).
\end{equation}
$|g_{Qu}(\boldsymbol{r}_{e\perp}, \omega)|^2$ is related to the optical absorption. Since the optical absorption should not exceed the optical extinction, which is the sum of absorption and scattering, the bound for the optical absorption can be found \cite{miller2016fundamental}. To maximize the absorption, the optimal total field is 
\begin{equation}
    \label{eq:Ee_opt}
    \boldsymbol{\bar{E}}_e^{\text{opt}}(\boldsymbol{r}, \omega) = i \frac{\chi^*(\boldsymbol{r}, \omega)}{\chi_I(\boldsymbol{r},\omega)} \boldsymbol{\bar{E}}_{e0}(\boldsymbol{r}, \omega),
\end{equation}
where $\chi$ is the susceptibility and $\chi_I$ is the imaginary part of the susceptibility.
Substitute the optimal field (Eq.\,\ref{eq:Ee_opt}) into Eq.\,\ref{eq:gQu_omega_4}.
\begin{equation}
    \label{eq:gQu_omega_bound}
    |g_{Qu}(\boldsymbol{r}_{e\perp}, \omega)|^2 \leq \frac{\epsilon_0}{\pi\hbar} \int d^3 \boldsymbol{r}' \frac{|\chi(\boldsymbol{r}', \omega)|^2}{\chi_I(\boldsymbol{r}', \omega)} \boldsymbol{\bar{E}}_{e0}^\dagger(\boldsymbol{r}', \omega) \boldsymbol{\bar{E}}_{e0}(\boldsymbol{r}', \omega),
\end{equation}
which is the upper bound for $|g_{Qu}(\boldsymbol{r}_{e\perp}, \omega)|^2$.
Substitute the form of $\boldsymbol{E}_{e0}$ using Eq.\,\ref{eq:Ee0_2}.
\begin{equation}
    \label{eq:gQu_omega_bound_2}
    |g_{Qu}(\boldsymbol{r}_{e\perp}, \omega)|^2 \leq \frac{q_e^2}{4\pi \hbar \epsilon_0 c} \frac{c}{\pi^2\omega^2} \int d^3 \boldsymbol{r}' \frac{|\chi(\boldsymbol{r}', \omega)|^2}{\chi_I(\boldsymbol{r}', \omega)} \big[\alpha_e^4 K_0^2(\alpha_e\rho) + k_e^2 \alpha_e^2 K_1^2(\alpha_e\rho) \big]
\end{equation}
Further, I assume the free-electron--light interaction length is $L$. I also take the maximal of the material related term:
\begin{equation}
    \label{eq:chi_m_omega}
    \Bigg[ \frac{|\chi|^2}{\chi_I}\Bigg]_\omega = \max_{\boldsymbol{r}} \frac{|\chi(\boldsymbol{r}, \omega)|^2}{\chi_I(\boldsymbol{r}, \omega)}.
\end{equation}
The upper bound becomes
\begin{equation}
    \label{eq:gQu_omega_bound_3}
    |g_{Qu}(\boldsymbol{r}_{e\perp}, \omega)|^2 \leq \frac{q_e^2}{4\pi \hbar \epsilon_0 c} \frac{2}{\pi\omega} \Bigg[ \frac{|\chi|^2}{\chi_I}\Bigg]_\omega \frac{kL}{2\pi}\int_{R_\perp} d^2 \boldsymbol{r}_\perp'  \Big[\frac{\alpha_e^4}{k^2} K_0^2(\alpha_e\rho) + \frac{k_e^2 \alpha_e^2}{k^2} K_1^2(\alpha_e\rho) \Big],
\end{equation}
where the subscript $R$ emphasizes that the integration is over the minimal region where the medium is located. This result has been shown in previous study \cite{yang2018maximal}.

\subsection{The upper bound of $|g_{Qu, m}|^2$}
\label{subsec:g_Qu_mode_bound}

I supplement the main text with a few steps in the derivation of the upper bound for the quantum coupling coefficient with a discrete mode ($|g_{Qu,m}|^2$), which is
\begin{equation}
    \label{eq:gQu_mode_m_2}
    |g_{Qu, m}|^2 = \frac{q_e^2}{2\hbar\omega_m \epsilon_0} \frac{ |\int dz e^{-i\frac{\omega_m}{v}z} E_{m,z}(\boldsymbol{r}_{e\perp}+z\hat{\boldsymbol{z}})|^2 }{ \int d^3 \boldsymbol{r} \boldsymbol{E}_m^{\dagger}(\boldsymbol{r}) \epsilon(\boldsymbol{r}) \boldsymbol{E}_m(\boldsymbol{r}) },
\end{equation}
where $\boldsymbol{E}_m$ is the eigen mode distribution without normalization. The permittivity $\epsilon(\boldsymbol{r})$ is treated as a Hermitian tensor here, such that the upper bound can be generalized to anisotropic media with negligible loss and dispersion. The electric field can be expressed from the polarization field ($\boldsymbol{P}$) using the free-space Green's function ($G_0$) \cite{miller2016fundamental}:
\begin{equation}
    \label{eq:E_G0_Einc}
    \boldsymbol{E}_m(\boldsymbol{r}) = \boldsymbol{E}_\text{inc}(\boldsymbol{r}) + \frac{\omega_m^2}{c^2 \epsilon_0} \int d^3\boldsymbol{r}' G_0(\boldsymbol{r},\boldsymbol{r}',\omega_m)\boldsymbol{P}_m(\boldsymbol{r}').
\end{equation}
Since $\boldsymbol{E}_m$ is the eigen mode, the incident field $\boldsymbol{E}_\text{inc}(\boldsymbol{r})=\boldsymbol{0}$. 
Substituting Eq.\,\ref{eq:E_G0_Einc} into the numerator in Eq.\,\ref{eq:gQu_mode_m_2}, one can get
\begin{equation}
    \label{eq:gQu_mode_nume_1}
    \begin{split}
        \Big| q_e \int dz e^{-i\frac{\omega_m}{v}z} E_{m,z}(\boldsymbol{r}_{e\perp}+z\hat{\boldsymbol{z}}) \Big|^2
        & = \Big| \frac{q_e \omega_m^2}{c^2 \epsilon_0} \int dz e^{-i\frac{\omega_m}{v}z} \int d^3 \boldsymbol{r}' G_{0, zi}(\boldsymbol{r}_{e\perp}+z\hat{\boldsymbol{z}}, \boldsymbol{r}', \omega_m) P_{m,i}(\boldsymbol{r}') \Big|^2 \\
        & = \Big|i\omega_m \int d^3 \boldsymbol{r}' \Big[- \frac{i q_e \omega_m}{\epsilon_0 c^2} \int dz G_{0, iz}(\boldsymbol{r}', \boldsymbol{r}_{e\perp}+z\hat{\boldsymbol{z}}, \omega_m) e^{-i\frac{\omega_m}{v}z} \Big] P_{m,i}(\boldsymbol{r}') \Big|^2 ,
    \end{split}
\end{equation}
where I have used the relations Eq.\,\ref{eq:G_symmetry}. Comparing with Eq.\,\ref{eq:Ee0_2}, the term in the bracket in Eq.\,\ref{eq:gQu_mode_nume_1} is the near field associated with the free electron with velocity $-v\hat{\boldsymbol{z}}$ in the free space. The near field associated with the free electron with velocity $v\hat{\boldsymbol{z}}$ is its complex conjugate, i.e., 
\begin{equation}
    \label{eq:Ee0}
    \begin{split}
    \boldsymbol{E}_{e0}(\boldsymbol{r}', \omega) & = [\bar{\boldsymbol{E}}_{e0}(\boldsymbol{r}', \omega)]^* \\ & = -\frac{q_e e^{ik_e z'}}{2\pi \epsilon_0 \omega} \big[ i\alpha_e^2 K_0(\alpha_e \rho)\hat{\boldsymbol{z}} - k_e \alpha_e K_1( \alpha_e \rho)\hat{\boldsymbol{\rho}} \big].
    \end{split}
\end{equation}
Thus,
\begin{equation}
    \label{eq:gQu_mode_m_3}
    |g_{Qu, m}|^2 = \frac{\omega_m}{2\hbar \epsilon_0} \frac{ \big|\int d^3 \boldsymbol{r} \boldsymbol{E}_{e0}^\dagger (\boldsymbol{r}, \omega_m) \boldsymbol{P}_m(\boldsymbol{r})\big|^2 }{ \int d^3 \boldsymbol{r} \boldsymbol{E}_m^{\dagger}(\boldsymbol{r}) \epsilon(\boldsymbol{r}) \boldsymbol{E}_m(\boldsymbol{r}) }.
\end{equation}
Using $\boldsymbol{P}_m(\boldsymbol{r}) = \epsilon_0 \chi(\boldsymbol{r}) \boldsymbol{E}_m(\boldsymbol{r})$, Eq.\,\ref{eq:gQu_mode_m_3} becomes
\begin{equation}
    \label{eq:gQu_mode_m_4}
    \begin{split}
        |g_{Qu, m}|^2 & = \frac{\omega_m \epsilon_0}{2\hbar} \frac{ \big|\int d^3 \boldsymbol{r} \boldsymbol{E}_{e0}^\dagger (\boldsymbol{r}, \omega_m) \chi(\boldsymbol{r}) \boldsymbol{E}_m(r)\big|^2 }{\int d^3 \boldsymbol{r} \boldsymbol{E}_m^{\dagger}(\boldsymbol{r}) \epsilon(\boldsymbol{r}) \boldsymbol{E}_m(\boldsymbol{r}) }\\ & = \frac{\omega_m \epsilon_0}{2\hbar} \frac{|\int d^3 \boldsymbol{r} \{[\epsilon(\boldsymbol{r})]^{-\frac{1}{2}} \chi(\boldsymbol{r}) \boldsymbol{E}_{e0}(\boldsymbol{r}, \omega_m)\}^\dagger\{[\epsilon(\boldsymbol{r})]^{\frac{1}{2}}\boldsymbol{E}_m(\boldsymbol{r})\}|^2}{\int d^3 \boldsymbol{r} \{[\epsilon(\boldsymbol{r})]^{\frac{1}{2}}\boldsymbol{E}_m(\boldsymbol{r})\}^\dagger \{[\epsilon(\boldsymbol{r})]^{\frac{1}{2}}\boldsymbol{E}_m(\boldsymbol{r})\}},
    \end{split}
\end{equation}
where $[\epsilon(\boldsymbol{r})]^\frac{1}{2}$ is a Hermitian matrix, since $\epsilon(\boldsymbol{r})$ is a Hermitian and positive definite tensor. The upper bound is obtained by applying the Cauchy-Schwarz inequality:
\begin{equation}
    \label{eq:cauchy-schwarz}
    \Big|\int d^3 \boldsymbol{r} [\epsilon^{-\frac{1}{2}} \chi \boldsymbol{E}_{e0}(\omega_m)]^\dagger [\epsilon^{\frac{1}{2}}\boldsymbol{E}_m]\Big|^2 \leq \int d^3 \boldsymbol{r} [\epsilon^{-\frac{1}{2}} \chi \boldsymbol{E}_{e0}(\omega_m)]^\dagger[\epsilon^{-\frac{1}{2}} \chi \boldsymbol{E}_{e0}(\omega_m)] \int d^3 \boldsymbol{r} [\epsilon^{\frac{1}{2}}\boldsymbol{E}_m]^\dagger [\epsilon^{\frac{1}{2}}\boldsymbol{E}_m],
\end{equation}
where the explicit $\boldsymbol{r}$ dependence is omitted to simplify the notation. Thus,
\begin{equation}
    \label{eq:gQu_mode_m_5}
    |g_{Qu, m}|^2 \leq \frac{\omega_m\epsilon_0}{2\hbar} \int d^3 \boldsymbol{r} \boldsymbol{E}_{e0}^\dagger(\boldsymbol{r}, \omega_m) \chi(\boldsymbol{r}) [\epsilon(\boldsymbol{r})]^{-1}\chi(\boldsymbol{r}) \boldsymbol{E}_{e0}(\boldsymbol{r}, \omega_m).
\end{equation}
The inequality is saturated under this condition:
\begin{equation}
    \label{eq:Cauchy_Schwarz_condition}
    \boldsymbol{E}_m(\boldsymbol{r}) = [\epsilon(\boldsymbol{r})]^{-1} \chi(\boldsymbol{r}) \boldsymbol{E}_{e0}(\boldsymbol{r}, \omega_m).
\end{equation}
Nevertheless, Eq.\,\ref{eq:Cauchy_Schwarz_condition} is typically not a solution to the Maxwell's equation. Then, the analytical upper bound is not saturable. 
To separate the material dependence and the structural dependence, one can take the material dependent term out of the integration. Define
\begin{equation}
    \label{eq:material_factor}
    \frac{|\chi|^2}{\epsilon} \equiv \max_{\boldsymbol{r}}\Big\{ \max_{\text{eigenvalue}}\big\{\chi(\boldsymbol{r}) [\epsilon(\boldsymbol{r})]^{-1}\chi(\boldsymbol{r})\big\}\Big\}.
\end{equation}
Then, 
\begin{equation}
    \label{eq:gQu_mode_m_6}
    |g_{Qu, m}|^2 \leq  \frac{\omega_m\epsilon_0}{2\hbar}  \frac{|\chi|^2}{\epsilon} \int_R d^3 \boldsymbol{r} \boldsymbol{E}_{e0}^\dagger (\boldsymbol{r}, \omega_m) \boldsymbol{E}_{e0} (\boldsymbol{r}, \omega_m),
\end{equation}
where the subscript $R$ denotes the region of the optical medium, since the polarization field is only nonzero within the region of the optical medium. Equation \ref{eq:gQu_mode_m_6} is the upper bound for $|g_{Qu, m}|^2$ shown in Eq.\,21 in the main text.

\subsection{The upper bound of $|g_{Qu,m}|^2$ with Lorentz dispersion}
\label{sec:Lorentz_dispersion}

In deriving the upper bound of $|g_{Qu,m}|^2$ in the main text, I assume that the material dispersion is negligible around the resonant frequency. This excludes important cases when the free electron interacts with the polariton, where the material dispersion plays an essential role. In this section, I extend the upper bound of $|g_{Qu,m}|^2$ to include materials with Lorentz dispersion near the considered resonant frequency. I derive the upper bound in the limit of negligible loss. Nevertheless, with the connection between $|g_{Qu,m}|^2$ and $\int d\omega |g_{Qu}(\omega)|^2$, which integrates over the mode bandwidth, as shown in Sec.\,\ref{sec:connections}, one can apply this upper bound to systems with low loss, which typically means that the mode bandwidth is smaller than the phase-matching bandwidth.  

The main difference from the derivation in the main text is that I adopt the method to formulate the eigenmode problem in a dispersive optical system as a Hermitian eigenvalue problem, as discussed in \cite{raman2010photonic}. Since some of the conventions used in \cite{raman2010photonic} are different from this manuscript, I repeat some key points of \cite{raman2010photonic} in the convention consistent with this manuscript. 

Suppose the relative permittivity of the optical medium can be described by a Lorentz dispersion model, which is the typical case for most materials:
\begin{equation}
    \label{eq:Lorentz_dispersion}
    \epsilon(\boldsymbol{r},\omega)=\epsilon_B(\boldsymbol{r}) + \epsilon_B(\boldsymbol{r}) \frac{\omega_p^2(\boldsymbol{r})}{\omega^2_0(\boldsymbol{r}) - \omega^2 -i\omega\gamma_L(\boldsymbol{r})}.
\end{equation}
I explicitly write the spatial dependence, since the optical medium can be spatially nonuniform. In the following derivation, the explicit spatial dependence is omitted to simplify the notation when the context is clear enough to indicate the spatial dependence. 
The first term in Eq.\,\ref{eq:Lorentz_dispersion} represents the non-dispersive background. The second term can be regarded as the response to the electric field from local oscillators. The microscopic picture of the second term is as following. The motion of the electrons, bounded by a harmonic potential and driven by the electric field, is
\begin{equation}
    \label{eq:Lorentz_microscopic}
    \frac{d^2 \boldsymbol{r}}{d t^2} + \gamma_L \frac{d \boldsymbol{r}}{dt} + \omega_0^2\boldsymbol{r} = \frac{q_e \boldsymbol{E}}{m}.
\end{equation}
The polarization originating from these oscillating electrons is therefore described by:
\begin{equation}
    \label{eq:Lorentz_microscopic_2}
    \frac{d^2 \boldsymbol{P}_L}{dt^2} + \gamma_L \frac{d \boldsymbol{P}_L}{dt} + \omega^2_0 \boldsymbol{P}_L = \omega_p^2 \epsilon_0 \epsilon_B \boldsymbol{E}.
\end{equation}
Thus, the total polarization is 
\begin{equation}
    \label{eq:polarization_Lorentz_total}
    \boldsymbol{P}(\boldsymbol{r}) = \boldsymbol{P}_B(\boldsymbol{r}) + \boldsymbol{P}_L(\boldsymbol{r}) = \epsilon_0 \chi_B(\boldsymbol{r})\boldsymbol{E} + \boldsymbol{P}_L(\boldsymbol{r}),
\end{equation}
where $\chi_B(\boldsymbol{r}) = \epsilon_B(\boldsymbol{r}) - 1$. The first part is referred to as the background polarization, and the second part is referred to as the oscillating polarization. 
To transform Eq.\,\ref{eq:Lorentz_microscopic_2} into first-order diffrentiation in time, an additional polarization velocity field $\boldsymbol{V}_L = \frac{d \boldsymbol{P}_L}{dt}$ is introduced \cite{raman2010photonic}. Then, the Maxwell's equations in such a dispersive medium are:
\begin{align}
    \begin{split}
        \frac{\partial \boldsymbol{H}}{\partial t} & = -\frac{1}{\mu_0}\nabla\times \boldsymbol{E}, \\
        \frac{\partial \boldsymbol{E}}{\partial t} & = \frac{1}{\epsilon_0 \epsilon_B}(\nabla\times \boldsymbol{H} - \boldsymbol{V}_L), \\
        \frac{\partial \boldsymbol{P}_L}{\partial t} & = \boldsymbol{V}_L, \\
        \frac{\partial \boldsymbol{V}_L}{\partial t} & = \omega_p^2 \epsilon_0 \epsilon_B \boldsymbol{E} - \omega_0^2 \boldsymbol{P}_L - \gamma_L \boldsymbol{V}_L.
    \end{split}
\end{align}
Suppose the system has a steady state solution with a time dependence $\exp(-i\omega t)$. Then, the following matrix equation must be satisfied:
\begin{equation}
    \label{eq:Lorentz_eigen_eq}
    \frac{\omega}{c} \begin{pmatrix}
        \tilde{\boldsymbol{H}} \\ \boldsymbol{E} \\ \boldsymbol{P}_L \\ \boldsymbol{V}_L 
    \end{pmatrix} = 
    \begin{pmatrix}
        0 & -i\nabla\times & 0 & 0 \\ \frac{i}{\epsilon_B}\nabla\times & 0 & 0 & -\frac{i}{\epsilon_0\epsilon_B c} \\ 0 & 0 & 0 & \frac{i}{c} \\ 0 & i\frac{\omega_p^2 \epsilon_0 \epsilon_B}{c} & -i\frac{\omega_0^2}{c} & -i\frac{\gamma_L}{c}
    \end{pmatrix} 
    \begin{pmatrix}
        \tilde{\boldsymbol{H}} \\ \boldsymbol{E} \\ \boldsymbol{P}_L \\ \boldsymbol{V}_L
    \end{pmatrix},
\end{equation}
where $\tilde{\boldsymbol{H}} = \sqrt{\frac{\mu_0}{\epsilon_0}}\boldsymbol{H}$. When $\omega_p = 0$, i.e., the oscillating dipole contribution is zero, $\boldsymbol{P}_L = \boldsymbol{0}$ and $\boldsymbol{V}_L = \boldsymbol{0}$ are a solution to Eq.\,\ref{eq:Lorentz_eigen_eq}, which relaxes to Eq.\,9 of the main text. In the limit of negligible loss ($\gamma_L\rightarrow 0$), Eq.\,\ref{eq:Lorentz_eigen_eq} is a Hermitian eigenvalue problem:
\begin{equation}
    \label{eq:Lorentz_eigen_eq_2}
    \frac{\omega}{c} 
    \underbrace{
    \begin{pmatrix}
        1 & & & \\ & \epsilon_B & & \\ & & \frac{\omega_0^2}{\omega_p^2 \epsilon_0^2 \epsilon_B} & \\ & & & \frac{1}{\omega_p^2\epsilon_0^2 \epsilon_B}
    \end{pmatrix} }_{\mathcal{A}}
    \begin{pmatrix}
        \tilde{\boldsymbol{H}} \\ \boldsymbol{E} \\ \boldsymbol{P}_L \\ \boldsymbol{V}_L 
    \end{pmatrix} = 
    \underbrace{
    \begin{pmatrix}
        0 & -i\nabla\times & 0 & 0 \\ i\nabla\times & 0 & 0 & -\frac{i}{\epsilon_0 c} \\ 0 & 0 & 0 & i\frac{\omega_0^2}{\omega_p^2 \epsilon_0^2 \epsilon_B c} \\ 0 & \frac{i}{\epsilon_0 c} & -i\frac{\omega_0^2}{\omega_p^2 \epsilon_0^2 \epsilon_B c} & 0
    \end{pmatrix} }_{\mathcal{B}}
    \begin{pmatrix}
        \tilde{\boldsymbol{H}} \\ \boldsymbol{E} \\ \boldsymbol{P}_L \\ \boldsymbol{V}_L
    \end{pmatrix},
\end{equation}
i.e., $\frac{\omega}{c} \mathcal{A} \boldsymbol{x} = \mathcal{B} \boldsymbol{x}$, where $\mathcal{A}$ is a Hermitian diagonal matrix and $\mathcal{B}$ is a Hermitian matrix, and $\boldsymbol{x} = [\tilde{\boldsymbol{H}}, \boldsymbol{E}, \boldsymbol{P}_L, \boldsymbol{V}_L]^T$. 
The eigenvectors corresponding to different eigenvalues are orthogonal such that $\boldsymbol{x}_m^\dagger \mathcal{A} \boldsymbol{x}_n = 0$ if $m\neq n$. Consistent with Eq.\,10 of the main text, the orthonormal condition for the fields of the eigenmode is 
\begin{equation}
    \label{eq:Lorentz_eigen_orthonomal}
    \int d^3\boldsymbol{r}\Big[\frac{1}{2}(\epsilon_B \boldsymbol{E}^\dagger_m \boldsymbol{E}_n + \tilde{\boldsymbol{H}}_m^\dagger \tilde{\boldsymbol{H}}_n) + \frac{1}{2\omega_p^2 \epsilon_0^2 \epsilon_B }(\boldsymbol{V}_{Lm}^\dagger \boldsymbol{V}_{Ln} + \omega_0^2 \boldsymbol{P}_{Lm}^\dagger \boldsymbol{P}_{Ln})\Big] = \delta_{mn}.
\end{equation}

With this normalization (Eq.\,\ref{eq:Lorentz_eigen_orthonomal}), the quantum coupling coefficient (Eq.\,16 of the main text) becomes
\begin{equation}
    \label{eq:gQu_mode_Lorentz_dispersion}
    |g_{Qu, m}|^2 = \frac{q_e^2}{2\hbar\omega_m \epsilon_0} \frac{ |\int dz e^{-i\frac{\omega_m}{v}z} E_{m,z}(\boldsymbol{r}_{e\perp}+z\boldsymbol{\hat{z}})|^2 }{ \int d^3\boldsymbol{r}\Big[\frac{1}{2}(\epsilon_B \boldsymbol{E}^\dagger_m \boldsymbol{E}_m + \tilde{\boldsymbol{H}}_m^\dagger \tilde{\boldsymbol{H}}_m) + \frac{1}{2\omega_p^2 \epsilon_0^2 \epsilon_B }(\boldsymbol{V}_{Lm}^\dagger \boldsymbol{V}_{Lm} + \omega_0^2 \boldsymbol{P}_{Lm}^\dagger \boldsymbol{P}_{Lm})\Big] }
\end{equation}
Similarly to the derivation before, the electric field can be expressed from the polarization field, which has a background and an oscillating contribution (Eq.\,\ref{eq:polarization_Lorentz_total}), using the free-space Green's function. Also, the polarization velocity field is related to the oscillating polarization field through $\boldsymbol{V}_{Lm} = -i\omega_m\boldsymbol{P}_{Lm}$. The quantum coupling coefficient takes the following form, which is an extension of Eq.\,18 of the main text.
\begin{equation}
    \label{eq:gQu_mode_Lorentz_dispersion_2}
    |g_{Qu, m}|^2 = \frac{\omega_m}{2\hbar\epsilon_0} \frac{ |\int d^3 \boldsymbol{r} \boldsymbol{E}_{e0}^\dagger (\omega_m)(\epsilon_0 \chi_B \boldsymbol{E}_m + \boldsymbol{P}_{Lm})|^2}{ \int d^3\boldsymbol{r}\Big[\frac{1}{2}(\epsilon_B \boldsymbol{E}^\dagger_m \boldsymbol{E}_m + \tilde{\boldsymbol{H}}_m^\dagger \tilde{\boldsymbol{H}}_m) + \frac{\omega_m^2 + \omega_0^2}{2\omega_p^2 \epsilon_0^2 \epsilon_B }(\boldsymbol{P}_{Lm}^\dagger \boldsymbol{P}_{Lm})\Big] }
\end{equation}
Further, the magnetic field in the denominator can be replaced, since $\boldsymbol{H}_m = \frac{1}{i\omega_m \mu_0}\nabla \times \boldsymbol{E}_m$ and $\nabla\times\boldsymbol{H}_m = -i\omega_m \epsilon_0\epsilon_B \boldsymbol{E}_m - i\omega_m \boldsymbol{P}_{Lm}$. 
\begin{equation}
\label{eq:magnetic_field_energy}
\begin{split}
    \int d^3 \boldsymbol{r} \tilde{\boldsymbol{H}}_m^\dagger \tilde{\boldsymbol{H}}_m & = \int d^3 \boldsymbol{r} \frac{\mu_0}{\epsilon_0} \boldsymbol{H}_m \cdot \boldsymbol{H}_m \\ & = \int d^3 \boldsymbol{r} \frac{\mu_0}{2\epsilon_0} \Big[\boldsymbol{H}_m\cdot \Big(\frac{1}{i\omega_m\mu_0}\nabla \times \boldsymbol{E}_m\Big) + \Big(\frac{1}{i\omega_m\mu_0}\nabla \times \boldsymbol{E}_m\Big)\cdot \boldsymbol{H}_m \Big] \\ & = \int d^3 \boldsymbol{r} \frac{\mu_0}{2\epsilon_0} \Big[\boldsymbol{E}_m\cdot \Big(\frac{i}{\omega_m\mu_0}\nabla \times \boldsymbol{H}_m\Big) + \Big(\frac{i}{\omega_m\mu_0}\nabla \times \boldsymbol{H}_m\Big)\cdot \boldsymbol{E}_m \Big] \\ & = \int d^3 \boldsymbol{r} \frac{1}{2} \Big[ \boldsymbol{E}_m \cdot \Big( \epsilon_B \boldsymbol{E}_m + \frac{\boldsymbol{P}_{Lm}}{\epsilon_0}\Big) + \Big( \epsilon_B \boldsymbol{E}_m + \frac{\boldsymbol{P}_{Lm}}{\epsilon_0}\Big) \cdot \boldsymbol{E}_m \Big] \\ & = \int d^3\boldsymbol{r} \Big[ \epsilon_B \boldsymbol{E}_m^\dagger \boldsymbol{E}_m + \frac{1}{2\epsilon_0} \Big(\boldsymbol{E}_m^\dagger \boldsymbol{P}_{Lm} + \boldsymbol{P}_{Lm}^\dagger \boldsymbol{E}_m\Big)\Big]
\end{split}
\end{equation}
From Eq.\,\ref{eq:magnetic_field_energy}, it is transparent to observe that the magnetic field energy is equal to the electric field energy when the optical medium is lossless and non-dispersive. On the other hand, $\int d^3 \boldsymbol{r} \tilde{\boldsymbol{H}}_m^\dagger \tilde{\boldsymbol{H}}_m \geq 0$. However, the form of the magnetic field energy (Eq.\,\ref{eq:magnetic_field_energy}), in terms of the electric field and polarization field, is not semi-definite. Thus, I relax this term in the denominator of Eq.\,\ref{eq:gQu_mode_Lorentz_dispersion_2} based on $\int d^3 \boldsymbol{r} \tilde{\boldsymbol{H}}_m^\dagger \tilde{\boldsymbol{H}}_m \geq 0$.
\begin{equation}
    \label{eq:gQu_mode_Lorentz_dispersion_3}
    \begin{split}
    |g_{Qu, m}|^2 & \leq \frac{\omega_m \epsilon_0}{\hbar} \frac{ \Big|\int d^3 \boldsymbol{r} \boldsymbol{E}_{e0}^\dagger (\omega_m)\Big(\chi_B \boldsymbol{E}_m + \frac{\boldsymbol{P}_{Lm}}{\epsilon_0}\Big)\Big|^2}{ \int d^3\boldsymbol{r}\Big[\epsilon_B \boldsymbol{E}^\dagger_m \boldsymbol{E}_m + \frac{\omega_m^2 + \omega_0^2}{\omega_p^2 \epsilon_B }\Big(\frac{\boldsymbol{P}_{Lm}}{\epsilon_0}\Big)^\dagger \Big(\frac{\boldsymbol{P}_{Lm}}{\epsilon_0}\Big)\Big] } 
    \end{split}
\end{equation}
Furthermore, as a solution to Eq.\,\ref{eq:Lorentz_eigen_eq}, the oscillating polarization ($\boldsymbol{P}_{Lm}$) is related to the electric field ($\boldsymbol{E}_m$) through
\begin{equation}
    \label{eq:relation_P_L_E}
    \frac{\boldsymbol{P}_{Lm}}{\epsilon_0} = \epsilon_B\frac{\omega_p^2}{\omega_0^2 - \omega_m^2} \boldsymbol{E}_m,
\end{equation}
where $\gamma_L\rightarrow 0$ is assumed. Substitute Eq.\,\ref{eq:relation_P_L_E} into Eq.\,\ref{eq:gQu_mode_Lorentz_dispersion_3}, one obtains
\begin{equation}
    \label{eq:gQu_mode_Lorentz_dispersion_3_2}
    \begin{split}
    |g_{Qu, m}|^2 & \leq \frac{\omega_m \epsilon_0}{\hbar} \frac{ \Big|\int d^3 \boldsymbol{r} \boldsymbol{E}_{e0}^\dagger (\omega_m)\Big(\chi_B + \epsilon_B\frac{\omega_p^2}{\omega_0^2 - \omega_m^2} \Big) \boldsymbol{E}_m\Big|^2}{ \int d^3\boldsymbol{r}\Big[\epsilon_B + \epsilon_B \frac{(\omega_0^2 + \omega_m^2) \omega_p^2}{(\omega_0^2 - \omega_m^2)^2} \Big] \boldsymbol{E}^\dagger_m \boldsymbol{E}_m } .
    \end{split}
\end{equation}
Equation \ref{eq:gQu_mode_Lorentz_dispersion_3_2} takes the form $|\boldsymbol{a}^\dagger \boldsymbol{b}|^2/{\boldsymbol{b}^\dagger\boldsymbol{b}}$, where
\begin{equation}
\begin{split}
    \boldsymbol{a} & = \Big[\epsilon_B + \epsilon_B \frac{(\omega_0^2 + \omega_m^2) \omega_p^2}{(\omega_0^2 - \omega_m^2)^2} \Big]^{-\frac{1}{2}}\Big(\chi_B + \epsilon_B\frac{\omega_p^2}{\omega_0^2 - \omega_m^2}\Big) \boldsymbol{E}_{e0}(\omega_m),\\
    \boldsymbol{b} & =\Big[\epsilon_B + \epsilon_B \frac{(\omega_0^2 + \omega_m^2) \omega_p^2}{(\omega_0^2 - \omega_m^2)^2} \Big]^{\frac{1}{2}} \boldsymbol{E}_m.
\end{split} 
\end{equation}
Applying the Cauchy-Schwarz inequality, i.e., $|\boldsymbol{a}^\dagger \boldsymbol{b}|^2/{\boldsymbol{b}^\dagger\boldsymbol{b}} \leq \boldsymbol{a}^\dagger \boldsymbol{a}$, one gets
\begin{equation}
    \label{eq:gQu_mode_Lorentz_dispersion_4}
    |g_{Qu, m}|^2 \leq \frac{\omega_m \epsilon_0}{\hbar} \int d^3 \boldsymbol{r} \frac{\big|\chi_B+\epsilon_B\frac{\omega_p^2}{\omega_0^2 - \omega_m^2}\big|^2}{\epsilon_B + \epsilon_B \frac{(\omega_0^2+\omega_m^2)\omega_p^2}{(\omega_0^2-\omega_m^2)^2}} \boldsymbol{E}_{e0}^\dagger(\omega_m) \boldsymbol{E}_{e0}(\omega_m).
\end{equation}
Comparing with Eq.\,\ref{eq:gQu_mode_m_5} (or Eq.\,21 of the main text), only the material-dependent part is different, as expected. 
The condition for the Cauchy-Schwarz inequality to take an `equal' sign is $\boldsymbol{b} = \boldsymbol{a}$, i.e.,
\begin{equation}
    \label{eq:Cauchy_Schwarz_equal_condition}
    \boldsymbol{E}_m = \Big[\epsilon_B + \epsilon_B \frac{(\omega_0^2 + \omega_m^2) \omega_p^2}{(\omega_0^2 - \omega_m^2)^2} \Big]^{-1}\Big(\chi_B + \epsilon_B\frac{\omega_p^2}{\omega_0^2 - \omega_m^2}\Big) \boldsymbol{E}_{e0}(\omega_m).
\end{equation}
In generic situations, Eq.\,\ref{eq:Cauchy_Schwarz_equal_condition} is not a solution of Maxwell's equations, and the upper bound is not saturable. Nevertheless, when the eigenmodes have a distribution close to the optimal condition (Eq.\,\ref{eq:Cauchy_Schwarz_equal_condition}), the quantum coupling coefficient can be close to the upper bound, which will be shown in Sec.\,\ref{sec:approach_upper_bounds}. Further, one can take the material dependence out of the volume integration to separate the material dependence and the geometrical dependence.
\begin{equation}
    \label{eq:gQu_mode_Lorentz_dispersion_5}
    |g_{Qu, m}|^2 \leq \frac{\omega_m \epsilon_0}{\hbar} \Bigg[ \frac{\big|\chi_B+\epsilon_B\frac{\omega_p^2}{\omega_0^2 - \omega_m^2}\big|^2}{\epsilon_B + \epsilon_B \frac{(\omega_0^2+\omega_m^2)\omega_p^2}{(\omega_0^2-\omega_m^2)^2}}\Bigg]_\text{max} \int d^3 \boldsymbol{r} \boldsymbol{E}_{e0}^\dagger(\omega_m) \boldsymbol{E}_{e0}(\omega_m),
\end{equation}
where 
\begin{equation}
   \Bigg[ \frac{\big|\chi_B+\epsilon_B\frac{\omega_p^2}{\omega_0^2 - \omega_m^2}\big|^2}{\epsilon_B + \epsilon_B \frac{(\omega_0^2+\omega_m^2)\omega_p^2}{(\omega_0^2-\omega_m^2)^2}}\Bigg]_\text{max} = \max_{\boldsymbol{r}}\Bigg[ \frac{\big|\chi_B(\boldsymbol{r})+\epsilon_B(\boldsymbol{r})\frac{\omega_p^2(\boldsymbol{r})}{\omega_0^2(\boldsymbol{r}) - \omega_m^2}\big|^2}{\epsilon_B(\boldsymbol{r}) + \epsilon_B(\boldsymbol{r}) \frac{(\omega_0^2(\boldsymbol{r})+\omega_m^2)\omega_p^2(\boldsymbol{r})}{(\omega_0^2(\boldsymbol{r}) -\omega_m^2)^2}}\Bigg]_\text{max}.
\end{equation}
With the explicit form for $\boldsymbol{E}_{e0}$ and assuming that the interaction length is $L$, the upper bound of $|g_{Qu,m}|^2$ becomes
\begin{equation}
    \label{eq:gQu_mode_Lorentz_dispersion_6}
    |g_{Qu, m}|^2 \leq \frac{q_e^2}{2\pi\hbar c \epsilon_0} \Bigg[ \frac{\big|\chi_B+\epsilon_B\frac{\omega_p^2}{\omega_0^2 - \omega_m^2}\big|^2}{\epsilon_B + \epsilon_B \frac{(\omega_0^2+\omega_m^2)\omega_p^2}{(\omega_0^2-\omega_m^2)^2}}\Bigg]_\text{max} \frac{kL}{2\pi} \int_{R_\perp} d^2 \boldsymbol{r}_\perp \;\Big[\frac{\alpha_e^4}{k^2} K_0^2(\alpha_e \rho)  + \frac{k_e^2 \alpha_e^2}{k^2} K_1^2(\alpha_e \rho)\Big].
\end{equation}
Equation \ref{eq:gQu_mode_Lorentz_dispersion_6} is the extension of Eq.\,22 of the main text, where the optical medium can be dispersive, as described by the Lorentz model. Furthermore, metals that can be described by the Drude model can be viewed as a special case of the Lorentz model, where $\omega_0 = 0$. Thus, for the dispersive optical medium described by the Drude model, the upper bound shown in Eq.\,\ref{eq:gQu_mode_Lorentz_dispersion_6} still holds.

\section{Consistency and connections between the discrete and continuous modes}
\label{sec:connections}
In this section, I show the connections and consistency between $g_{Qu}(\omega)$ and $g_{Qu, m}$ when the loss and dispersion are small.
I use the model decomposition of the Green's function, when the loss and dispersion becomes negligible, to show the connection between $g_{Qu, m}$ and $g_{Qu}(\omega)$.

\subsection{Model decomposition of the Green's function}
\label{sec:model decomposition}
The Green's function can be decomposed using the eigen modes when the loss and dispersion are negligible. With a small decay rate $\gamma_d$, the frequency domain Maxwell's equation becomes
\begin{equation}
    \label{eq:Maxwell_freq_loss}
    \Big[ -\frac{\omega^2}{c^2} \epsilon(\boldsymbol{r}) -\frac{i\omega \gamma_d}{c^2} \epsilon(\boldsymbol{r}) + \nabla\times\nabla\times \Big] \boldsymbol{E}(\boldsymbol{r}, \omega) = i\omega \mu_0 \boldsymbol{J}(\boldsymbol{r}, \omega).
\end{equation}
Suppose the field can be decomposed as a frequency dependent sum of the eigen modes,
\begin{equation}
    \label{eq:E_freq_decomp}
    \boldsymbol{E}(\boldsymbol{r}, \omega) = \sum_m a_m(\omega) \boldsymbol{U}_m(\boldsymbol{r}),
\end{equation}
where $\boldsymbol{U}_m$ are defined as
\begin{equation}
    \label{eq:Um_1}
    \nabla\times\nabla\times \boldsymbol{U}_m(\boldsymbol{r})  - \frac{\omega_m^2}{c^2} \epsilon(\boldsymbol{r}, \omega_m) \boldsymbol{U}_m(\boldsymbol{r}) = 0,
\end{equation}
\begin{equation}
    \label{eq:Um_normalizatin}
    \int d^3 \boldsymbol{r}\epsilon(\boldsymbol{r}, \omega_m) \boldsymbol{U}_{m,i}(\boldsymbol{r})\boldsymbol{U}_{n,i}^*(\boldsymbol{r}) = \delta_{mn}.
\end{equation}
Then,
\begin{equation}
    \label{eq:E_freq_decomp_2}
    \sum_m a_m(\omega) \Big[ -\frac{\omega^2}{c^2} \epsilon(\boldsymbol{r}) -\frac{i\omega \gamma_d}{c^2} \epsilon(\boldsymbol{r}) + \frac{\omega_m^2}{c^2} \epsilon(\boldsymbol{r}) \Big] \boldsymbol{U}_m(\boldsymbol{r}) = i\omega \mu_0 \boldsymbol{J}(\boldsymbol{r}, \omega).
\end{equation}
With the orthogonal relation of $\{\boldsymbol{U}_m\}$,
\begin{equation}
    \label{eq:E_freq_decomp_3}
    a_m(\omega) = \frac{i\omega \mu_0 c^2}{-\omega^2 - i\omega \gamma_d + \omega_m^2} \int d^3 \boldsymbol{r} \boldsymbol{U}_m^\dagger(\boldsymbol{r}) \boldsymbol{J}(\boldsymbol{r}, \omega).
\end{equation}
Thus,
\begin{equation}
    \label{eq:E_freq_decomp_4}
    \boldsymbol{E}(\boldsymbol{r}, \omega) = i\omega\mu_0 \int d^3 \boldsymbol{r}' \sum_m \frac{c^2}{-\omega^2 - i\omega\gamma_d + \omega_m^2}\boldsymbol{U}_m(\boldsymbol{r}) \boldsymbol{U}_m^\dagger(\boldsymbol{r}') \boldsymbol{J}(\boldsymbol{r}', \omega)
\end{equation}
Comparing Eq.\,\ref{eq:E_freq_decomp_4} with Eq.\,\ref{eq:E_operator_G_freq}, one can get the model decomposition of the Green's function in the limit of low loss.
\begin{equation}
    \label{eq:G_freq_mode_decomp}
    G(\boldsymbol{r}, \boldsymbol{r}',\omega) = \sum_m \frac{c^2}{-\omega^2 - i\omega\gamma_d + \omega_m^2}\boldsymbol{U}_m(\boldsymbol{r}) \boldsymbol{U}_m^\dagger(\boldsymbol{r}') 
\end{equation}

\subsection{Connections between $|g_{Qu,m}|^2$ and $|g_{Qu}(\omega)|^2$}
\label{sec:consistency_gQu_gQu_omega}
Using the model decomposition of the Green's function, one can find the connections between  $|g_{Qu,m}|^2$ and $|g_{Qu}(\omega)|^2$.
I take the following approximation for the frequency dependent term in Eq.\,\ref{eq:G_freq_mode_decomp}, since the decay rate is much smaller than the resonant frequency, i.e., $\gamma_d \ll \omega_m$. 
\begin{equation}
    \label{eq:frequency_dependence}
    \begin{split}
        \frac{1}{-\omega^2-i\omega\gamma_d+\omega_m^2} & = \frac{1}{-(\omega - \omega_m)(\omega + \omega_m) - i\omega\gamma_d} \\ & \approx \frac{1}{-2\omega_m(\omega - \omega_m + i\frac{\gamma_d}{2})} \\ & = -\frac{1}{2\omega_m}\frac{\omega-\omega_m - i\frac{\gamma_d}{2}}{(\omega-\omega_m)^2 + \frac{\gamma_d^2}{4}}
    \end{split}
\end{equation}
With this approximation, the model decomposition of the Green's function becomes
\begin{equation}
    \label{eq:G_freq_mode_decomp_2}
    G(\boldsymbol{r}, \boldsymbol{r}',\omega) = - \sum_m \frac{c^2}{2\omega_m} \frac{\omega-\omega_m - i\frac{\gamma_d}{2}}{(\omega-\omega_m)^2 + \frac{\gamma_d^2}{4}} \boldsymbol{U}_m(\boldsymbol{r}) \boldsymbol{U}_m^\dagger(\boldsymbol{r}').
\end{equation}

Substitute this model-decomposed Green's function into the expression for $g_{Qu}(\omega)$ (Eq.\,\ref{eq:gQu_omega_2}),
\begin{equation}
    \label{eq:gQu_omega_mode}
    \begin{split}
    |g_{Qu}(\omega)|^2 = &\; \frac{q_e^2}{\hbar \pi \epsilon_0} \int dz \int dz' \text{Re}\Big[  \sum_m \frac{1}{2\omega_m} \frac{i(\omega-\omega_m) + \frac{\gamma_d}{2}}{(\omega-\omega_m)^2 + \frac{\gamma_d^2}{4}} e^{i\frac{\omega}{v}(z'- z)} U_{m,z}(\boldsymbol{r}_{e\perp}+z\boldsymbol{\hat{z}}) U_{m,z}^*(\boldsymbol{r}_{e\perp}+z'\boldsymbol{\hat{z}})\Big] \\ = & \; \frac{q_e^2}{\hbar \pi \epsilon_0 } \sum_m \frac{1}{2\omega_m} \frac{\frac{\gamma_d}{2}}{(\omega-\omega_m)^2 + \frac{\gamma_d^2}{4}} \Big| \int dz e^{-i\frac{\omega}{v}z}U_{m,z}(\boldsymbol{r}_{e\perp}+z\boldsymbol{\hat{z}})\Big|^2.
    \end{split}
\end{equation}
At resonant frequency $\omega_m$, I assume that only mode m contributes and all other modes are negligible. Then,
\begin{equation}
    \label{eq:gQu_omega_m}
    |g_{Qu}(\omega_m)|^2 = \frac{q_e^2}{\hbar \pi \epsilon_0 \omega_m \gamma_d} \Big| \int dz e^{-i\frac{\omega_m}{v}z}U_{m,z}(\boldsymbol{r}_{e\perp}+z\boldsymbol{\hat{z}})\Big|^2.
\end{equation}
Thus, the on-resonant $|g_{Qu}(\omega_m)|^2$ has a resonant enhancement scaling with $1/\gamma_d$.

Next, I consider the integration of $|g_{Qu}(\omega)|^2$ over a bandwidth $\delta\omega$ that can cover the resonance. I assume that the resonant bandwidth is smaller than the mode frequency separation, such that only one mode m has a contribution within the bandwidth.
\begin{equation}
    \label{eq:gQu_freq_integral}
    \begin{split}
    \int_{\omega_m-\frac{\delta\omega}{2}}^{\omega_m+\frac{\delta\omega}{2}} d\omega |g_{Qu}(\omega)|^2 = \frac{q_e^2}{\hbar \pi \epsilon_0 2\omega_m} \int_{\omega_m-\frac{\delta\omega}{2}}^{\omega_m+\frac{\delta\omega}{2}} d\omega \frac{\frac{\gamma_d}{2}}{(\omega-\omega_m)^2 + \frac{\gamma_d^2}{4}} \Big| \int dz e^{-i\frac{\omega}{v}z}U_{m,z}(\boldsymbol{r}_{e\perp}+z\boldsymbol{\hat{z}})\Big|^2
    \end{split}
\end{equation}
Since the decay rate is infinitesimal, the last term is almost independent of the frequency. The integration of the frequency dependent term is
\begin{equation}
    \label{eq:frequency_integral}
    \int_{\omega_m-\frac{\delta\omega}{2}}^{\omega_m+\frac{\delta\omega}{2}} d\omega \frac{\frac{\gamma_d}{2}}{(\omega-\omega_m)^2 + \frac{\gamma_d^2}{4}} \approx \int d\omega \frac{\frac{\gamma_d}{2}}{(\omega-\omega_m)^2 + \frac{\gamma_d^2}{4}} = \pi.
\end{equation}
Thus,
\begin{equation}
    \label{eq:gQu_freq_integral_2}
    \int_{\omega_m-\frac{\delta\omega}{2}}^{\omega_m+\frac{\delta\omega}{2}} d\omega  |g_{Qu}(\omega)|^2 \approx \frac{q_e^2}{2\hbar  \epsilon_0 \omega_m} \Big| \int dz e^{-i\frac{\omega}{v}z}U_{m,z}(\boldsymbol{r}_{e\perp}+z\boldsymbol{\hat{z}})\Big|^2.
\end{equation}
Comparing Eq.\,\ref{eq:gQu_freq_integral_2} with Eq.\,\ref{eq:gQu_mode_m_2}, I find the connection between $|g_{Qu,m}|^2$ and $|g_{Qu}(\omega)|^2$
\begin{equation}
    \label{eq:gQu_omega_gQu_relation}
     \int_{\omega_m-\frac{\delta\omega}{2}}^{\omega_m+\frac{\delta\omega}{2}} d\omega |g_{Qu}(\omega)|^2 \approx |g_{Qu,m}|^2.
\end{equation}

\section{Influence of the transverse distribution of the free-electron wave function}
\label{sec:trans}

In this section, I discuss the influence of the finite transverse spread of the free-electron wave function on the quantum coupling coefficient and the upper bound. I extend the expression of the quantum coupling coefficient, as well as the upper bounds, with the consideration of the free-electron transverse wave function. Then, I show numerical examples when the free-electron wave function has a Gaussian transverse distribution. I also study the upper bound with the estimated maximal interaction length, which originate from the diffraction of the free electron with a finite transverse spread.

\subsection{The quantum coupling coefficient with a finite free-electron transverse spread}

With the explicit consideration of the transverse free-electron wave function $\phi_e(r_\perp)$, the scattering matrix shown in Eq.\,\ref{eq:S_3} can be reformulated into
\begin{equation}
    \label{eq:S_6}
    \hat{S} = \exp({i\hat{\chi}})\exp\Big\{{\int_0^{+\infty} d\omega \big [g_{Qu}(\omega) \hat{b}^\dagger_{\frac{\omega}{v}} \hat{a}_\omega - g_{Qu}^*(\omega) \hat{b}_{\frac{\omega}{v}} \hat{a}^\dagger_\omega \big]}\Big\},
\end{equation}
where 
\begin{equation}
    \label{eq:gQu_a_2}
    g_{Qu}(\omega)\hat{a}_\omega = \frac{i q_e}{\hbar}\int dz e^{-i\frac{\omega}{v}z} \int d^2 \boldsymbol{r}_\perp \phi_e^*(\boldsymbol{r}_\perp) \phi_e(\boldsymbol{r}_\perp)\hat{A}_z(\boldsymbol{r}_\perp+z\boldsymbol{\hat{z}},\omega).
\end{equation}
Following the same procedure outline in Sec.\,\ref{sec:gQu_general} and the main text, one can obtain the expression of the quantum coupling coefficient with the consideration of the transverse free-electron wave function, coupling with photonic excitations in a general optical medium ($|g_{Qu}(\omega)|^2$), or coupling with discrete photonic modes ($|g_{Qu, m}|^2$). 
\begin{equation}
    \label{eq:gQu_omega_trans}
    \begin{split}
    |g_{Qu}(\omega)|^2 = & \; \frac{q_e^2\omega^2}{\hbar \pi \epsilon_0 c^4} \int dz \int dz' e^{i\frac{\omega}{v}(z'- z)} \int d^3 \boldsymbol{s} \Big[\int d^2 \boldsymbol{r}_\perp \phi_e^*(\boldsymbol{r}_\perp) \phi_e(\boldsymbol{r}_\perp) G_{zi}(\boldsymbol{r}_\perp + z\boldsymbol{\hat{z}}, s, \omega)\Big] \\ & \; \Big[ \int d^2 \boldsymbol{r}_\perp' \phi_e^*(\boldsymbol{r}_\perp') \phi_e(\boldsymbol{r}_\perp')  G^*_{zi}(\boldsymbol{r}_\perp' + z'\boldsymbol{\hat{z}}, \boldsymbol{s}, \omega) \Big] \epsilon_I(\boldsymbol{s}, \omega) \\ = & \; \frac{q_e^2}{\hbar \pi \epsilon_0 c^2}  \int dz \int dz' \text{Re}\Big[ -i e^{i\frac{\omega}{v}(z'- z)} \int d^2 \boldsymbol{r}_\perp \int d^2 \boldsymbol{r}_\perp' \phi_e^*(\boldsymbol{r}_\perp) \phi_e(\boldsymbol{r}_\perp)  \phi_e^*(\boldsymbol{r}_\perp') \phi_e(\boldsymbol{r}_\perp') \\ & \; G_{zz}(\boldsymbol{r}_\perp + z\boldsymbol{\hat{z}}, \boldsymbol{r}_\perp' + z'\boldsymbol{\hat{z}},\omega) \Big] ,
    \end{split}
\end{equation}
which is an extension of Eqs.\,\ref{eq:gQu_omega} and \ref{eq:gQu_omega_2}.
\begin{equation}
    \label{eq:gQu_mode_m_trans}
    |g_{Qu, m}|^2 = \frac{q_e^2}{2\hbar\omega_m \epsilon_0} \frac{ |\int dz e^{-i\frac{\omega_m}{v}z} \int d^2 \boldsymbol{r}_\perp \phi_e^*(\boldsymbol{r}_\perp) \phi_e(\boldsymbol{r}_\perp) E_{m,z}(\boldsymbol{r}_\perp+z\boldsymbol{\hat{z}})|^2 }{ \int d^3 \boldsymbol{r} \epsilon(\boldsymbol{r}) \boldsymbol{E}_m^{\dagger}(\boldsymbol{r}) \boldsymbol{E}_m(\boldsymbol{r}) },
\end{equation}
which is an extension of Eq.\,\ref{eq:gQu_mode_m_2}.

In summary, the influence of the free-electron transverse wave function is a spatial average of the Green's function (or the eigenmode), with respect to the free-electron transverse density distribution.

\subsection{The upper bound of the quantum coupling coefficient with a finite free-electron transverse spread}

From the expressions of the quantum coupling coefficient, it is straightforward to obtain the upper bound of the quantum coupling coefficient with a finite free-electron transverse spread, following the derivations in Sec.\,\ref{sec:gQu_omega_bound}. In the derivation, the main modification originates from the electromagnetic field carried by the free electron with a finite transverse distribution, which is an extension of Eqs.\,\ref{eq:Ee0_2} and \ref{eq:Ee0}.
\begin{equation}
    \label{eq:Ee0_trans}
    \begin{split}
    \boldsymbol{E}_{e0}(\boldsymbol{r}', \omega) = &\; [\boldsymbol{\bar{E}}_{e0}(\boldsymbol{r}', \omega)]^* \\ = &\; -\frac{q_e}{2\pi \epsilon_0 \omega} \exp({ik_e z'}) \Big[ i\alpha_e^2 \int d^2 \boldsymbol{r}_\perp |\phi_e(\boldsymbol{r}_\perp)|^2 K_0(\alpha_e |\boldsymbol{r}_\perp' - \boldsymbol{r}_\perp|)\boldsymbol{\hat{z}} \\ & \;- k_e \alpha_e \int d^2 \boldsymbol{r}_\perp |\phi_e(\boldsymbol{r}_\perp)|^2 K_1( \alpha_e |\boldsymbol{r}_\perp' - \boldsymbol{r}_\perp|)\frac{\boldsymbol{r}_\perp' - \boldsymbol{r}_\perp}{|\boldsymbol{r}_\perp' - \boldsymbol{r}_\perp|} \Big].
    \end{split}
\end{equation}
Substitute Eq.\,\ref{eq:Ee0_trans} into Eqs.\,\ref{eq:gQu_omega_bound} and \ref{eq:gQu_mode_m_6}, one can obtain the upper bound of the quantum coupling coefficient with a finite free-electron transverse spread. The explicit forms are as follows:
\begin{equation}
    \label{eq:gQu_omega_bound_trans}
    \begin{split}
    |g_{Qu}(\omega)|^2 \leq &\; \frac{q_e^2}{4\pi \hbar \epsilon_0 c} \frac{2}{\pi\omega} \Bigg[ \frac{|\chi|^2}{\chi_I}\Bigg]_\omega \frac{kL}{2\pi}\int_{R_\perp} d^2 \boldsymbol{r}_\perp' \Big\{\frac{\alpha_e^4}{k^2} \Big[\int d^2 \boldsymbol{r}_\perp |\phi_e(\boldsymbol{r}_\perp)|^2 K_0(\alpha_e |\boldsymbol{r}_\perp' - \boldsymbol{r}_\perp|)\Big]^2 \\ & \; + \frac{k_e^2 \alpha_e^2}{k^2} \Big|\int d^2 \boldsymbol{r}_\perp |\phi_e(\boldsymbol{r}_\perp)|^2 K_1( \alpha_e |\boldsymbol{r}_\perp' - \boldsymbol{r}_\perp|)\frac{\boldsymbol{r}_\perp' - \boldsymbol{r}_\perp}{|\boldsymbol{r}_\perp' - \boldsymbol{r}_\perp|} \Big|^2 \Big\},
    \end{split}
\end{equation}
which is an extension of Eq.\,\ref{eq:gQu_omega_bound_3}; and 
\begin{equation}
    \label{eq:gQu_mode_m_bound_trans}
    \begin{split}
    |g_{Qu, m}|^2 < g_\text{ub}^2 \equiv & \; \frac{q_e^2}{4\pi\hbar c \epsilon_0} \frac{|\chi|^2}{\epsilon} \frac{kL}{2\pi} \int_{R_\perp} d^2 \boldsymbol{r}_\perp' \Big\{\frac{\alpha_e^4}{k^2} \Big[\int d^2 \boldsymbol{r}_\perp |\phi_e(\boldsymbol{r}_\perp)|^2 K_0(\alpha_e |\boldsymbol{r}_\perp' - \boldsymbol{r}_\perp|)\Big]^2 \\ & \; + \frac{k_e^2 \alpha_e^2}{k^2} \Big|\int d^2 \boldsymbol{r}_\perp |\phi_e(\boldsymbol{r}_\perp)|^2 K_1( \alpha_e |\boldsymbol{r}_\perp' - \boldsymbol{r}_\perp|)\frac{\boldsymbol{r}_\perp' - \boldsymbol{r}_\perp}{|\boldsymbol{r}_\perp' - \boldsymbol{r}_\perp|} \Big|^2 \Big\},
    \end{split}
\end{equation}
which is an extension of Eq.\,22 in the main text. The geometric factor is
\begin{equation}
    \label{eq:g_geo}
    \begin{split}
    g_\text{geo}^2 = & \; \int_{R_\perp} d^2 \boldsymbol{r}_\perp' \Big\{\frac{\alpha_e^4}{k^2} \Big[\int d^2 \boldsymbol{r}_\perp |\phi_e(\boldsymbol{r}_\perp)|^2 K_0(\alpha_e |\boldsymbol{r}_\perp' - \boldsymbol{r}_\perp|)\Big]^2 \\ & \; + \frac{k_e^2 \alpha_e^2}{k^2} \Big|\int d^2 \boldsymbol{r}_\perp |\phi_e(\boldsymbol{r}_\perp)|^2 K_1( \alpha_e |\boldsymbol{r}_\perp' - \boldsymbol{r}_\perp|)\frac{\boldsymbol{r}_\perp' - \boldsymbol{r}_\perp}{|\boldsymbol{r}_\perp' - \boldsymbol{r}_\perp|} \Big|^2 \Big\}.
    \end{split}
\end{equation}

\subsection{Numerical demonstration with a finite free-electron transverse spread}


\begin{figure}
    \centering
    \includegraphics[width=12cm]{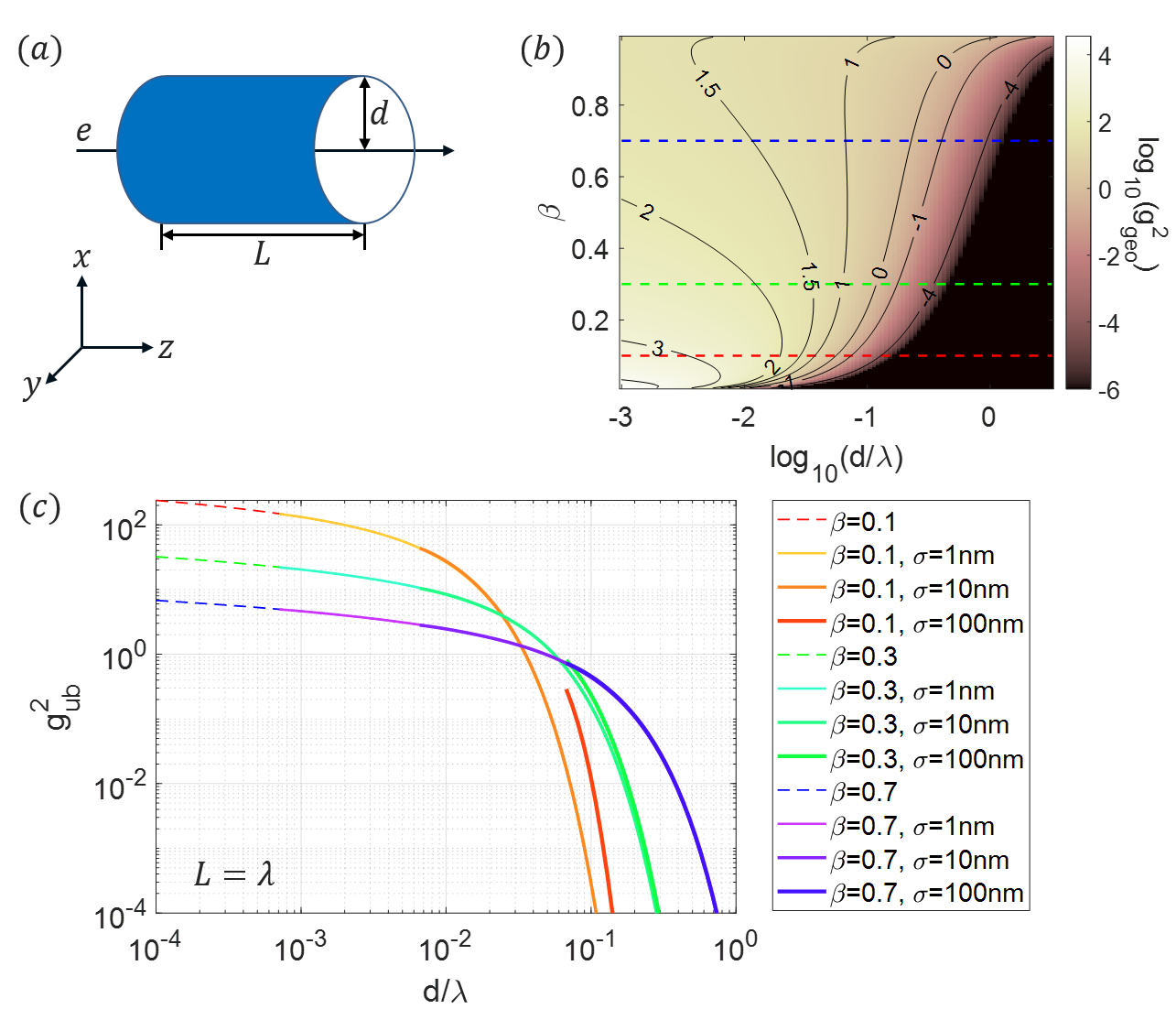}
    \caption{(a) A schematic showing that the photonic medium is separated from the free-electron trajectory by a minimal distance $d$. (b) Log scale $g^2_\text{geo}$ as a function of $d$ and $\beta$. Only $g_\text{geo}^2 > 10^{-6}$ is shown. The red, green and blue dashed lines represent $\beta=0.1$, $\beta=0.3$, and $\beta=0.7$ respectively. (c) The upper bound of $|g_{Qu, m}|^2$ as a function of $d$ for silicon structures ($\epsilon=12$), where $\beta=0.1$ (red color group), $\beta=0.3$ (green color group), or $\beta=0.7$ (blue color group). The interaction length $L=\lambda$. The dashed curves represent a free electron with a $\delta$-function transverse distribution. The solid curves represent free electrons with different Gaussian waists, where only the part of the curve with $d>\sigma$ is shown. The free-space wavelength of the photonic mode is assumed to be $\lambda=1550$ nm.}
    \label{fig:g_geo_cyl}
\end{figure}

In the main text, I show numerical examples of the upper bound where the photonic medium and the free electron can be separated by planes with a distance $d$. In this section, I show another generic configuration where the photonic medium is outside a cylinder concentric with the free electron trajectory with radius $d$ (Fig.\,\ref{fig:g_geo_cyl}(a)) \cite{wang2020coherent, adamo2009light}. Figure \ref{fig:g_geo_cyl}(b) shows the geometric factor $g_\text{geo}^2$ as a function of $d$ and the free-electron normalized velocity $\beta$. The color shows $g^2_\text{geo}$ in log scale if $g^2_\text{geo}>10^{-6}$. The trend in the geometric factor is similar to that shown in Fig.\,2 in the main text.

\begin{figure}
    \centering
    \includegraphics[width=12cm]{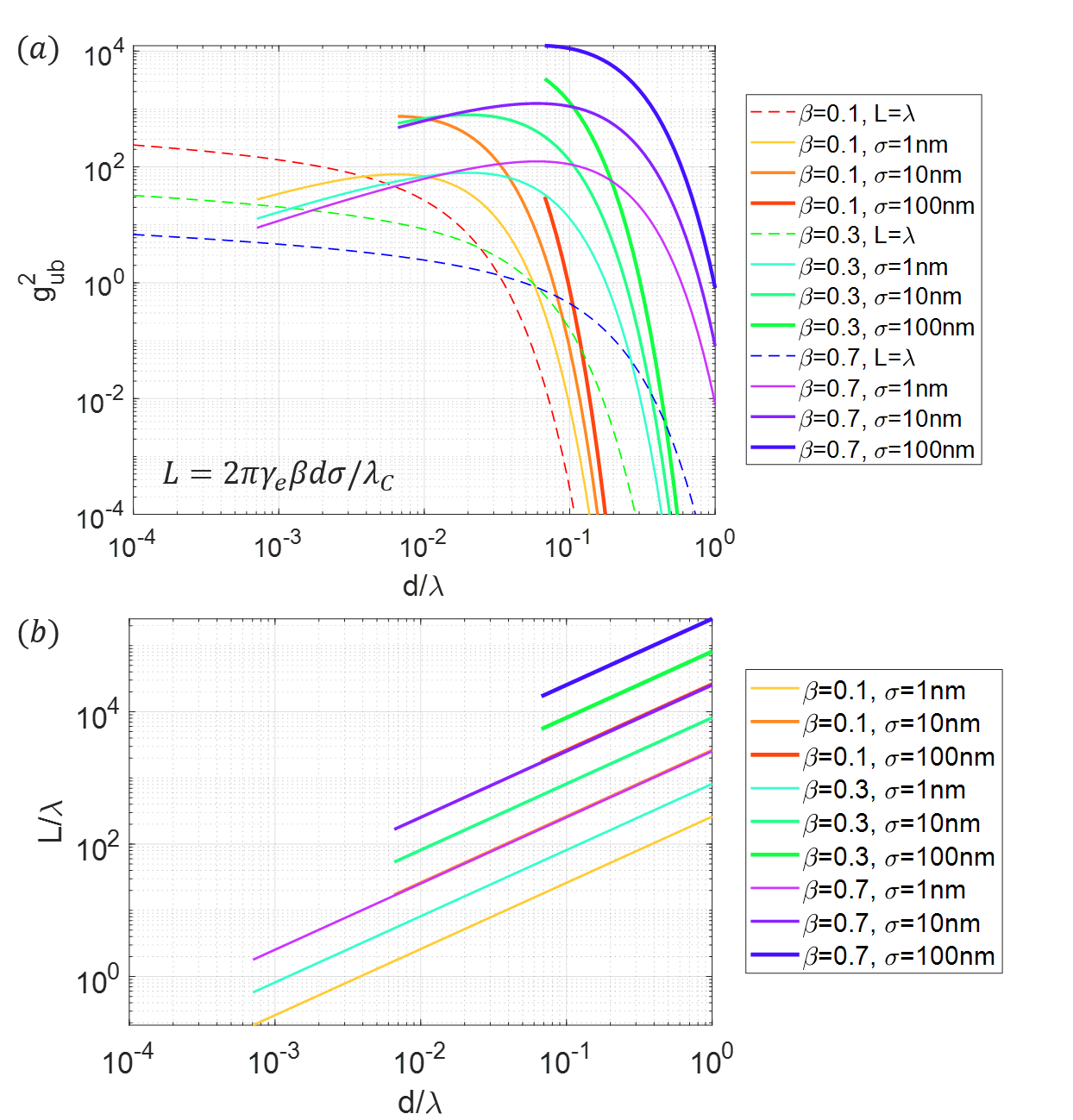}
    \caption{(a) The upper bound of $|g_{Qu,m}|^2$ as a function of $d$ for silicon structures ($\epsilon=12$), where $\beta=0.1$ (red color group), $\beta=0.3$ (green color group), or $\beta=0.7$ (blue color group), and $\sigma=1$ nm (thin solid line), $\sigma=10$ nm (moderate thick line), or $\sigma=100$ nm (thick solid line). Only the part of the curve where $d>\sigma$ is shown. The interaction length is $L=2\pi\gamma\beta d\sigma/\lambda_C$, where $\lambda_C$ is the Compton wavelength, except the dashed curves, which are the same as the dashed curves in Fig.\,\ref{fig:g_geo_cyl}(c). The explicit values of the interaction length $L$ as a function of $d$ is shown in (b). (The curve for $\beta=0.1$ and $\sigma=10$ nm and the curve for $\beta=0.7$ and $\sigma=1$ nm almost overlap. The curve for $\beta=0.1$ and $\sigma=100$ nm and the curve for $\beta=0.7$ and $\sigma=10$ nm almost overlap.) $\lambda=1550$ nm.}
    \label{fig:g2_ub}
\end{figure}

To show the influence of the free-electron transverse spread on the upper bound of the quantum coupling coefficient, I consider a typical case where the free-electron transverse distribution is a Gaussian function centered at $\boldsymbol{r}_{e\perp}$:
\begin{equation}
    \label{eq:phi_e_Gaussian}
    |\phi_e(\boldsymbol{r}_\perp)|^2 = \frac{2}{\pi\sigma^2} \exp\Big( -\frac{2 |\boldsymbol{r}_\perp - \boldsymbol{r}_{e\perp}|^2}{\sigma^2}\Big),
\end{equation}
where $\sigma$ represents the transverse spread of the free-electron wave function. Now, $d$ represents the minimal distance between the center of the free-electron transverse distribution ($\boldsymbol{r}_{e\perp}$) and the optical medium. I choose the following parameters for the numerical demonstration of the upper bound $g^2_\text{ub}$: The free-space wavelength of the photonic mode is $\lambda=1550$ nm; The optical material is silicon with relative permittivity $\epsilon=12$; And the electron normalized velocity is represented by 3 cases $\beta=0.1$ (kinetic energy 2.58 keV), $\beta=0.3$ (kinetic energy 24.7 keV), and $\beta=0.7$ (kinetic energy 205 keV), as respectively indicated by the dashed red, green and blue curves in Fig.\,\ref{fig:g_geo_cyl}(b). 

To study the influence on the upper bound per interaction length, I choose the interaction length $L=\lambda$ and plot the upper bound of $|g_{Qu,m}|^2$ as a function of $d$ in Fig.\,\ref{fig:g_geo_cyl}(c). The red, green, and blue color groups represent $\beta=0.1$, $\beta=0.3$, and $\beta=0.7$ respectively. Within each color group, solid curves with slightly different colors and thickness represent free electrons with different transverse spread $\sigma=1$ nm, $\sigma=10$ nm, and $\sigma=100$ nm, as shown in the legend in Fig.\,\ref{fig:g_geo_cyl}(c). Only the part of the curve where $d>\sigma$ is shown. The condition $d>\sigma$ means that the minimal separation $d$ should be larger than the half Gaussian waist $\sigma$, otherwise the free electron has a large probability to collide with the optical medium. The dashed curves represent the ``simplified'' case, where the free-electron transverse distribution is a $\delta$ function. From Fig.\,\ref{fig:g_geo_cyl}(c), I find that the influence of the finite free-electron transverse spread on the upper bound is negligible when the transverse spread is relatively small. One important length scale is the decay length of the near field carried by the free electron, represented by $2\pi/\alpha_e = \gamma_e\beta\lambda$, where $\gamma_e = 1/\sqrt{1 - \beta^2}$. When the transverse spread is large, i.e., comparable or even larger than $\gamma_e\beta\lambda$, the upper bound will increase compared with the simplified case. This is demonstrated by a large deviation of the red curve representing $\beta=0.1$ and $\sigma=100$ nm and a small deviation of the green curve representing $\beta=0.3$ and $\sigma=100$ nm.

On the other hand, a finite transverse spread of the free electron results in a finite divergence, which can limit the interaction length. To estimate the maximal interaction, I assume that the free electron occupies a minimal phase space constrained by the uncertainty principle, and the divergence angle can be derived from the Gaussian beam waist, i.e.,  
\begin{equation}
    \label{eq:theta_gaussian}
    \theta = \frac{\lambda_e}{\pi \sigma},
\end{equation}
where $\lambda_e$ is the wavelength of the free electron.
\begin{equation}
    \label{eq:lambda_e}
    \lambda_e = \frac{h}{\gamma_e m c \beta} = \frac{1}{\gamma_e\beta} \lambda_C,
\end{equation}
where $\lambda_C$ is the Compton wavelength. The estimated maximal interaction length is \cite{karnieli2024strong}
\begin{equation}
    \label{eq:L_max}
    L_\text{max} = \frac{2d}{\theta} = 2\pi\gamma_e\beta\frac{d\sigma}{\lambda_C},
\end{equation}
beyond which the free electron has a large probability to collide with the optical medium. This maximal interaction length increases with the electron speed, the separation distance and the transverse spread. This can set the ultimate upper bound on the coupling coefficient, without transversely guiding the free electron \cite{karnieli2024strong}.

With this estimated maximal interaction length (Eq.\,\ref{eq:L_max}), I calculate the upper bound of $|g_{Qu.m}|^2$, where the rest of the parameters are the same as in Fig.\,\ref{fig:g_geo_cyl}(c). The upper bounds are shown in Fig.\,\ref{fig:g2_ub}(a) and the corresponding interaction lengths are shown in Fig.\,\ref{fig:g2_ub}(b). The legend is consistent with that in Fig.\,\ref{fig:g_geo_cyl}(c). Since the slow electrons generally have shorter interaction length, the advantage of achieving a large coupling coefficient using slow electrons at the deep sub-wavelength region becomes marginal or even unreachable. Nevertheless, Fig.\,\ref{fig:g2_ub}(a) indicates that a large range of parameters, which are achievable in scanning electron microscopes or transmission electron microscopes, have the potential to reach the strong coupling regime of the  free-electron--photon interaction.

\section{Coupling between the free electron and an extended optical system}
\label{sec:extended_system}

\begin{figure}
    \centering
    \includegraphics[width=13cm]{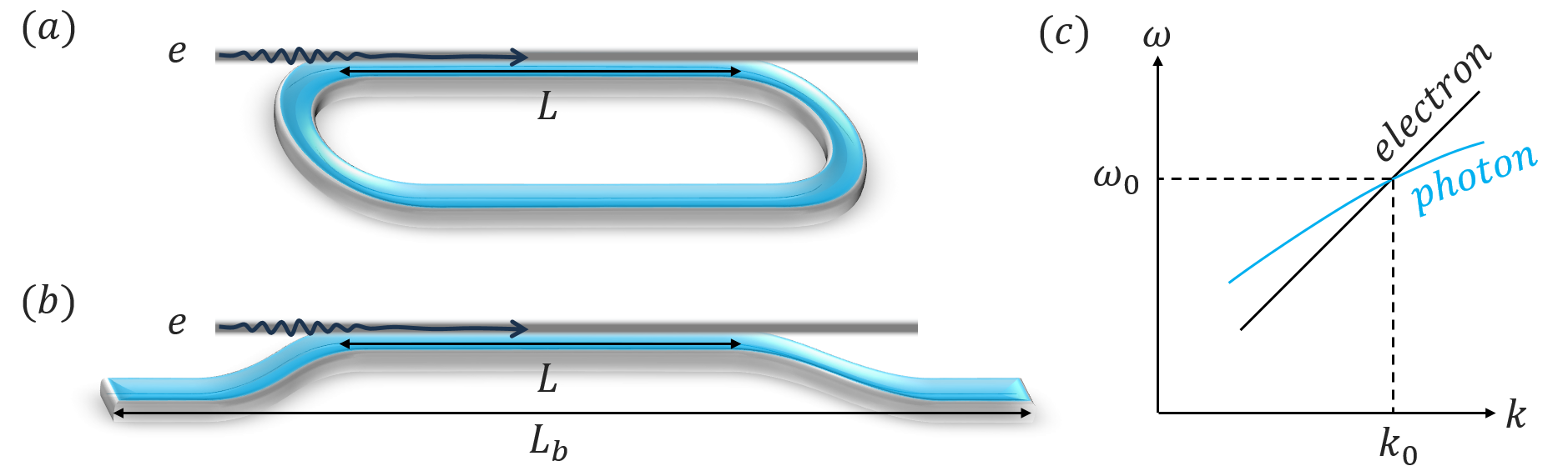}
    \caption{Schematics of the coupling between the free electron and a ring resonator (a) or a waveguide (b). (c) A sketch of the electron (black) and photon (blue) dispersion relation near the phase matching condition.}
    \label{fig:ring_wg}
\end{figure}

In this section, I discuss the interaction between the free electron and an extended optical system, where a representative extended optical system is a waveguide (Fig.\,\ref{fig:ring_wg}(b)). In contrast to the case when the optical medium is confined spatially, such as a ring resonator (Fig.\,\ref{fig:ring_wg}(a)), and the optical modes can have a discrete spectrum, the waveguide modes are typically continuous in frequency, described by a dispersion relation, as illustrated in Fig.\,\ref{fig:ring_wg}(c). Here, I take the waveguide as an example of an extended optical system to illustrate the quantum coupling coefficient between the free electron and a spatial-temporal mode in the waveguide. 

I assume the total length $L_b$ of the waveguide is much longer than the interaction length $L$ (Fig.\,\ref{fig:ring_wg}(b)) and take a periodic boundary condition to discretize the modes in the waveguide. The coupling between the free electron and a spatial-temporal mode supported by the waveguide is equivalent to summing up the coupling between the free electron and the discretized modes, i.e., 
\begin{equation}
    \label{eq:gQu_wg}
    |g_{Qu, wg}|^2 = \sum_m |g_{Qu,m}|^2.
\end{equation}
Here, I assume that the interaction length is large, such that the interaction time is much longer than the optical periodicity. Thus, only the modes that are close to the phase matching condition ($k_0 = \omega_0/v$) contributes to the coupling. Moreover, if the dispersion relation of the optical system has multiple branches, such that the phase-matching condition can be satisfied at multiple frequencies, I assume that these phase-matching frequencies are well separated and there is one spatial-temporal mode for each phase-matching frequency. Therefore, in considering the coupling between the free electron and one spatial-temporal mode, the sum on the right hand side of Eq.\,\ref{eq:gQu_wg} include only the discretized modes that are close to the considered phase-matching frequency. 

The discretized eigenmodes supported by the waveguide have the form
\begin{equation}
    \label{eq:modes_waveguide}
    \boldsymbol{E}_m(\boldsymbol{r}) = \boldsymbol{E}_m(\boldsymbol{r}_\perp)e^{ik_m z}.
\end{equation}
The longitudinal momemtum $k_m$ takes discretized values due to the periodic boundary condition assumption, i.e., $k_m = \frac{2\pi m_z}{L_b}$, where $m_z$ is an integer.
Thus, the quantum coupling coefficient for one of such discretized eigenmodes is
\begin{equation}
    \label{eq:gQu_wg_mode_1}
    \begin{split}
        |g_{Qu,m}|^2 & =\frac{q_e^2}{2\hbar\omega_m\epsilon_0} \frac{\Big|\int_{L/2}^{L/2} dz \exp(-i \frac{\omega_m}{v} z + i k_m z) E_{m,z}(\boldsymbol{r}_{e\perp})\Big|^2}{\int d^2\boldsymbol{r}_\perp \epsilon(\boldsymbol{r}_\perp) \boldsymbol{E}_m^\dagger(\boldsymbol{r}_\perp) \boldsymbol{E}_m(\boldsymbol{r}_\perp) L_b}\\
        & = \frac{q_e^2}{2\hbar\omega_m\epsilon_0} \frac{|E_{m,z}(\boldsymbol{r}_{e\perp})|^2}{\int d^2\boldsymbol{r}_\perp \epsilon(\boldsymbol{r}_\perp) \boldsymbol{E}_m^\dagger(\boldsymbol{r}_\perp) \boldsymbol{E}_m(\boldsymbol{r}_\perp)} \frac{L^2}{L_b} \text{sinc}^2\Big[\Big(k_m - \frac{\omega_m}{v}\Big)\frac{L}{2}\Big].
    \end{split}
\end{equation}
The quantum coupling coefficient for a spatial-temporal mode is therefore
\begin{equation}
    \label{eq:gQu_wg_2}
    \begin{split}
    |g_{Qu, wg}|^2 & = \sum_m \frac{q_e^2}{2\hbar\omega_m\epsilon_0} \frac{|E_{m,z}(\boldsymbol{r}_{e\perp})|^2}{\int d^2\boldsymbol{r}_\perp \epsilon(\boldsymbol{r}_\perp) \boldsymbol{E}_m^\dagger(\boldsymbol{r}_\perp) \boldsymbol{E}_m(\boldsymbol{r}_\perp)} \frac{L^2}{L_b} \text{sinc}^2\Big[\Big(k_m - \frac{\omega_m}{v}\Big)\frac{L}{2}\Big] \\ & \approx \frac{q_e^2}{2\hbar\omega_0\epsilon_0} \frac{|E_{wg,z}(\boldsymbol{r}_{e\perp}, \omega_0)|^2}{\int d^2\boldsymbol{r}_\perp \epsilon(\boldsymbol{r}_\perp) \boldsymbol{E}_{wg}^\dagger(\boldsymbol{r}_\perp, \omega_0) \boldsymbol{E}_{wg}(\boldsymbol{r}_\perp, \omega_0)} \frac{L^2}{L_b} \sum_m \text{sinc}^2\Big[\Big(k_m - \frac{\omega_m}{v}\Big)\frac{L}{2}\Big],
    \end{split}
\end{equation}
where, in the last step, I assume that the transverse profile ($\boldsymbol{E}_m(\boldsymbol{r}_\perp)$) of the discretized eigenmodes are close to the transverse profile ($\boldsymbol{E}_{wg}(\boldsymbol{r}_\perp)$) of the waveguide mode at the phase-matching frequency, and the eigenfrequencies are close to the phase-matching frequency, such that they are taken out of the summation. In the limit of $L_b\rightarrow \infty$, the summation becomes an integration, with the exchange $\sum_m \rightarrow \frac{L_b}{2\pi}\int dk$. 
\begin{equation}
    \label{eq:gQu_wg_3}
    |g_{Qu, wg}|^2 \approx \frac{q_e^2}{2\hbar\omega_0\epsilon_0} \frac{|E_{wg,z}(\boldsymbol{r}_{e\perp}, \omega_0)|^2}{\int d^2\boldsymbol{r}_\perp \epsilon(\boldsymbol{r}_\perp) \boldsymbol{E}_{wg}^\dagger(\boldsymbol{r}_\perp, \omega_0) \boldsymbol{E}_{wg}(\boldsymbol{r}_\perp, \omega_0)} \frac{L^2}{L_b} \frac{L_b}{2\pi}\int \text{sinc}^2\Big[\Big(k - \frac{\omega(k)}{v}\Big)\frac{L}{2}\Big] dk
\end{equation}
The total length of the waveguide drops in Eq.\,\ref{eq:gQu_wg_3} as expected. To compare with the quantum coupling coefficient formula for a discrete mode, as shown in Eq.\,16 of the maintext, I reformulate Eq.\,\ref{eq:gQu_wg_3} as 
\begin{equation}
    \label{eq:gQu_wg_4}
    |g_{Qu, wg}|^2 \approx \Bigg\{ \frac{q_e^2}{2\hbar\omega_0\epsilon_0} \frac{|E_{wg,z}(\boldsymbol{r}_{e\perp}, \omega_0)|^2 L}{\int d^2\boldsymbol{r}_\perp \epsilon(\boldsymbol{r}_\perp) \boldsymbol{E}_{wg}^\dagger(\boldsymbol{r}_\perp, \omega_0) \boldsymbol{E}_{wg}(\boldsymbol{r}_\perp, \omega_0)} \Bigg\} \underbrace{\Bigg\{ \frac{L}{2\pi}\int \text{sinc}^2\Big[\Big(k - \frac{\omega(k)}{v}\Big)\frac{L}{2}\Big] dk \Bigg\}}_{N_\text{eff}},
\end{equation}
where the first part is the same as the coupling coefficient for a discrete mode confined longitudinally within the interaction length $L$ satisfying the phase-matching condition (Eq.\,16 of the main text). The second part is a unitless number that can be regarded as the \textit{effective number of modes} ($N_\text{eff}$). In another word, in comparing with the coupling with a discrete mode, the coupling with a spatial-temporal mode in a waveguide also depends on this effective number of modes. Thus, the upper bound of the quantum coupling between the free electron and a spatial-temporal mode in a waveguide ($|g_{Qu, wg}|^2$) can be larger than the upper bound for the quantum coupling between the free electron and a discrete mode ($|g_{Qu, m}|^2$) when the effective number of modes is larger than 1. The effective number of modes can be controlled by engineering the dispersion relation of the photonic mode near the phase-matching condition \cite{karnieli2024strong}.

To calculate the effective number of modes, one can expend the dispersion relation near the phase-matching condition.
\begin{equation}
    \label{eq:dispersion_expansion}
    \begin{split}
    k - \frac{\omega(k)}{v} & = k_0 - \frac{\omega_0}{v} + \Big(1-\frac{1}{v}\frac{d\omega}{dk}\Big)(k-k_0) - \frac{1}{2v}\frac{d^2\omega}{dk^2}(k-k_0)^2 - \frac{1}{6v}\frac{d^3\omega}{dk^3}(k - k_0)^3 + ...\\ & = \Big(1-\frac{v_g}{v}\Big)(k-k_0) - \frac{1}{2v}\frac{d^2\omega}{dk^2}(k-k_0)^2 - \frac{1}{6v}\frac{d^3\omega}{dk^3}(k - k_0)^3 + ...,
    \end{split}
\end{equation}
where $v_g$ is the group velocity of the photonic mode. When $v_g \neq v$, the linear dispersion term dominates. Neglecting higher order terms, the effective number of modes is:
\begin{equation}
    \label{eq:Neff_linear}
    N_\text{eff} = \frac{1}{|1 - \frac{v_g}{v}|}, \text{ when linear dispersion dominates.}
\end{equation}
In this case, the effective number of modes is independent of the interaction length.
When $v_g=v$ and $\frac{d^2\omega}{dk^2}\neq 0$, the quadratic dispersion term dominates. Neglecting higher order terms, the effective number of modes is \cite{karnieli2024strong}
\begin{equation}
    \label{eq:Neff_quadratic}
    N_\text{eff} = \frac{4}{3\sqrt{\pi}}\Big|\frac{1}{v}\frac{d^2\omega}{dk^2}\Big|^{-\frac{1}{2}} L^\frac{1}{2}, \text{ when quadratic dispersion dominates.}
\end{equation}
This is consistent with recent study \cite{karnieli2024strong}. Similarly, when $v_g=v$, $\frac{d^2\omega}{dk^2} = 0$, and $\frac{d^3\omega}{dk^3} \neq 0$, the cubic dispersion dominates. Neglecting  higher order terms, the effective number of modes is
\begin{equation}
    \label{eq:Neff_cubic}
    N_\text{eff} = 0.8 \Big|\frac{1}{v}\frac{d^3\omega}{dk^3}\Big|^{-\frac{1}{3}} L^\frac{2}{3}, \text{ when cubic dispersion dominates.}
\end{equation}
Generally, when the n-th order dispersion dominates in Eq.\,\ref{eq:dispersion_expansion}, the effective number of modes is
\begin{equation}
    \label{eq:Neff_n}
    N_\text{eff} = \text{const} \Big|\frac{1}{v}\frac{d^n\omega}{dk^n}\Big|^{-\frac{1}{n}} L^\frac{n-1}{n}, \text{ when n-th order dispersion dominates.}
\end{equation}
Thus, when higher order dispersion ($n>1$) dominates, the effective number of modes depend on the interaction length, which provides a pathway to achieve strong coupling between the free electron and the spatial-temporal mode \cite{karnieli2024strong}. The above discussion also applies to guided modes in periodic structures, such as photonic crystal waveguides. 

As a side note, in the case where the free electron interacts with a ring resonator (Fig.\,\ref{fig:ring_wg}(a)), it is possible that the free electron can interact with multiple discrete modes, and the measurement cannot differentiate the coupling to individual modes \cite{huang2023electron}. Then, one also needs to sum up the coupling with all the interacting modes to account for the total effect. The analysis is very similar to the analysis with the discretized modes in a long waveguide, except that the total length $L_b$ should be the circumference of the ring resonator. The effective number of modes in this case is $N_\text{eff} = \frac{L}{L_b}\sum_m \text{sinc}^2 \Big[\Big(k_m - \frac{\omega_m}{v}\Big)\frac{L}{2}\Big]$, which is close to $N_\text{eff}$ for the spatial-temporal mode in a waveguide.

\section{Numerical examples of dielectric optical systems}
\label{sec:numerical_example}

In this section, I show numerical examples of the quantum coupling coefficient between the free electron and a photonic mode in a dielectric waveguide or in a dielectric sub-wavelength grating (SWG). 

\subsection{Free-electron--light interaction near a dielectric waveguide}

A generic configuration of the free-electron--light interaction consists of a free electron propagating in the vicinity of a waveguide \cite{kfir2019entanglements, henke2021integrated, feist2022cavity}. 
Thus, I numerically calculate $g_{Qu, m}$ for the interaction between the free electron and a photonic mode in a waveguide with length $L$. The results are presented in Fig.\,2(b) of the main text. Here, I provide the details of the numerical calculation.

The guided mode in a uniform waveguide has the following form:
\begin{equation}
    \label{eq:E_guided_mode}
    E_j(x,y,z) = W_j(x,y)\exp(ik_z z),
\end{equation}
where $k_z$ is the wave vector, and $W_j$ is the guided mode profile in the waveguide cross section. Substituting Eq.\,\ref{eq:E_guided_mode} into Eq.\,\ref{eq:gQu_mode_m_2}, one can calculate $|g_{Qu,m}|$. To have a nonzero integration in Eq.\,\ref{eq:gQu_mode_m_2} when the interaction length is large, it is necessary that 
\begin{equation}
    \label{eq:phase_matching_wg}
    k_z = \frac{\omega_m}{v},
\end{equation}
which is referred to as the phase matching condition. If the phase matching condition is satisfied, the quantum coupling coefficient is
\begin{equation}
    \label{eq:gQu_waveguide_mode}
    |g_{Qu, m}|^2 =  \frac{q_e^2}{2\hbar\omega_0\epsilon_0} \frac{|W_z(\boldsymbol{r}_{e\perp}, \omega_0)|^2 L}{\int d^2\boldsymbol{r}_\perp \epsilon(\boldsymbol{r}_\perp) \boldsymbol{W}^\dagger(\boldsymbol{r}_\perp, \omega_0) \boldsymbol{W}(\boldsymbol{r}_\perp, \omega_0)}.
\end{equation}
Since the guided mode profile $\boldsymbol{W}(\boldsymbol{r}, \omega_0)$ can be numerically calculated with waveguide mode solver, it is straightforward to get $|g_{Qu,m}|$ thereafter. Here, I do not consider the influence of the effective number of modes ($N_\text{eff}$). Eventually, one can normalize $|g_{Qu,m}|$ with respect to the interaction length and get a length normalized quantum coupling coefficient $|g_{Qu,m}|/\sqrt{L/\lambda}$, as shown in Fig.\,2(b) of the main text.

\begin{figure}
    \centering
    \includegraphics[width=7cm]{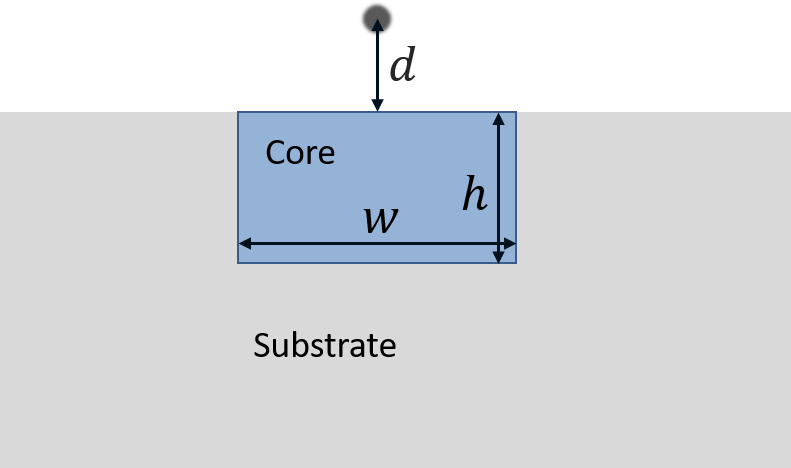}
    \caption{Illustration of the cross-section view of the free-electron--light interaction near a waveguide. The blue rectangle represents the core of the waveguide, with width $w$, height $h$, and relative permittivity $\eps_c$. The gray surrounding medium represents the substrate, with relative permittivity $\eps_s$. It is vacuum above the waveguide. The electron trajectory is separated from the waveguide top surface by a distance $d$.}
    \label{fig:wg_cross_section}
\end{figure}

The configuration of silicon (Si) waveguides and silicon nitride (SiN) waveguides to demonstrate $|g_{Qu,m}|$ is shown in Fig.\,\ref{fig:wg_cross_section}. The free electron propagates above the top surface of the dielectric grating with a separation distance $d$. The waveguide core is surrounded by vacuum from the top and by a substrate from other sides. The waveguide core is rectangular, with width $w$ and height $h$. The electron trajectory is on the midline above the waveguide core. The waveguide core is Si ($\eps_c=12$) for Si waveguide, or SiN ($\eps_c=4$) for SiN waveguide.
The considered mode frequency is 193 THz ($\lambda=1.55$ $\mu$m). Three separation distances ($d=0.02\lambda$, $d=0.05\lambda$, and $d=0.1\lambda$) are numerically studied. 

To get $|g_{Qu,m}|$ for different electron velocities, I scan the width ($w$) and height ($h$) of the rectangular waveguide core, and require that the electron velocity matches the phase velocity of the waveguide mode (Eq.\,\ref{eq:phase_matching_wg}). Since only TM modes have non-zero $E_z$ fields at the electron trajectory, I only consider the coupling with TM modes. 
For the Si waveguides, I scan the waveguide core width ($w$) from 120 nm to 600 nm, and the waveguide height ($h$) from 100 nm to 600 nm, and I choose the substrate dielectric among vacuum ($\eps_s=1$), SiO2 ($\eps_s=2$), or SiN ($\eps_s=4$). After I obtain $g_{Qu,m}$ for different Si waveguide configurations, I find the largest $|g_{Qu,m}|$ for each electron velocity for the three studied separation distances. The results are plotted in squares in Fig.\,2(b) of the main text. 
For the SiN waveguides, I scan the waveguide core  width ($w$) from 200 nm to 1000 nm, and the waveguide height ($h$) from 200 nm to 1000 nm, and I choose the substrate dielectric between vacuum ($\eps_s=1$) and SiO2 ($\eps_s=2$). Then, I apply the same procedure, as for the Si waveguides, to find the largest $|g_{Qu,m}|$ for each electron velocity for the three studied separation distances. The results are plotted in dots in Fig.\,2(b) of the main text. 
I use Lumreical MODE solver \cite{lumerical} to calculate the waveguide mode profile. When the phase velocity of the waveguide mode approaches the speed of light, the mode is weakly guided and the numerical error increases. Therefore, I only show $|g_{Qu,m}|$ results for $\beta\leq0.93$. Moreover, since the largest $|g_{Qu,m}|$ for each $\beta$ is of interest, I find that considering the coupling between free electrons and high-order TM modes does not increase the Pareto front of the $|g_{Qu,m}|$--$\beta$ relation where only fundamental TM modes are considered.

\subsection{Free-electron--light interaction near a silicon sub-wavelength grating}

\begin{figure}
     \centering
     \includegraphics[width=16cm]{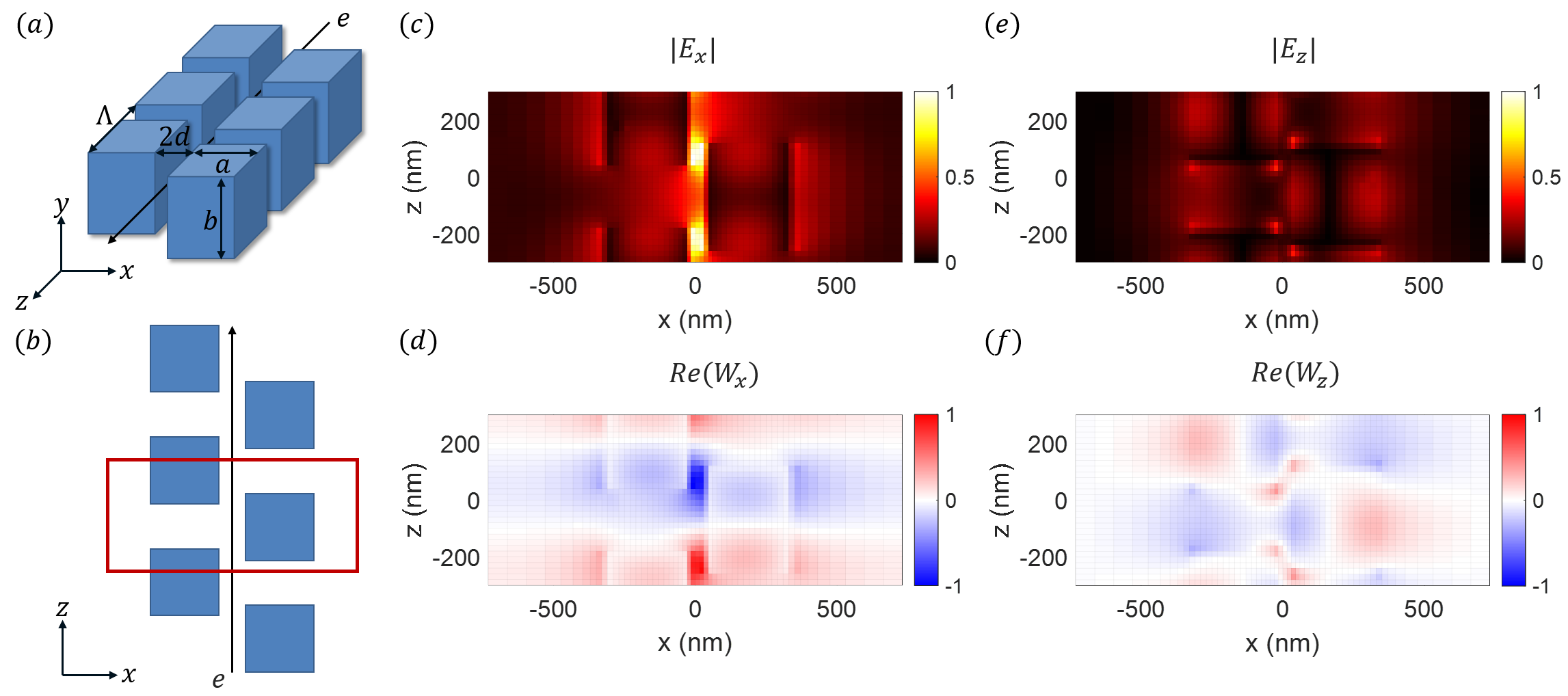}
     \caption{(a) An electron propagating in the slot of the SWG. The SWG consists of periodic rectangular pillars with a width of $a=300$ nm, a height of $b=360$ nm, and a duty cycle along z-direction of 0.6. The periodicity in z-direction is $\Lambda=600$ nm. The slot width is $2d=60$ nm. A top view of the sub-wavelength grating is shown in (b), where the red box highlights the unit cell. (c-f) The fields of the guided mode ($\lambda=1.55$ $\mu$m, $k_z = -0.94 \pi/\Lambda$) in the y-cut plane passing the center of the grating. (c) and (e) are the magnitude of the dominant transverse field $E_x$ and the longitudinal field $E_z$ respectively. (d) and (f) show the real part of the periodic fields $W_x$ and $W_z$ respectively. The simulation is performed with Lumerical FDTD \cite{lumerical}. }
     \label{fig:numerical_example}
\end{figure}

Although the study of the upper bound indicates that slow electrons are preferred when the separation between the electron trajectory and the structure is deep sub-wavelength, it is empirically hard to achieve a large coupling coefficient with the slow electrons ($\beta\lesssim 0.3$), especially with a dielectric medium. The challenge is two-fold. Firstly, since the quantum coupling coefficient increases with interaction length, it is preferred that the free electron interacts with an extended guided mode supported by the dielectric structure, and the phase matching condition is satisfied, which means that the free-electron velocity equals the phase velocity of the guided mode. Slow electrons ($\beta\lesssim 0.3$) typically cannot satisfy the phase matching condition with the guided modes in longitudinally uniform dielectric waveguides, due to the limited refractive index of low-loss dielectric media. To achieve the phase matching condition, a generic structure is a dielectric sub-wavelength grating (SWG) \cite{halir2015waveguide}. Secondly, the optimal field distribution to achieve the quantum coupling upper bound is tightly confined for slow electrons. With typical dielectric media, it is generally hard to tightly confine optical modes to the deep sub-wavelength. Although such tightly confined modes exist in metallic structures, the high loss limits the usage of such metallic structures in many applications, for instance, for high fidelity quantum optics. Therefore, I focus on dielectric optical media.
In this section, I present a numerical example of an electron interacting with the guided mode in a Si SWG and discuss how far the quantum coupling coefficient in such an intuitive design is from the upper bound.

\begin{figure}
    \centering
    \includegraphics[width=12cm]{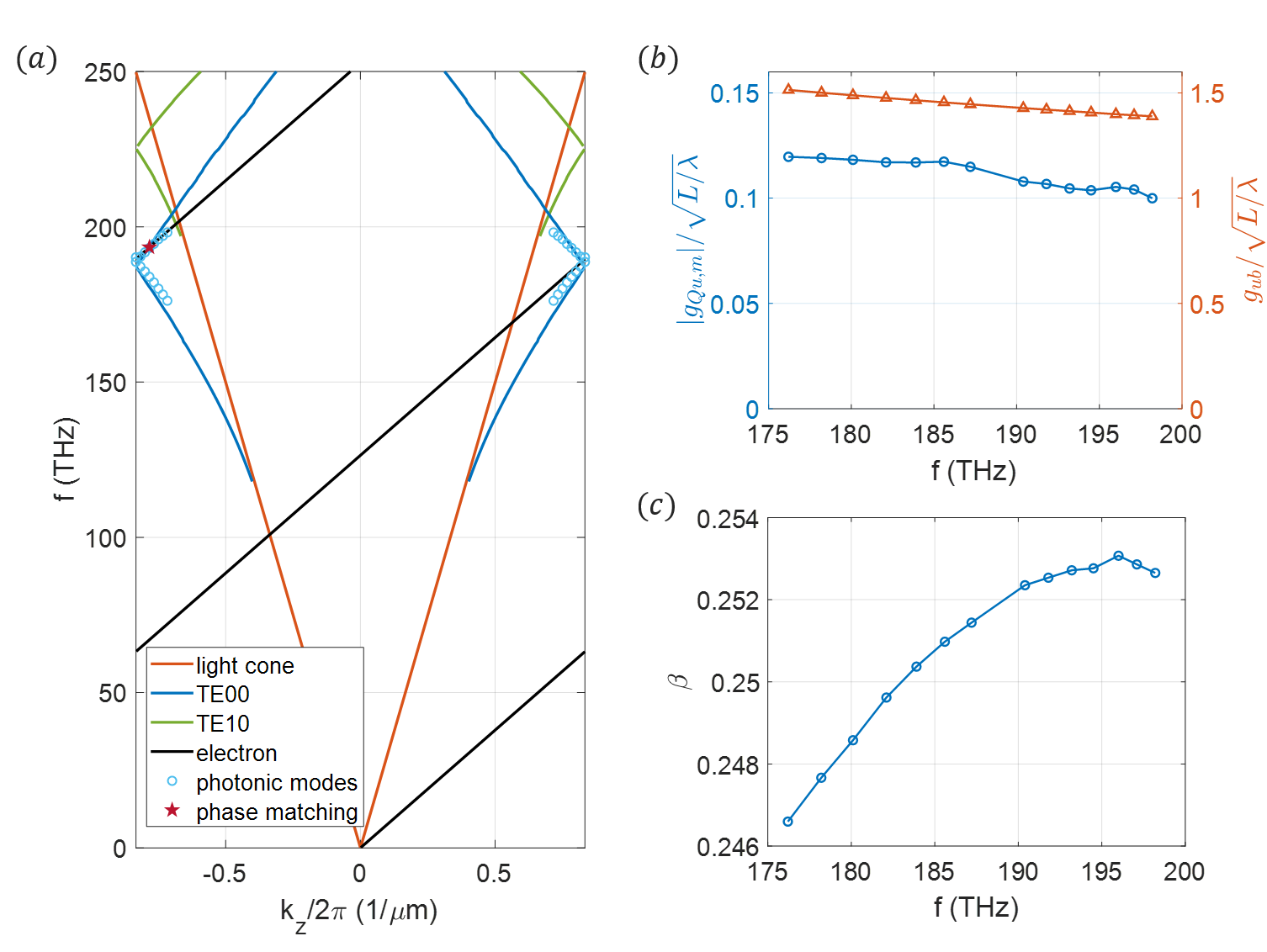}
    \caption{(a) The dispersion relation of the guided modes in the Si SWG, whose parameters are shown in Fig.\,\ref{fig:numerical_example}. The orange curves represent the light cone ($k_z=\omega/c$). The blue and green curves are the dispersion relation of the fundamental and first excited TE modes in a longitudinally uniform slot waveguide, with a width of 300 nm, a height of 360 nm, a slot width of 60 nm, and a core permittivity of $\eps_c=6$, which approximates the dispersion relation of the Si SWG near the considered frequency (193 THz). The blue circles represent the dispersion relation of the guided modes in the Si SWG. The red star highlights the mode at 193 THz and $k_z=-0.94 \pi/\Lambda$. The black curves show the dispersion relation of the free electron that is phase-matched with the highlighted mode. (b) The interaction length normalized quantum coupling coefficient (blue) for the modes shown by blue circles in (a), and the corresponding upper bounds (orange). (c) The corresponding free-electron velocity that satisfies the phase matching condition.}
    \label{fig:SWG_dispersion}
\end{figure}

The schematic of the Si SWG is shown in Figs.\,\ref{fig:numerical_example}(a-b). The grating consists of periodic rectangular pillars with a width of $a=300$ nm, a height of $b=360$ nm, and a duty cycle along z-direction of 0.6. The periodicity in z-direction is $\Lambda=600$ nm. The slot width is $2d=60$ nm, and the electron is assumed to propagate along the center of the slot. The structure has a glide reflection symmetry with respect to the center of the slot \cite{black2020operating}. The phase matching condition is
\begin{equation}
    \label{eq:phase_match}
    k_z + \frac{2\pi m_p}{\Lambda} = \frac{\omega_m}{\beta c},
\end{equation}
where $k_z$ is the Bloch wave vector of the guided mode in the first Brillouin  zone, and $m_p$ is an integer. 
The dispersion relation of the guided TE modes near the frequency of interest ($f=193$ THz, $\lambda=1.55\,\mu$m) is shown in Fig.\,\ref{fig:SWG_dispersion}(a). To qualitatively understand the dispersion of the guided modes, I also study the dispersion relation of a longitudinally uniform slot waveguide, where the two rectangular cores share the same dimensions as the Si SWG, i.e., the width is $a = 300$ nm, the height is $b = 360$ nm, and the slot width $2d = 60$ nm. And the refractive index of the core is a weighted average of those of Si and vacuum weighted by the duty cycle, i.e., $\eps_c = (0.6\sqrt{\eps_{\text{Si}}} + 0.4\sqrt{\eps_{\text{Vacuum}}})^2\approx 6$. The dispersion relation of the fundamental TE mode (TE00) and the first excited TE mode (TE10) are shown respectively in blue and green in Fig.\,\ref{fig:SWG_dispersion}(a). The dispersion relation of the guided modes in the Si SWG is close to the dispersion relation of the fundamental TE mode in the longitudinal uniform slot waveguide with the weighted average refractive index. 

The studied mode is at $f=193$ THz, as highlighted by the red star in Fig.\,\ref{fig:SWG_dispersion}(a). The corresponding wave vector is $k_z = -0.94 \pi/\Lambda$. 
Since this mode is below the light line, i.e., $\omega_m < c|k_z|$, this mode is non-radiative. The transverse electric field of this mode is dominantly polarized in the x-direction (TE mode). The field distributions are shown in Figs.\,\ref{fig:numerical_example}(c-f). The modes supported by a periodic grating satisfy the Bloch theorem, such that
\begin{equation}
    \label{eq:Bloch_field}
    E_j(x, y, z) =W_j(x, y, z)\exp(i k_z z),
\end{equation}
where $W_j$ is periodic, i.e., $W_j(x,y,z) = W_j(x,y,z+\Lambda)$. The real parts of $W_x$ and $W_z$ of the guided mode are shown in Figs.\,\ref{fig:numerical_example}(d) and \ref{fig:numerical_example}(f). 
With $m_p=2$ in Eq.\,\ref{eq:phase_match}, this guided mode is phase-matched with the free electron with $\beta=0.253$.

The quantum coupling coefficient can be calculated from the field distribution of the guided mode, using Eq.\,\ref{eq:gQu_mode_m_2}. With interaction length $L=\lambda$, the quantum coupling coefficient is $|g_{Qu,m}|=0.10$. The corresponding upper bound of $|g_{Qu,m}|$ in this configuration ($d=30$ nm, $\lambda=1.55$ $\mu$m, with optical medium on both sides and a duty cycle of 0.6) is $g_\text{ub}=1.42$. The quantum coupling coefficient in this realistic system is about one order of magnitude lower than the upper bound. 
Nevertheless, with interaction length $L>95\lambda$, it can achieve strong coupling, i.e., $|g_{Qu,m}|>1$. Using the analysis in Sec.\,\ref{sec:trans} and assuming the free-electron transverse spread $\sigma=10$ nm, the maximal interaction length $L_\text{max}=130\lambda>95\lambda$, which indicates that this system can reach strong coupling, with a high-quality electron beam, for slow electrons.

Modes with different frequencies and wave vectors can be phased-matched with free electrons with different velocities. For the guided modes shown in blue circles in Fig.\,\ref{fig:SWG_dispersion}(a), I calculate the corresponding quantum coupling coefficients, as well as the corresponding upper bounds, as shown in Fig.\,\ref{fig:SWG_dispersion}(b). The free-electron velocities to satisfy the phase matching condition are shown in Fig.\,\ref{fig:SWG_dispersion}(c). From Fig.\,\ref{fig:SWG_dispersion}(c), the electron velocity satisfying the phase-matching condition for the guided modes with frequency between 190 THz and 198 THz is almost constant, which indicates that the group velocity of the guided mode is close to the electron velocity and the effective number of modes ($N_{\text{eff}}$) can be significantly larger than 1 \cite{karnieli2024strong}, if one consider the interaction between the free electron and the spatial-temporal mode in the SWG.

The guided mode profiles are calculated using Lumerical FDTD solver \cite{lumerical}, where the unit cell of the Si SWG with perfect match layers in the x- and y-direction and the Bloch boundary condition in the z-direction is excited by random dipole sources. The guided modes have long lifetime and can be extracted using the Fourier transform of the time-domain fields after the initial transient time. 

The intuition of this design is as follows. With duty cycle approaches 1, this sub-wavelength grating becomes a slot waveguide. The fundamental TE mode of the slot waveguide is tightly confined in the slot \cite{xu2004experimental}. Such a confined mode is preferred to achieve a large quantum coupling coefficient. However, the fundamental TE mode of the slot waveguide has zero longitudinal component at the slot center, due to the reflection symmetry with respect to the slot center \cite{zhao2018design}. In order to couple to the free electron, this reflection symmetry should be broken in the SWG. Thus, I design the SWG to have a glide reflection symmetry instead. The field distributions shown in Fig.\,\ref{fig:numerical_example}(e) verifies that the $E_z$ component along the slot center is non-zero. Moreover, the $W_z$ distribution in Fig.\,\ref{fig:numerical_example}(f) shows a prominent spatial harmonic of order 2 along the slot center, which is suitable for the coupling with a free electron with $\beta=0.253$ ($m_p=2$ in Eq.\,\ref{eq:phase_match}). With inverse design \cite{hughes2017method, haeusler2022boosting}, it is possible to achieve higher quantum coupling coefficient to approach the upper bound, but an extensive optimization is beyond the scope of this study. The SWG example presented here is based on physical intuition, and it is simple enough for future experimental demonstrations.

\section{Numerical examples approaching the upper bounds}
\label{sec:approach_upper_bounds}
An important question about the analytical upper bound is that how tight the upper bounds are. To answer this question, I show two pedagogical examples of free-electron--light coupling: (1) with guided modes in a dielectric hollow-core waveguide, and (2) with surface plasmon polariton (SPP) modes in a metallic hole. With certain choices of parameters, the free-electron--photon coupling coefficient ($|g_{Qu,m}|$) can reach about 70\% and over 99\% of the corresponding upper bounds, with dielectric hollow-core waveguides and with metallic holes, respectively. Although a rigorous proof of the tightness of the analytical upper bound is beyond the scope of this study, these numerical examples indicate that the analytical upper bound for the free-electron--photon coupling is likely to be tight. 

\subsection{Free-electron--light interaction in a dielectric hollow-core waveguide}

\begin{figure}
    \centering
    \includegraphics[width=13cm]{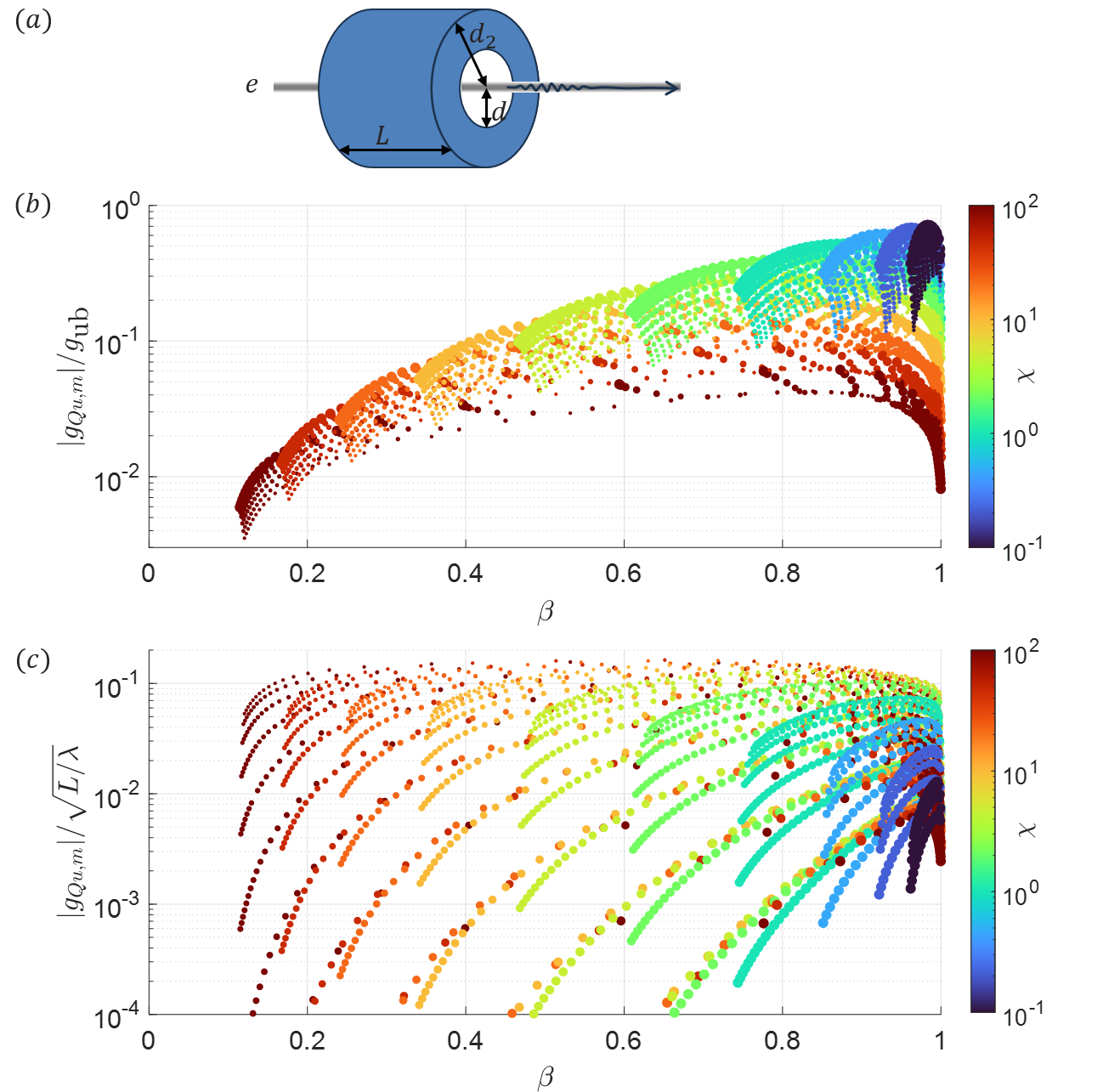}
    \caption{(a) Schematic of the interaction between the free electron and the guided mode in a dielectric hollow-core waveguide. The annular cylindrical waveguide is concentric with the free-electron trajectory. The inner and outer radius of the hollow-core waveguide are $d$ and $d_2$ respectively. The interaction length is $L$. (b) The ratio between the quantum coupling coefficient $|g_{Qu,m}|$ and the upper bound $g_\textrm{ub}$, as a function of the normalized free-electron velocity $\beta$. (c) The interaction length normalized quantum coupling coefficient. The phase-matching condition requires that the phase velocity of the guided mode matches the free-electron velocity. The color represents the susceptibility of the dielectric medium. The size of the dots represents $d$, which varies from $0.01\lambda$ (smallest dots) to $\lambda$ (largest dots). For the upper bound, the design region is the same annular cylinder, as depicted in (a), such that the dielectric medium (with susceptibility $\chi$) can take any structures enclosed in the design region. }
    \label{fig:hollow_core_wg}
\end{figure}

The free-electron--light interaction with guided modes in a dielectric hollow-core waveguide is shown schematically in Fig.\,\ref{fig:hollow_core_wg}(a). The annular cylindrical waveguide is concentric with the free-electron trajectory. The inner and outer radius of the hollow-core waveguide are $d$ and the $d_2$ respectively. The interaction length is $L$. The susceptibility of the dielectric material is $\chi = \epsilon-1$. In such a dielectric hollow-core waveguide, the guided mode that couples efficiently with the free electron traveling on axis is the transverse magnetic mode ($\textrm{TM}_{01}$). The nonzero components of the $\textrm{TM}_{01}$ mode $(E_\rho, \tilde{H}_\phi, E_z)$, where $\tilde{\boldsymbol{H}} = \sqrt{\frac{\mu_0}{\epsilon_0}}\boldsymbol{H}$, take the form:
\begin{equation}
    \label{eq:hollow_core_mode_1}
    (E_\rho, \tilde{H}_\phi, E_z) = (e_\rho(\rho), h_\phi(\rho), e_z(\rho))\exp(ik_z z).
\end{equation}
By solving the Maxwell's equation, one can obtain the explicit form of the guided mode \cite{pollock1995fundamentals}.
\begin{equation}
    \label{eq:hollow_core_e_z}
    e_z(\rho) = 
    \begin{cases} 
    \mathcal{A} I_0(\alpha\rho)/I_0(\alpha\rho) & \rho \leq d, \\
    \mathcal{B} Y_0(\kappa \rho)/Y_0(\kappa d) + \mathcal{C} J_0(\kappa \rho)/J_0(\kappa d_2) & d < \rho \leq d_2, \\
    \mathcal{D} K_0(\alpha\rho)/K_0(\alpha d_2) & \rho > d_2,
    \end{cases}
\end{equation}
where $\alpha = \sqrt{k_z^2 - k^2}$, $\kappa = \sqrt{\epsilon k^2 - k_z^2}$, $J_0$ and $Y_0$ are the 0-th order Bessel functions of the first and second kind, $I_0$ and $K_0$ are the 0-th order modified Bessel functions of the first and second kind, and $\mathcal{A}$, $\mathcal{B}$, $\mathcal{C}$, and $\mathcal{D}$ are coefficients. The transverse fields ($e_\rho$ and $h_\phi$) can be derived from the longitudinal field ($e_z$).
\begin{equation}
    \label{eq:hollow_core_h_phi}
    h_\phi(\rho) = 
    \begin{cases}
    -i \frac{k}{\alpha} \mathcal{A} \frac{I_0'(\alpha \rho)}{I_0(\alpha d)} & \rho \leq d, \\
    i\frac{k}{\kappa}\epsilon \Big[\mathcal{B} \frac{Y_0'(\kappa \rho)}{Y_0(\kappa d)} + \mathcal{C} \frac{J_0'(\kappa \rho)}{J_0(\kappa d_2)} \Big] & d<\rho\leq d_2, \\
    -i \frac{k}{\alpha} \mathcal{D} \frac{K_0'(\alpha \rho)}{K_0(\alpha d_2)} & \rho > d_2.
    \end{cases}
\end{equation}
\begin{equation}
    \label{eq:hollow_core_e_rho}
    e_\rho(\rho) = 
    \begin{cases}
    -i \frac{k_z}{\alpha} \mathcal{A} \frac{I_0'(\alpha \rho)}{I_0(\alpha d)} & \rho \leq d, \\
    i\frac{k_z}{\kappa} \Big[\mathcal{B} \frac{Y_0'(\kappa \rho)}{Y_0(\kappa d)} + \mathcal{C} \frac{J_0'(\kappa \rho)}{J_0(\kappa d_2)} \Big] & d<\rho\leq d_2, \\
    -i \frac{k_z}{\alpha} \mathcal{D} \frac{K_0'(\alpha \rho)}{K_0(\alpha d_2)} & \rho > d_2.
    \end{cases}
\end{equation}
Since $e_z$ and $h_\phi$ are continuous across the boundaries, one can get the equations to solve for $\mathcal{A}$, $\mathcal{B}$, $\mathcal{C}$,  $\mathcal{D}$, and the dispersion relation. Using the relations $J_0'(x) = -J_1(x)$, $Y_0'(x) = -Y_1(x)$, $I_0'(x) = I_1(x)$, and $K_0'(x) = -K_1(x)$, the dispersion relation is the solution of the following equation:
\begin{equation}
    \label{eq:hollow_core_dispersion_eq}
    \frac{\frac{\alpha}{\kappa}\epsilon [Y_1(\kappa d)J_1(\kappa d_2) - J_1(\kappa d) Y_1(\kappa d_2)] + \frac{I_1(\alpha d)}{I_0(\alpha d)}[-Y_0(\kappa d)J_1(\kappa d_2) + J_0(\kappa d) Y_1(\kappa d_2)]}{[-J_1(\kappa d)Y_0(\kappa d_2) + Y_1(\kappa d)J_0(\kappa d_2)] + \frac{\kappa}{\epsilon \alpha}\frac{I_1(\alpha d)}{I_0(\alpha d)} [J_0(\kappa d)Y_0(\kappa d_2) - Y_0(\kappa d)J_0(\kappa d_2)]} = -\frac{K_1(\alpha d_2)}{K_0(\alpha d_2)}.
\end{equation}
The solution $k_z$ would determine the corresponding electron velocity under the phase-matching condition, i.e., $\beta = k/k_z$. With the guided mode distribution, one can obtain the quantum coupling coefficient using Eq.\,\ref{eq:gQu_mode_m_2}.

For the upper bound, I choose the design region ($R$) to be the same annular cylinder, as depicted in Fig.\,\ref{fig:hollow_core_wg}(a). The dielectric medium with suscpetibility $\chi$ can take any structures enclosed within the design region. Thus, the upper bound of the quantum coupling coefficient can be calculated using Eq.\,\ref{eq:gQu_mode_m_6} (or Eq.\,22 of the main text).

Figures \ref{fig:hollow_core_wg}(b) and \ref{fig:hollow_core_wg}(c) show the ratio between the quantum coupling coefficient and the upper bound, i.e., $|g_{Qu,m}|/g_\textrm{ub}$, and the interaction length normalized quantum coupling coefficient, i.e., $|g_{Qu,m}|/\sqrt{L/\lambda}$, respectively. I scan the susceptibility $\chi$ of the optical medium (10 sampling points from $10^{-1}$ to $10^{2}$ with log-scale sampling), the inner radius $d$ of the hollow core waveguide (11 sampling points from $0.01\lambda$ to $\lambda$ with log-scale sampling), and the outer radius $d_2$ of the hollow core waveguide (61 sampling points from $d+0.1\lambda/\sqrt{\epsilon}$ to $d+\lambda/\sqrt{\chi}$ with log-scale sampling). In Figs.\,\ref{fig:hollow_core_wg}(b) and \ref{fig:hollow_core_wg}(c), the color of the dots represents the susceptibility of the dielectric medium. The size of the dots is related to $d$, where the smaller dot represents smaller $d$. 

Although larger susceptibility and smaller separation distance lead to larger quantum coupling coefficient, the ratio between the quantum coupling coefficient and the upper bound typically decreases with larger susceptibility and smaller separation distance. On the other hand, with small susceptibility $\chi=0.1$ and relatively large separation distance $d=\lambda$, the quantum coupling coefficient $|g_{Qu,m}|$ reaches 72\% of the corresponding upper bound. It indicates that, with small susceptibility and relatively large separation distance, the hollow-core waveguide is close to the optimal structure to maximize the coupling between free electrons and photos. However, with large susceptibility and small separation distance, such simple hollow-core waveguide structures are far from optimum, while complex structures might improve the quantum coupling coefficient further.

\subsection{Free-electron--light interaction in a metallic hole}

\begin{figure}
    \centering
    \includegraphics[width=13cm]{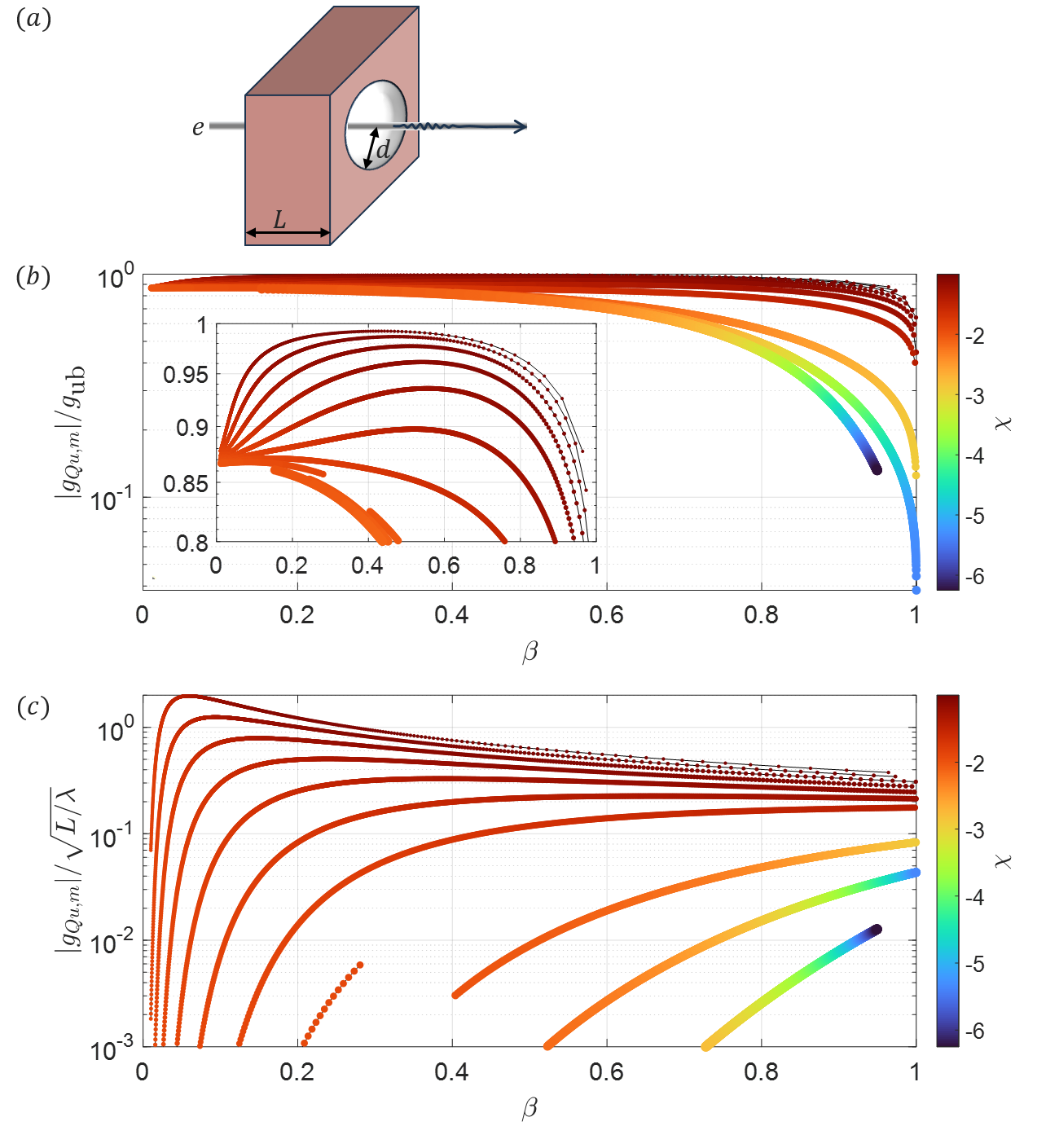}
    \caption{(a) Schematic of the interaction between the free electron and the SPP mode in a metallic hole. The cylindrical hole is concentric with the free-electron trajectory. The radius of the hole is $d$. The interaction length is $L$. (b) The ratio between the quantum coupling coefficient $|g_{Qu,m}|$ and the upper bound $g_\textrm{ub}$, as a function of the normalized free-electron velocity $\beta$. The insert is an enlarged view for the ratio between 0.8 and 1. (c) The interaction length normalized quantum coupling coefficient. The phase-matching condition requires that the phase velocity of the SPP mode matches the free-electron velocity. The color represents the susceptibility of the metallic medium. The size of the dots represents $d$, which varies from $0.01\lambda$ (smallest dots) to $\lambda$ (largest dots). Dots representing the same $d$ are connected by thin black lines to guide the eyes. For the upper bound, the design region is the same region outside the cylinder, as depicted in (a), such that the metallic medium (with susceptibility $\chi$) can take any structures enclosed in the design region. }
    \label{fig:metallic_hole}
\end{figure}

The free-electron--light interaction with the SPP modes in a metallic hole is shown schematically in Fig.\,\ref{fig:metallic_hole}(a). The cylindrical hole is concentric with the free-electron trajectory. The radius of the hole is $d$. The interaction length is $L$. I assume that the interaction length is much larger than the hole radius and neglect the effects of the entrance and the exit of the hole. The dispersion of the metal is described by the Drude model, which can be regarded as a limiting case of the Lorentz model. The relative permittivity is $\epsilon(\omega) = 1 - \omega_p^2/(\omega^2 + i\omega\gamma_L)$. For simplicity, I assume the material absorption is negligible, i.e., $\gamma_L\rightarrow 0$. In this limit, the susceptibility is $\chi(\omega) = \epsilon(\omega) - 1 = -\omega_p^2/\omega^2$. The metallic hole can support a SPP mode, with transverse magnetic (TM) polarization, which can couple efficiently with the free electron traveling on axis. The nonzero components of the SPP mode, which are $E_\rho$, $\tilde{H}_\phi$, and $E_z$, take the form:
\begin{equation}
    \label{eq:metallic_hole_mode_1}
    (E_\rho, \tilde{H}_\phi, E_z) = (e_\rho(\rho), h_\phi(\rho), e_z(\rho))\exp(ik_z z).
\end{equation}
By solving the Maxwell's equation, one can obtain the explicit form of the guided mode.
\begin{equation}
    \label{eq:metallic_hole_e_z}
    e_z(\rho) = 
    \begin{cases} 
    I_0(\alpha\rho)/I_0(\alpha\rho) & \rho \leq d, \\
    K_0(\kappa\rho)/K_0(\kappa d) & \rho > d,
    \end{cases}
\end{equation}
where $\alpha = \sqrt{k_z^2 - k^2}$, $\kappa = \sqrt{k_z^2 - \epsilon k^2}$. The definition of $\kappa$ here is different from that in the dielectric hollow-core waveguide. Since the SPP mode exists when $\epsilon < 0$, i.e., $\omega < \omega_p$, the $\kappa$ defined here is real. The transverse fields ($e_\rho$ and $h_\phi$) can be derived from the longitudinal field ($e_z$). 
\begin{equation}
    \label{eq:metallic_hole_h_phi}
    h_\phi(\rho) = 
    \begin{cases}
    -i\frac{k}{\alpha}\frac{I_0'(\alpha \rho)}{I_0(\alpha d)} & \rho \leq d, \\
    -i\frac{k}{\kappa} \epsilon \frac{K_0'(\kappa \rho)}{K_0(\kappa d)}  & \rho > d.
    \end{cases}
\end{equation}
\begin{equation}
    \label{eq:metallic_hole_e_rho}
    e_\rho(\rho) = 
    \begin{cases}
    -i\frac{k_z}{\alpha} \frac{I_0'(\alpha \rho)}{I_0(\alpha d)} & \rho \leq d, \\
    -i \frac{k_z}{\kappa} \frac{K_0'(\kappa \rho)}{K_0(\kappa d)} & \rho > d.
    \end{cases}
\end{equation}
The form of $e_z$ (Eq.\,\ref{eq:metallic_hole_e_z}) already utilizes the constraint that $E_z$ is continuous across the boundary. Furthermore, by requiring $h_\phi$ to be continuous across the boundary, one can get the equation to solve for the dispersion relation:
\begin{equation}
    \label{eq:metallic_hole_dispersion}
    \frac{1}{\alpha} \frac{I_1(\alpha d)}{I_0(\alpha d)} = -\frac{\epsilon}{\kappa} \frac{K_1(\kappa d)}{K_0(\kappa d)}.
\end{equation}
For certain choices of parameters, including the mode frequency ($\omega$), the plasma frequency ($\omega_p$), and the hole radius ($d$), there exists a real solution to Eq.\,\ref{eq:metallic_hole_dispersion}. Then, the metallic hole supports such a SPP mode. This solution $k_z$ would determine the corresponding electron velocity under the phase-matching condition, i.e., $\beta = k/k_z$. With the SPP mode distribution, one can obtain the quantum coupling coefficient using Eq.\,\ref{eq:gQu_mode_Lorentz_dispersion}.

For the upper bound, I choose the design region ($R$) to be the same region outside the cylinder, as depicted in Fig.\,\ref{fig:metallic_hole}(a). The metallic medium with suscpetibility $\chi(\omega)= - \omega_p^2/\omega^2$ can take any structures enclosed within the design region. Thus, the upper bound of the quantum coupling coefficient can be calculated using Eq.\,\ref{eq:gQu_mode_Lorentz_dispersion_6} with $\omega_0 = 0$.

Figures \ref{fig:metallic_hole}(b) and \ref{fig:metallic_hole}(c) show the ratio between the quantum coupling coefficient and the upper bound, i.e., $|g_{Qu,m}|/g_\textrm{ub}$, and the interaction length normalized quantum coupling coefficient, i.e., $|g_{Qu,m}|/\sqrt{L/\lambda}$, respectively.
To scan the parameter space, I fix the mode frequency ($\omega_m = 2\pi c/\lambda$) and scan the hole radius ($d$) and the plasma frequency ($\omega_p$), which determines the metal susceptibility.
The sampling of the radius $d$ is the same as that in Fig.\,\ref{fig:hollow_core_wg} (11 sampling points from $0.01\lambda$ to $\lambda$ with log-scale sampling). The plasma frequency ($\omega_p$) is sampled from $\omega_m$ to $2.5\omega_m$ with 3000 sampling points and a log-scale sampling. In Figs.\,\ref{fig:metallic_hole}(b) and \ref{fig:metallic_hole}(c), the color of the dots represents the susceptibility of the metal. The size of the dots is related to $d$, where the smaller dot represents smaller $d$. 

From Fig.\,\ref{fig:metallic_hole}(b), the ratio $|g_{Qu,m}|/g_\textrm{ub}$ is high  for a large range of electron velocity ($\beta$) and separation distance ($d$). When the hole radius ($d$) is below about $0.25\lambda$, the quantum coupling coefficient can reach above 50\% of the upper bound almost for arbitrary free-electron velocities, as long as the phase-matching condition is satisfied. Furthermore, as presented in the enlarged view in Fig.\,\ref{fig:metallic_hole}(b), the ratio $|g_{Qu,m}|/g_\textrm{ub}$ is above 99\% for $d=0.01\lambda$ and $\beta$ around 0.4. 
This suggests that the simple structure as a cylindrical hole in a metal is close to the optimum for the purpose of approaching the upper bound of the free-electron--light coupling with such a metallic medium. This also implies the tightness of the analytical upper bound. 

Figure \ref{fig:metallic_hole}(c) demonstrates the interaction length normalized quantum coupling coefficient. For a fixed hole radius and tunable plasma frequency, the quantum coupling coefficient, as a function of free-electron velocity, shows a subrelativistic peak when the hole radius is small ($d < 0.1\lambda$) and no subrelativistic peak when the hole radius is large ($d > 0.1\lambda$). The qualitative behaviour is consistent with the $d$ and $\beta$ dependence of the geometric factor ($g_\textrm{geo}$), as illustrated in Fig.\,\ref{fig:g_geo_cyl}(b) and Figs.\,2(a-b) of the main text. This suggests that the subrelativistic peak of $|g_{Qu,m}|$ for small separation distance, as shown in Fig.\,2(b) of the main text, is not an artifact of the derivation of the upper bound, but has a physical ground.   
Moreover, the quantum coupling coefficients with SPP modes in a metallic hole can be much higher than those in a dielectric hollow-core waveguide for small separation distance ($d<0.1\lambda$). This suggests that the metallic hole can be a valuable platform for strong free-electron--light interaction, especially when the material loss is low. 

The large free-electron--light coupling with SPP modes in a metallic hole is consistent with previously reported large free-electron--light coupling with SPP modes on planar metals \cite{adiv2023observation}. However, in contrast to SPP modes at a planar metal-vacuum interface, where the SPP modes only exist when the frequency is below the surface plasmon frequency ($\omega < \omega_\textrm{sp}=\omega_p/\sqrt{2}$), the SPP modes in a metallic hole can exist above the surface plsamon frequency ($\omega_\textrm{sp}$) and below the plasma frequency ($\omega_p$), when the hole radius is small (more explicitly, $d < 0.27 \lambda$).  
Figure \ref{fig:metallic_hole}(b) shows that smaller $d$ can reach a higher percentage of the upper bound, where the corresponding SPP mode frequency is between the surface plasmon frequency ($\omega_p/\sqrt{2}$) and the plasma frequency ($\omega_p$). Interestingly, as the SPP mode frequency approaches the surface plasmon frequency ($\omega_\textrm{sp} = \omega_p/\sqrt{2}$), and the SPP mode wave vector approaches infinity, corresponding to $\beta\rightarrow 0$, $|g_{Qu,m}|/g_\textrm{ub} \rightarrow \sqrt{3}/2$. 
Moreover, with the hole radius as a design parameter, the metallic hole system has the potential to satisfy the phase-matching condition with a large range of free-electron velocities, making it a versatile platform to study free-electron--light interaction. 

\bibliography{DLA_quantum.bib}{}
\bibliographystyle{apsrev4-1}